\documentclass[reprint,amsmath,amssymb,aps,onecolumn]{revtex4-2}

\usepackage{amssymb}
\usepackage{amsmath}

\usepackage{todonotes}

\usepackage{blindtext}
\usepackage{multirow}
\usepackage{graphicx}

\newcommand{\norm}[1]{\left\lVert#1\right\rVert}

\usepackage{ulem}
\usepackage{hyperref}
\usepackage{booktabs}

\begin{document}

\title{Multilayer Network Science: from Cells to Societies}

\author{Oriol Artime$^1$, Barbara Benigni$^1$, Giulia Bertagnolli$^1$, Valeria d'Andrea$^1$, Riccardo Gallotti$^1$, Arsham Ghavasieh$^1$, Sebastian Raimondo$^1$, Manlio De Domenico$^{1,2}$}

\affiliation{$^1$Complex Multilayer Networks Lab, Fondazione Bruno Kessler, Via Sommarive 18, 38123 Povo (TN), Italy}
\affiliation{$^2$Physics and Astronomy Department, University of Padova, via Marzolo 8, 35131, Padova, Italy}

\date{23 August 2022}

\begin{abstract}
Networks are convenient mathematical models to represent the structure of complex systems, from cells to societies. In the past decade, multilayer network science -- the branch of the field dealing with units interacting in multiple distinct ways, simultaneously -- was demonstrated to be an effective modeling and analytical framework for a wide spectrum of empirical systems, from biopolymer networks (such as interactome and metabolomes) to neuronal networks (such as connectomes), from social networks to urban and transportation networks. In this Element, a decade after the publication of one of the most seminal papers on this topic, we review the most salient features of multilayer network science, covering both theoretical aspects and direct applications to real-world coupled/interdependent systems, from the point of view of multilayer structure, dynamics, and function. We discuss potential frontiers for this topic and the corresponding challenges in the field for the future.
\end{abstract}

\maketitle

%%%%%%%%%%%%%%%%%%%%%%%%%%%%%
%%%%%%%%%%%%%%%%%%%%%%%%%%%%%
%%%%%%%%%%%%%%%%%%%%%%%%%%%%%
\section{Introduction}
What is a complex system? It is a \emph{network} of actors or units related by special types of relationships or interactions which, together, form a whole. Being two proteins within a cell or two individuals within a social group, relationships and interactions tie together the units in such a way that ``the whole is larger than the sum of its parts'', a concept initially introduced by the Greek philosopher Aristotle and later exploited by Gestalt psychologists, at the end of 19th century, to explain human perception beyond the traditional atomistic view.

In fact, the ``whole'' exhibits features that each actor or unit, in isolation, does not (and also, could not) exhibit: therefore, it is usually difficult, if not impossible, to understand a system from the analysis of its components alone, like in atomistic or other reductionist theories. The framework required to study such relationships and interactions is known as Network Science\footnote{We refer the reader to this interesting, non-technical, and recent introduction to the basic concepts characterizing complex systems~\citep{complexity_explained_2019}.}.

The foundations of Network Science can be found in the pioneering work by Leonhard Euler in 1736, when the famous mathematician provided the first mathematically grounded proof to solve, definitively, the problem of the Seven Bridges of K\"{o}nigsberg. He mapped the empirical problem of traversing the city of K\"{o}nigsberg -- under the constraint that one should use each one of its seven bridges only one time -- to the abstract problem of performing a special walk through a graph. Since Euler's solution, graph theory quickly developed in the successive two centuries, culminating in the pioneering works by Paul Erd\H{o}s and Alfr\'{e}d R\'{e}nyi on random graphs and their statistical analysis at the end of the 50's of the past century.

\begin{figure}[!ht]
\centering
\includegraphics[width=1.1\textwidth]{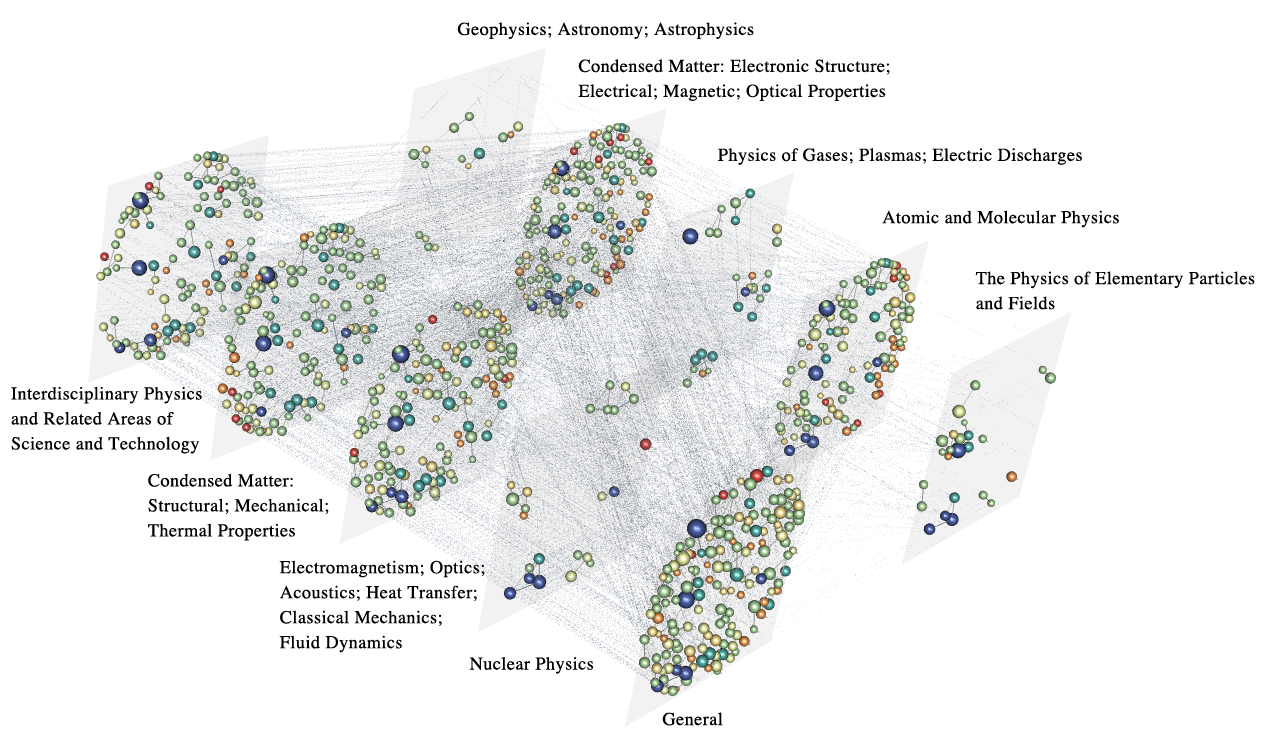}
\caption{\label{fig:chap_hum_concepts} Multilayer representation of a co-authorship network. Nodes represent authors publishing papers in the journals of the American Physical Society, links connect two authors if they have published a paper together. Layers encode distinct sub-topics of physics (e.g., Geophysics or Nuclear Physics): links within the same layer represent co-authorship of one paper about the same topic, while links between layers indicate co-authorship of one paper categorized simultaneously across distinct topics. Figure from~\citep{de2020illustrations} under Creative Commons Attribution-ShareAlike 4.0 International License.}
\end{figure}

For decades, graph theory has been widely used by social scientists and (systems) biologists to map connections between individuals and biological units, respectively, to gain novel insights about the \emph{properties of a system}, the relevance of a \emph{unit within the system} and the \emph{organization of units within the system}. In 1974, Fran\c{c}ois Jacob, Nobel Prize in Physiology or Medicine in 1965, described biology as the science dealing, effectively, with systems within systems~\citep{trewavas2006brief}, well before the age of genomics and large-scale biology. He recognized that biological systems can be mapped as well to units of systems at a larger scale: in fact, protein interact with each other to make the cell function, cells interact with each other to make tissues and organs, which in turn interact with each other to build an organism. Finally, at the top of this hierarchical web of interactions, organisms interact with each other to define a population, like our society. In the same decade, similar ideas regarding the non-trivial interdependencies between scales were laid out by the 1977 Nobel Laureate in Physics Phillip Anderson, in the context of natural sciences~\citep{anderson1972more}.

Social scientists, as biologists, were among the first to face the existence of multiple levels (or scales) as well as multiple \emph{layers} of descriptions for the units of a social system. In the early 70's of the 20th century, Wayne W. Zachary observed the interactions within a group of individuals belonging to a Karate club, over 3 years~\citep{zachary1977information}, to understand the dynamics of conflicts which allowed him to predict the outcome of the group split, happened later. 
He annotated interactions across eight distinct contexts, from ``the association in and between academic classes at the university'' to ``attendance at intercollegiate karate tournaments held at local universities''. However, at that time, the mathematical framework required to study a network with multiple \emph{layers of complexity} -- such as the eight contexts --  was not yet developed, and Zachary opted for an approximation: aggregating the multiple interactions across contexts into a single representative number denoting the intensity of the relationship between a pair of actors. 

\begin{figure}[!ht]
\centering
\includegraphics[width=0.6\textwidth]{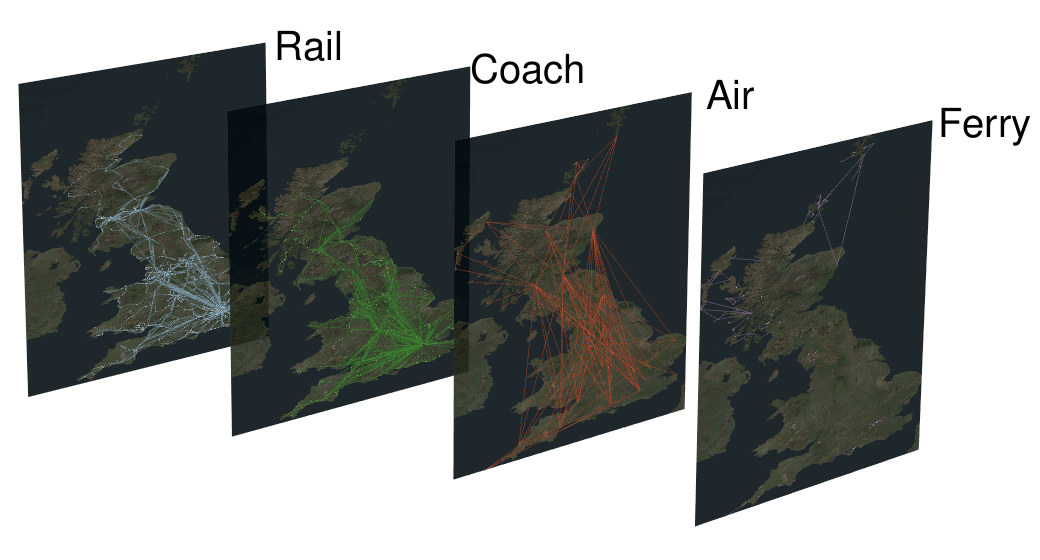}
\caption{\label{fig:chap_uk_transport} A multilayer transportation network, where connections using a particular means of transport are associated with intra-layer links and inter-modal exchanges are represented the inter-layer links. Here, the national public transportation network for Great Britain~\citep{gallotti2014anatomy,gallotti2015multilayer} as rendered by MuxViz~\citep{de2015muxviz}. (Figure from \citep{gallotti2015multilayer}). }
\end{figure}

\begin{figure}[!ht]
\centering
\includegraphics[width=0.65\textwidth]{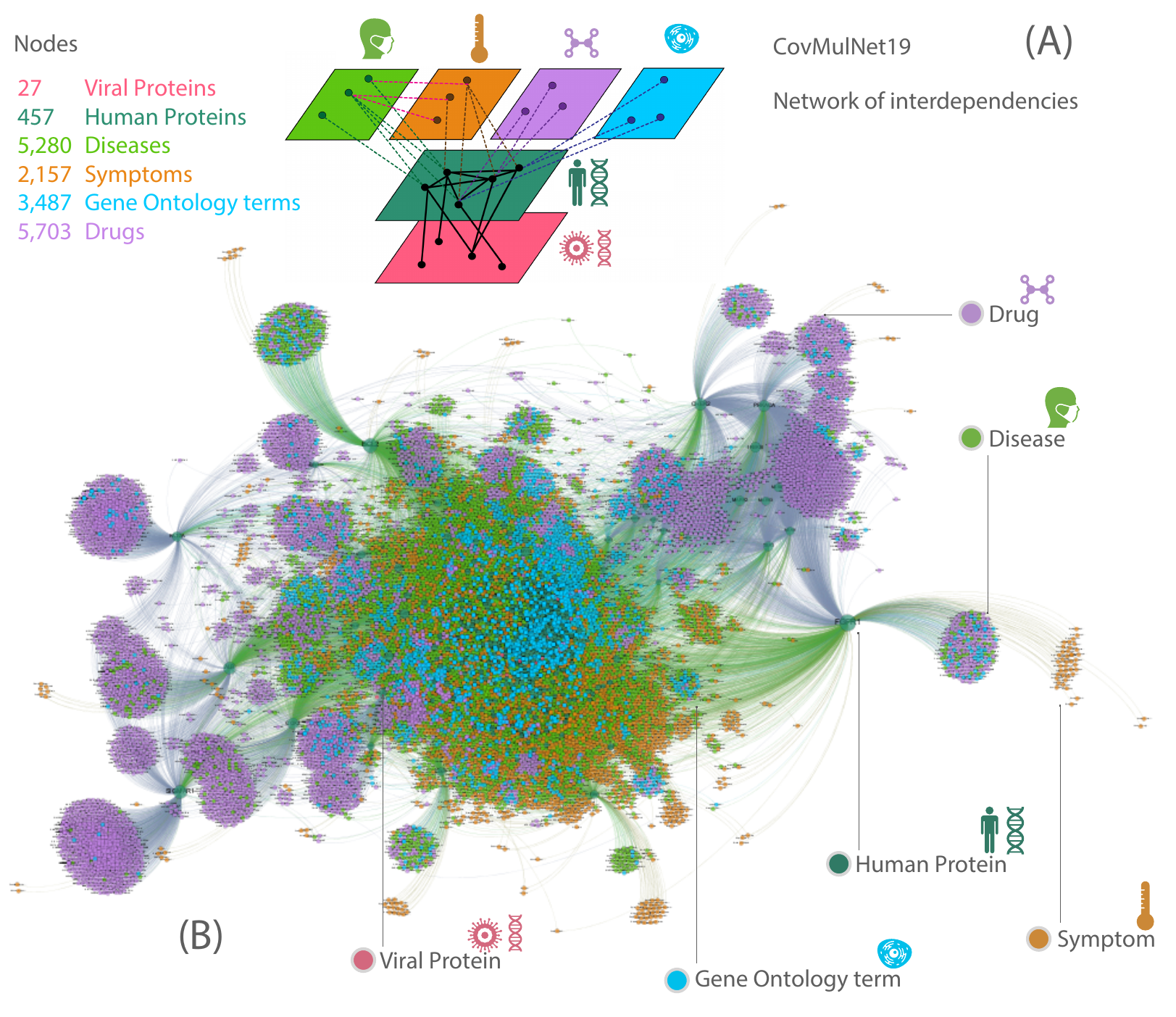}
\caption{\label{fig:chap_covmulnet19} 
Illustration of CovMulNet19, the multilayer network encoding COVID--19 Genotype-Phenotype-Drug interactions. A schematic map of intra- and inter-layer interdependencies between diseases, symptoms, drugs, Gene Ontology terms, human proteins and viral proteins of SARS-CoV-2, the virus of COVID--19. Figure from~\citep{Verstraete2020}.}
\end{figure}

In this work, we will better understand the challenges faced by systems biologists and social scientists between the 70's and the past decade, while introducing the basic concepts required to define the framework of \emph{multilayer network science} with an interdisciplinary language which should be familiar to biologists, social scientists, computer scientists, applied mathematicians and physicists. Therefore, it will become clear that, for instance, Zachary's approach was a possible model to study the Karate club network, but likely neither the most accurate nor the most predictive one. We will discuss under which conditions a system admits a multilayer representation, providing examples such as the ones shown in Figs.~\ref{fig:chap_hum_concepts} and \ref{fig:chap_uk_transport}, where units are individuals and geographic areas, respectively, and interactions represent co-authorship of scientific papers and transportation routes, respectively. Another emblematic example, accounting for the temporal and socio-spatial interdependence typical of many systems, concerns the organization of ecological systems~\citep{pilosof2017multilayer}. Finally, very recently, multilayer modeling in systems biology and medicine has been used to integrate information about biological processes, drug-target, genotype and phenotype to the subset of the human interactome targeted by SARS-CoV-2, the virus of the COVID-19~\citep{Verstraete2020} (see Fig.~\ref{fig:chap_covmulnet19}). 

This work is full of examples like the ones above and we hope to make clear the broad spectrum of potential interdisciplinary applications of the multilayer framework.

Our ultimate goal is to guide the reader through the potential applications of multilayer modeling, which nowadays provides a well-established paradigm for the analysis of systems characterized by multiple levels and multiple layers of description, including systems whose structure changes over time, with the aim of providing the reader with the tools required to model and analyze their systems in terms of coupled layers, as well as with the conditions under which this approach is plausible or not.

It is worth remarking here that this work should be considered as an extended introduction to the field but, at the same time, not the most complete one. For this reason, we point the reader to the first reviews~\citep{kivela2014multilayer,boccaletti2014structure,wang2015evolutionary,de2016physics,battiston2017new} and recent books~\citep{bianconi2018multilayer,cozzo2018multiplex} on this topic or more specifically on analysis and visualization of multilayer networks~\citep{de2021multilayer}, which, taken together with our work, will provide a more comprehensive view of the field.

In Chap.~2 we will introduce the representation of multilayer networks based on the tensorial formulation~\citep{de2013mathematical}, providing the mathematical ground for the analytical techniques for structure (Chap.~3) and dynamics (Chap.~4), allowing the reader to find a reference for the analysis of versatility (or multilayer centrality) and mesoscale organization (or community detection), as well as for percolation, synchronization, competition and modeling of intertwined phenomena. Towards the end (Chap.~5), we will discuss a few selected advances in network science -- namely the latent geometry of a complex network based on network-driven processes and the statistical theory of information dynamics leading to the formalism of network density matrices -- and will discuss their recent generalization and application to multilayer networks. Finally (Chap.~6), we will show how multilayer networks are ubiquitous and can be used for modeling complex systems, from cells to societies.

%%%%%%%%%%%%%%%%%%%%%%%%%%%%%
%%%%%%%%%%%%%%%%%%%%%%%%%%%%%
%%%%%%%%%%%%%%%%%%%%%%%%%%%%%
\section{Representation of multilayer systems} 
\subsection{Tensorial representation of a complex network}\label{sec:representation}

One convenient way to represent, mathematically, a complex network is by means of its adjacency matrix~\citep{barrat2008dynamical,newman2018networks,estrada2012structure,barabasi2016network,latora2017complex}. However, to deal with multilayer networks, it might be more convenient to introduce first the more general concept of tensor, a multilinear function which maps objects defined in a vector space into other objects of the same type, regardless of the choice of a coordinate system. For instance, a simple scalar $x$ is also a rank-0 tensor, a vector $x_i$ is a rank-1 tensor and a matrix $X_{ij}$ is a rank-2 tensor. More generally, given a vector space $\mathcal{V}$ with algebraic dual space\footnote{This is the space of all the possible linear transformations that map an object of $\mathcal{V}$ into a real number. For instance, think about $\mathcal{V}=\mathbb{R}^2$ and the linear functional $f: \mathbb{R}^2\longrightarrow \mathbb{R}$: it follows that $f(x,y)=ax+by$, with $a,b$ two integer numbers, is an element of $\mathcal{V}^{\star}$.} 
$\mathcal{V}^\star$ over the real numbers $\mathbb{R}$, one can define the tensor $M$ as the multilinear function
\begin{eqnarray}
M: \mathcal{V}^{\star}\times\mathcal{V}^{\star}\times...\mathcal{V}^{\star} \times \mathcal{V}\times\mathcal{V}\times...\mathcal{V} \longrightarrow \mathbb{R},
\end{eqnarray}
where the number of products is $m$ for the vector space and $n$ for its dual. This definition formally characterizes a rank-$mn$ tensor $M^{i_1 i_2 ... i_n}_{j_1 j_2 ... j_m}$ which is $m$-covariant and $n$-contravariant: in fact, under a change of basis $B$, $m$ components transform as the same linear mapping of the change of basis ($B$), whereas $n$ components transform as the inverse one ($B^{-1}$). Therefore, in general, there are two types of canonical basis: the covariant basis denoted by $e_{i}(a)$ ($a=1,2,...,m$) which is defined in $\mathcal{V}$, and the contravariant (or dual) basis denoted by $e^{i}(b)$ ($b=1,2,...,n$) which is defined in $\mathcal{V}^{\star}$. If the vector space is Euclidean, the coordinates of the canonical vectors and their duals are the same, whereas this is not the case in general. In the following, to define an adjacency matrix, or a rank-2 adjacency tensor, we will work in the Euclidean space but we will keep the covariant and contravariant notation, since it will allow us to generalize the results to the case of non-Euclidean spaces. The interested reader can find more about the tensorial framework in any good linear algebra textbook, while for the purpose of this work it is sufficient to understand how we can use tensors in practice in a few key situations.

Let us start by better defining the canonical vectors in the case of networks. For a graph with $N$ nodes, the canonical covariant vectors $e_{i}(a)$ defined in the space of nodes $\mathbb{R}^{N}$ are $N$ rank-1 tensors of dimension $N$ with all entries equal to 0 except for the $a$-th entry, which is equal to 1. Similarly for canonical contravariant vectors. The product of canonical vectors gives canonical matrices, e.g., $E_{ij}(ab)=e_{i}(a)e_{j}(b)$ is a rank-2 covariant tensor with all components equal to zero except for the one corresponding to the $a$-th row and the $b$-th column, equal to 1. Similarly, we can build contravariant tensors and mixed tensors, i.e., tensors obtained by the product between the covariant and contravariant vectors. 

The careful reader has noticed, at this point, that we have defined the \emph{outer} product of two canonical vectors, also known as the Kronecker product, which gives a rank-2 tensor as a result. This result is general: the outer product of two tensors $X$ and $Y$ is a new tensor $Z$ with a number of covariant (contravariant) indices given by the sum of the number of covariant (contravariant) indices of $X$ and $Y$. Therefore, the outer product of two tensors is always a tensor of higher order than the original ones: e.g., $X_{ij}^{k}Y_{l}^{mn}=Z_{ijl}^{kmn}$.

It is possible to define also an \emph{inner} product: in this case, we talk about a contraction because the rank of resulting tensor is reduced by two units. For instance, this is the case in the product $X_{ij}^{k}Y_{k}^{mn}=Z_{ij}^{mn}$, where the index $k$ is covariant for $X$ and contravariant for $Y$. This operation corresponds to summing over the components of $X$ and $Y$ identified by the index $k$. The careful reader has noticed that we have omitted the summation symbol: this choice -- known as Einstein summation convention -- is optional and often adopted for sake of simplicity. In the following, we will make use of this convention.

At this point, we are ready to define the adjacency tensor of a complex network in terms of canonical vectors~\citep{de2013mathematical} as
\begin{eqnarray}
W^{i}_{j}=\sum_{a,b=1}^{N}w_{ab}e^{i}(a)e_{j}(b)=\sum_{a,b=1}^{N}w_{ab}E^{i}_{j}(ab),
\end{eqnarray}
where $w_{ab}$ is a real number, usually non-negative, used to encode the intensity of the interaction between nodes $a$ and $b$, while $E^{i}_{j}(ab)\in\mathbb{R}^{N\times N}$ are the mixed canonical rank-2 tensors. We might wonder if $W^{i}_{j}$ is a true tensor, or just a matrix. To this end, it is enough to understand how it transforms under a change of basis
\begin{eqnarray}
B^{i}_{j}=\sum_{a=1}^{N}e'^{i}(a)e_{j}(a),
\end{eqnarray}
a linear function which transforms the basis vector set $\{e^{i}(a)\}$ into a second set $\{e'^{i}(a)\}$. By noting that $w_{ab}$ must be invariant with respect to the change of basis, we have: 
\begin{eqnarray}
{W'}^{k}_{l}&=&\sum_{a,b=1}^{N}w_{ab}e'^{k}(a)e'_{l}(b)=\sum_{a,b=1}^{N}w_{ab}B^{k}_{i}e^{i}(a)e_{j}(b)(B^{-1})^{j}_{l}\nonumber\\
&=&B^{k}_{i}\left[\sum_{a,b=1}^{N}w_{ab}e^{i}(a)e_{j}(b)\right](B^{-1})^{j}_{l}=B^{k}_{i}W^{i}_{j}(B^{-1})^{j}_{l},
\end{eqnarray}
i.e., the adjacency object $W^{i}_{j}$ transforms like a tensor~\citep{de2015ranking}. This result is important since a tensor is an object with features that, in general, are not shared by a matrix or, at higher orders, a hypermatrix. In fact, the components of a tensor can be always arranged into hypermatrices, while the opposite is not necessarily true.

Since we work in the Euclidean space, we might wonder why we use this notation and not a simpler one. In general, this is convenient because of the presence of directed relationships between nodes: to distinguish between incoming and outgoing directions, it is sufficient to map this information into covariant and contravariant indices in such a way that the adjacency tensor $W^{i}_{j}$ represents a linear transformation which maps nodes into a function of their incoming or outgoing flow. For instance, node $a$ is represented by $e_i(a)$ in the space of nodes and $W^{i}_{j}e_{i}(a)=w_{j}(a)$ provides a rank-1 tensor encoding the set of nodes linked by $a$, while $W^{i}_{j}u_{i}=s_{j}$, with $u_{i}$ the rank-1 tensor with all components equal to 1, provides a rank-1 tensor encoding the outgoing strength of all nodes. Similarly, $W^{i}_{j}e^{j}(a)=w^{i}(a)$ gives the set of nodes linking to $a$, while $W^{i}_{j}u^{j}=s^{i}$ gives the incoming strength of nodes.

Before moving to the next section, it is useful to define some tensors used throughout this work. We have just seen the rank-1 1--tensor in action: similarly we can define the rank-2 1--tensor $U^{i}_{j}=u^{i}u_{j}$ or higher-order tensors. Another fundamental tensor is the Kronecker one, defined by $\delta^{i}_{j}$, with components equal to 1 if $i=j$ and equal to 0 otherwise.

\begin{figure}[!t]
\centering
\begin{minipage}[c]{0.4\textwidth}
\includegraphics[width=\textwidth]{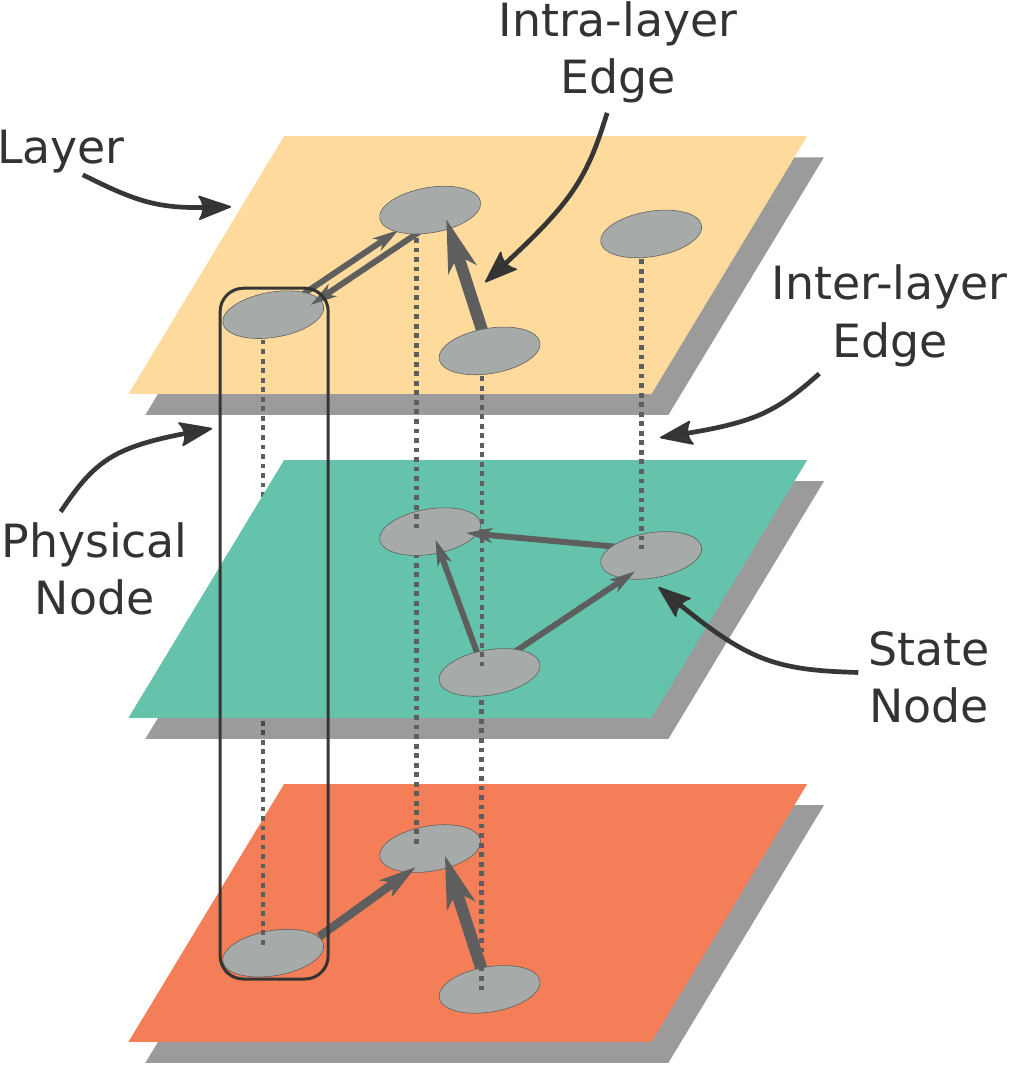}
\end{minipage}
\quad
\begin{minipage}[c]{0.55\textwidth}
\caption{\label{fig:chap_intro_mux_struct_terminology} A system where nodes are characterized by three distinct types of interactions, encoded by colored layers. Overall, the system is a multilayer network, because to describe relationships we need to specify more than one network. Units are physical nodes: each one is a set of \emph{state nodes} or \emph{replicas}, each one encoding the identity of the corresponding physical node in each layer separately. \emph{Intra-layer} edges define connectivity within each layer, whereas \emph{inter-layer} edges define connectivity across layers. Reproduced with permission from~\citep{de2020illustrations}.}
\end{minipage}
\end{figure}

\subsection{Tensorial representation of a multilayer network}

In the previous section we have introduced the fundamental procedure required to build an adjacency tensor to represent a classical network (a monoplex). Using a similar procedure, we can build a \emph{multilayer adjacency tensor} to represent a multilayer network, as the one shown in Fig.~\ref{fig:chap_intro_mux_struct_terminology}. A multilayer system is characterized by $N$ physical nodes interacting in $L$ distinct ways simultaneously: each type of interaction defines a \emph{layer}. At variance with single-layer networks, there are more edges sets to encode: as many as the number ($L$) of layers and, in general, as many as the number ($L(L-1)$) of directed pairwise connections between layers, since we have to specify which node $i$ in a layer $\alpha$ is connected to which node $j$ in a layer $\beta$ ($i,j=1,2,...,N$, $\alpha,\beta=1,2,...,L$)\footnote{This simple observation suggests that a good candidate for multilayer adjacency tensor should be a rank-4 tensor.}. Note that for simplicity, we are indicating with Greek letters the indices related to layers and with Latin letters the indices related to nodes.

\begin{figure}[!ht]
\centering
\includegraphics[width=0.85\textwidth]{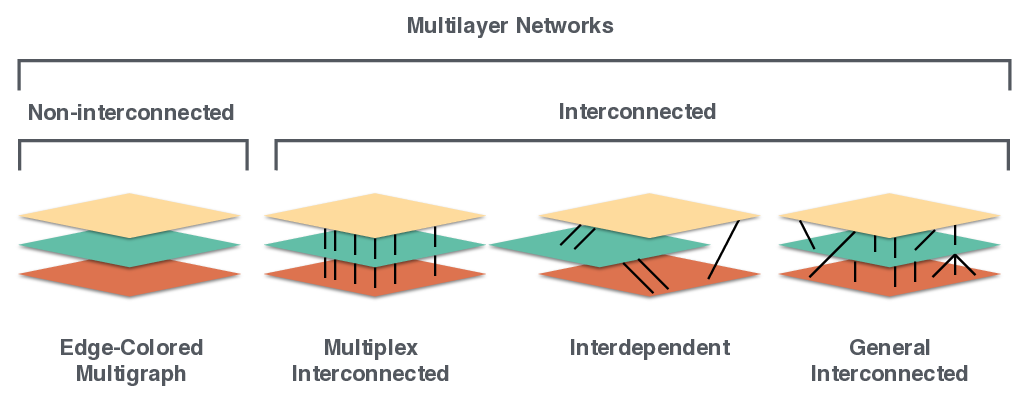}
\caption{\label{fig:chap_intro_muxmodels} Multilayer networks include a broad spectrum of possible models. Edge-colored networks are useful models when inter-layer connectivity is not well defined: this is the case of a social network where edges can represent different types of social relationships (e.g., trust, family, business, etc.)~\citep{cardillo2013emergence,nicosia2015measuring,de2015structural}. Conversely, in interconnected networks inter-layer connectivity is well defined and allows us to model a variety of systems~\citep{de2014navigability,de2015ranking,osat2017optimal,radicchi2017redundant}, including those with interdependencies where nodes control and/or are controlled by nodes in another network~\citep{rosato2008modelling,buldyrev2010catastrophic,vespignani2010complex,gao2012networks,radicchi2015percolation}. Reproduced with permission from~\citep{de2020illustrations}.}
\end{figure}

There are different types of multilayer networks, depending on the presence or absence of links between layers and on the way nodes are defined (see Fig.~\ref{fig:chap_intro_muxmodels}). In the following, we will mostly deal with the class of systems characterized by inter-layer connectivity, since it is not possible to define a meaningful multilayer adjacency tensor for the class of edge-colored multigraphs\footnote{Note that, instead, it is possible to define a valid hypermatrix encoding this object, and this hypermatrix can be thought of as an array of matrices~\citep{bianconi2013statistical}.}. 

Let us introduce the canonical rank-1 vectors $e^{\alpha}(p)$ ($\alpha,p=1,\ldots,L$) in the space of layers $\mathbb{R}^{L}$, and the corresponding canonical rank-2 tensors $E^{\alpha}_{\beta}(pq)=e^{\alpha}(p)e_{\beta}(q)$, similarly to what we have done for monoplexes. It is straightforward to show~\citep{de2013mathematical} that the linear combination of 
\begin{eqnarray}
M^{i\alpha}_{j\beta}=\sum_{a,b=1}^{N}\sum_{p,q=1}^{L}w_{ab}(pq)e^{i}(a)e_{j}(b)e^{\alpha}(p)e_{\beta}(q)
\end{eqnarray}
fully characterizes a multilinear object in the space $\mathbb{R}^{N\times L \times N \times L}$. This object is, in fact, the desired multilayer adjacency tensor since, under a change of coordinates, it transforms like a tensor:
\begin{eqnarray}
M'^{i\alpha}_{j\beta}&=& \sum_{a,b=1}^{N}\sum_{p,q=1}^{L}w_{ab}(pq)B^{i}_{k}e^{k}(a)(B^{-1})^{l}_{j}e_{l}(b){\tilde{B}}^{\alpha}_{\gamma}e^{\gamma}(p)({\tilde{B}}^{-1})^{\delta}_{\beta}e_{\delta}(q)\nonumber\\
&=& B^{i}_{k}{\tilde{B}}^{\alpha}_{\gamma}M^{k\gamma}_{l\delta}(B^{-1})^{l}_{j}({\tilde{B}}^{-1})^{\delta}_{\beta}.
\end{eqnarray}

By indicating with $E^{i\alpha}_{j\beta}(ab;pq)=E^{i}_{j}(ab)E^{\alpha}_{\beta}(pq)$ the canonical rank-4 tensors, we can simply reduce the definition of the multilayer adjacency tensor to
\begin{eqnarray}
M^{i\alpha}_{j\beta}=\sum_{a,b=1}^{N}\sum_{p,q=1}^{L}w_{ab}(pq)E^{i\alpha}_{j\beta}(ab;pq),
\end{eqnarray}
where $w_{ab}(pq)$ encodes the intensity of the interaction between node $a$ in layer $p$ and node $b$ in layer $q$. Note that $w_{ab}(pp)$ indicates the weights of the links in layer $p$.

It is worth noticing that, as for the space of nodes, in the space of layers, we can define multilayer 1--tensors and Kronecker tensors as $U^{i\alpha}_{j\beta}=U^{i}_{j}U^{\alpha}_{\beta}$ and $\delta^{i\alpha}_{j\beta}$, respectively. Another important tensor, representing a complete multilayer network without self-edges, will be used later in this work to characterize multilayer triadic closure: for consistency, we prefer to introduce it here as $F^{i\alpha}_{j\beta}=U^{i\alpha}_{j\beta}-\delta^{i\alpha}_{j\beta}$.

At this point, the reader should be familiar enough with tensors to note that different decompositions are possible. Here, we are not referring to operations like Tucker decomposition -- the higher-order generalization of singular value decomposition (SVD)~\citep{tucker1966some} -- but to a linear decomposition to highlight the fundamental components of a multilayer system. In fact, we can identify four tensors which encode distinct structural information:
\begin{eqnarray}
\label{eq:sexi-decomp-structure}
	m^{j\beta}_{i\alpha} &= \underbrace{m^{j\beta}_{i\alpha}\delta_{\alpha}^{\beta}\delta_{i}^{j} +m^{j\beta}_{i\alpha}\delta_{\alpha}^{\beta}(1-\delta_{i}^{j})}_{\text{intra-layer relationships}}+\underbrace{m^{j\beta}_{i\alpha}(1-\delta_{\alpha}^{\beta})\delta_{i}^{j} +m^{j\beta}_{i\alpha}(1-\delta_{\alpha}^{\beta})(1-\delta_{i}^{j})}_{\text{inter-layer relationships}}\nonumber\\
&= \underbrace{m^{i\alpha}_{i\alpha}}_{\text{self-relationships}} + \underbrace{m_{i\alpha}^{j\alpha}}_{\text{endogenous}}  + \underbrace{m_{i\alpha}^{j\beta}}_{\text{exogenous}} + \underbrace{m_{i\alpha}^{i\beta}}_{\text{intertwining}}\nonumber\\
&=\mathbb{S}_{i\alpha}(M) + \mathbb{N}^{j}_{i\alpha}(M) + \mathbb{X}^{j\beta}_{i\alpha}(M)  + \mathbb{I}_{i\alpha}^{\beta}(M)\,.
\end{eqnarray}
Here, the components of the tensor are indicated by $m_{i\alpha}^{j \beta}$ ($i,j =1,2,\dots,N$ and $\alpha,\beta =1,2,\dots,L$), while $\delta^{j}_{i}$ and $\delta_{\alpha}^{\beta}$ indicate the Kronecker delta function in the space of nodes and layers, respectively. The four tensors encode the following relationships:
\begin{itemize}
\item \textbf{Intra-layer interactions}: 
\begin{itemize}
\item \textbf{self-interactions} ($\mathbb{S}$): from a node to itself;
\item \textbf{endogeneous interactions} ($\mathbb{N}$): between distinct nodes belonging to the same layer;
\end{itemize}
\item \textbf{Inter-layer interactions}: 
\begin{itemize}
\item \textbf{exogenous interactions} ($\mathbb{X}$): between distinct nodes belonging to distinct layers;
\item \textbf{intertwining} ($\mathbb{I}$): from a node to its replicas in other layers.
\end{itemize}
\end{itemize}

Equation~(\ref{eq:sexi-decomp-structure}) characterizes the ``structural $\mathbb{S}\mathbb{N}\mathbb{X}\mathbb{I}$ decomposition'' of the multilayer adjacency tensor $M$: the models for interconnected systems described in Fig.~\ref{fig:chap_intro_muxmodels} can be characterized by which $\mathbb{S}\mathbb{N}\mathbb{X}\mathbb{I}$ components contribute to the tensor. In particular, we identify:

\begin{itemize}
\item \textbf{Interconnected multiplex networks}: type $\mathbb{S}\mathbb{N}\mathbb{I}$;
\item \textbf{Interdependent networks}: type $\mathbb{S}\mathbb{N}\mathbb{X}$;
\item \textbf{General multilayer networks}: $\mathbb{S}\mathbb{N}\mathbb{X}\mathbb{I}$.
\end{itemize}

As previously mentioned, edge-colored networks do not admit a meaningful representation in terms of multilayer adjacency tensor, but according to this classification they would define type $\mathbb{S}\mathbb{N}$.

\begin{figure}[!ht]
\centering
\begin{minipage}[c]{.5\textwidth}
\includegraphics[width=\textwidth]{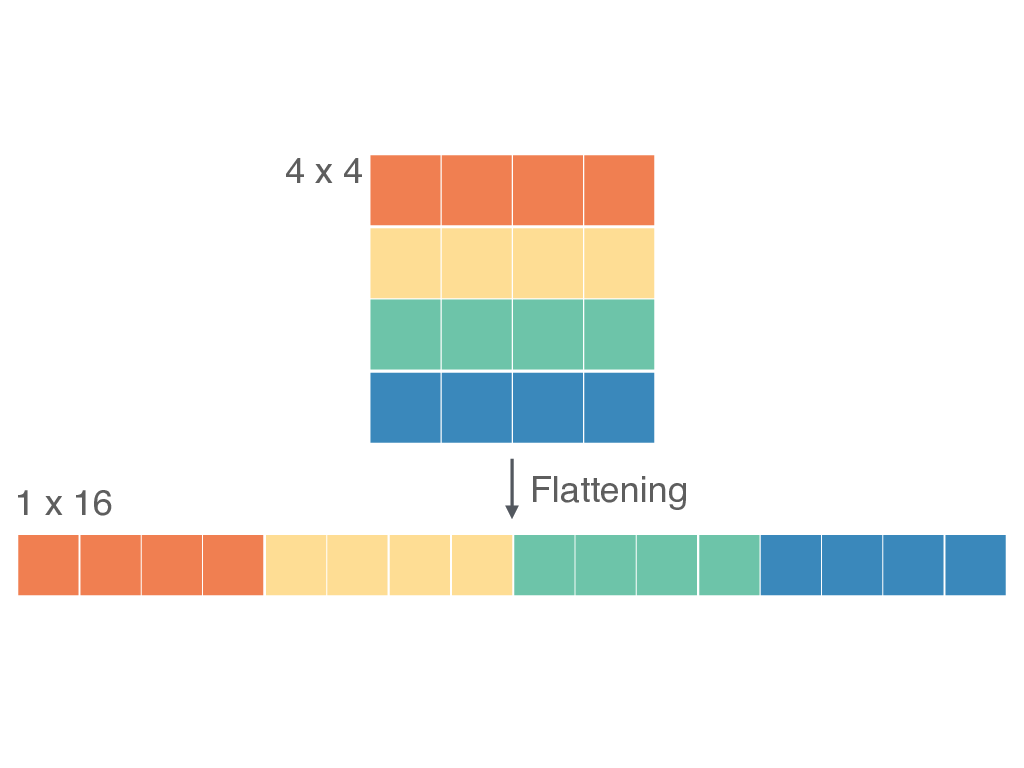}
\end{minipage}
\quad
\begin{minipage}[c]{.45\textwidth}
\caption{\label{fig:chap_gui_rank2_flattening} A rank-2 tensor defined in $\mathbb{R}^{4\times4}$ is flattened into a rank-1 tensor defined in $\mathbb{R}^{1\times16}$, without loss of information. Reproduced with permission from~\citep{de2020illustrations}.}
\end{minipage}
\end{figure}

\begin{figure}[!bt]
\centering
\begin{minipage}[c]{.5\textwidth}
\includegraphics[width=\textwidth]{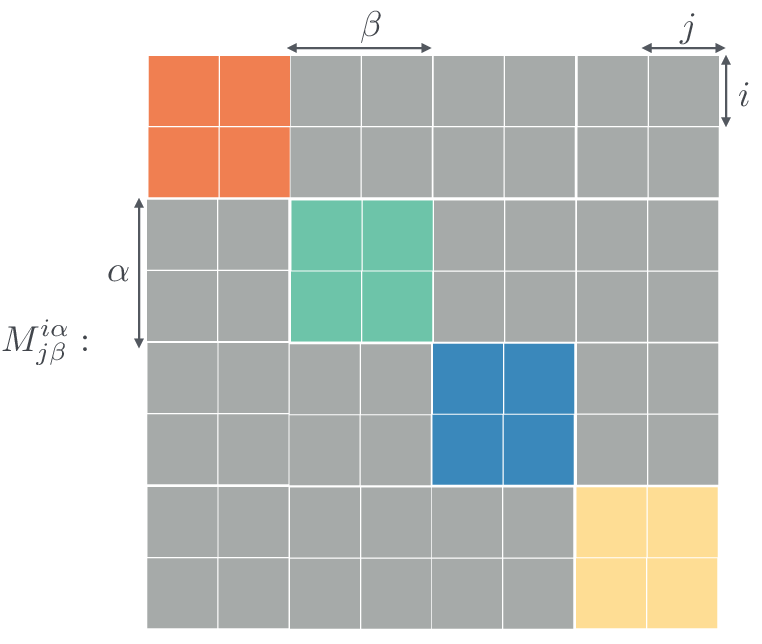}
\end{minipage}
\quad
\begin{minipage}[c]{.45\textwidth}
\caption{\label{fig:chap_gui_rank4_flattening} As in Fig.~\ref{fig:chap_gui_rank2_flattening} but applied to a multilayer adjacency tensor $M^{i\alpha}_{j\beta}$ in $\mathbb{R}^{2\times4\times2\times4}$ flattened into a supra-adjacency matrix in $\mathbb{R}^{8\times 8}$. Colors highlight the diagonal blocks where adjacency matrices corresponding to layers are encoded. Off-diagonal blocks encode inter-layer connectivity. Reproduced with permission from~\citep{de2020illustrations}.}
\end{minipage}
\end{figure}

From an operational perspective, working with tensors might be complicated and cumbersome. Nevertheless, from a theoretical perspective, the tensorial formulation allows us to write complex equations in a very compact and handy way, and it can be used to guide our intuition about generalizing existing network descriptors for multilayer analysis, as we will see in the next chapters. One widely adopted approach is based on the operation known as \emph{matricization} (or \emph{flattening})~\citep{kolda2009tensor}, which maps a complex object like a high-order tensor into a lower-order object while preserving the information content (see Fig.~\ref{fig:chap_gui_rank2_flattening} for an emblematic example). In the case of the multilayer adjacency tensor
$M^{i\alpha}_{j\beta}$, which is defined in the space $\mathbb{R}^{N\times L\times N \times L}$, this operation corresponds to flatten entries into a rank-2 object defined in the space $\mathbb{R}^{NL\times NL}$, as shown in Fig.~\ref{fig:chap_gui_rank4_flattening}. It is manifest that there is no loss of information, although care is needed when dealing with this new object, which is known in the literature as \emph{supra-adjacency matrix}~\citep{gomez2013diffusion,de2013mathematical,sole2013spectral}.

\begin{figure}[!t]
\centering
\includegraphics[width=0.65\textwidth]{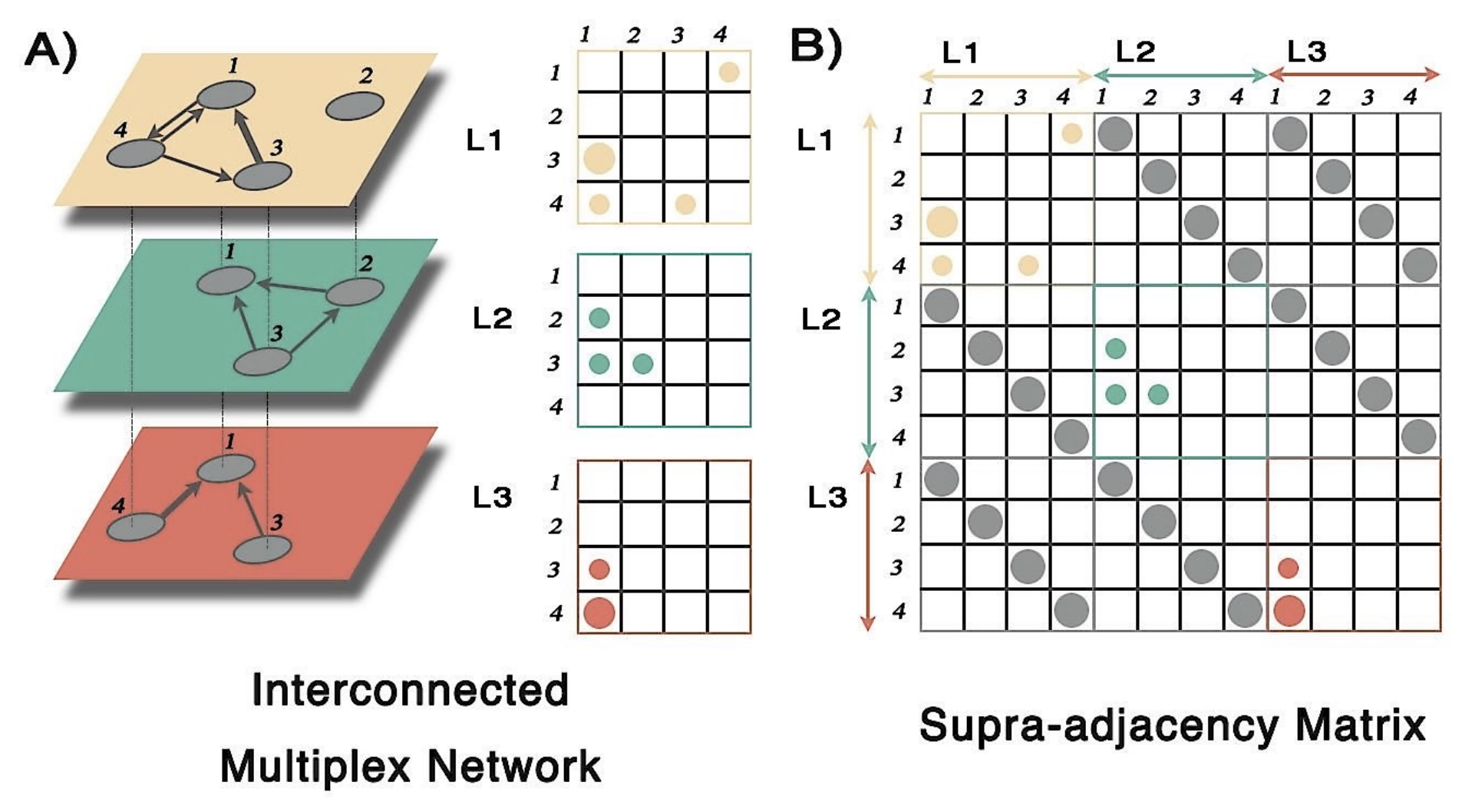}
\caption{\label{fig:chap_intro_multi2supra} A multilayer network, consisting of 3 layers, flattened into a supra-adjacency matrix~\citep{gomez2013diffusion,de2013mathematical}. (A) The interconnected multiplex network (left) with $N = 4$ nodes and $L = 3$ layers, the latter ones are color-coded. Connectivity is weighted (see edge thickness) and directed (see arrows): the adjacency matrix of each layer is shown on the right-hand side of this panel. (B) Result of the flattening procedure: the original rank--4 tensor is now encoded into a rank--2 supra-adjacency matrix with a block structure. Intra-layer connectivity is encoded into diagonal blocks (according to colors), whereas inter-layer interactions are represented into off-diagonal blocks. Reproduced with permission from~\citep{baggio2016multiplex}.}
\end{figure}

\begin{figure}[!t]
  \centering
  \includegraphics[width=0.55\textwidth]{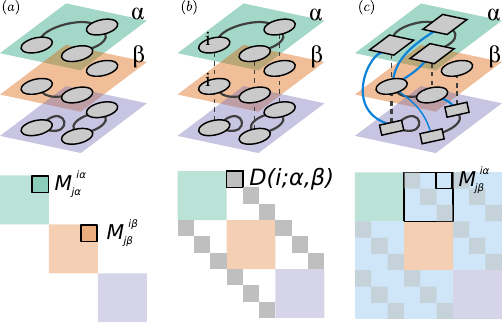}
  \caption{Supra-adjacency matrix representation of three distinct types of multilayer networks~\citep{gomez2013diffusion,de2013mathematical}, where layers are encoded by colors: (a) edge-colored multigraph (no inter-layer connectivity); (b) interconnected multiplex network (diagonal coupling in the off-diagonal blocks), and (c) general interconnected case, where inter-layer connections are not restricted to replicas (exogenous interactions). Latin letters denote nodes, while Greek letters are used for layers. $D(i; \alpha \beta)$ indicates the intensity of the connection between state nodes $(i, \alpha)$ and $(i, \beta)$. Figure from~\citep{bertagnolli2020diffusion}.}
  \label{fig:mux_rw_multilayers}
\end{figure}

Working with a supra-adjacency matrix comes with several computational advantages and notational disadvantages. For practical purposes, the flattening allows for a visual inspection of the multilayer adjacency tensor, as shown in Fig.~\ref{fig:chap_intro_multi2supra}. The supra-adjacency matrix is, in fact, a block matrix with intra-layer connectivity encoded into diagonal blocks and inter-layer connectivity encoded into off-diagonal ones (see also Fig.~\ref{fig:chap_intro_multi2supra}). However, it is worth remarking that this is a matter of convention: in fact, this arrangement of blocks is not unique and other arrangements are also valid, although some ones are less convenient than others for the design of algorithms. This non-uniqueness of the supra-adjacency matrix makes it less suitable for theoretical calculations but still an alternative to higher-order tensors. Fig.~\ref{fig:mux_rw_multilayers} shows the supra-adjacency matrix representation for three widely used multilayer network models.

In the next chapters we will use the tensorial formulation, when possible, for the analysis of the topology of multilayer networks and to define dynamical processes.

%%%%%%%%%%%%%%%%%%%%%%%%%%%%%
%%%%%%%%%%%%%%%%%%%%%%%%%%%%%
%%%%%%%%%%%%%%%%%%%%%%%%%%%%%
\section{Multilayer structural analysis} 
In this chapter, we will operationally define some of the most important theoretical and computational tools for the analysis of a multilayer network. We start with walks (Sec.~\ref{sec:basicdefs}) to define distinct types of connected components (Sec.~\ref{sec:conncomp}) and we will quickly move to describe several measures of node importance within the system, i.e., multilayer versatility, which is the generalization of the concept of node centrality (Sec.~\ref{sec:versatility}). 

It will follow an overview of methods to characterize the mesoscale organization of a system (Sec.~\ref{sec:mesoscale}), by identifying how nodes group together into small clusters (Sec.~\ref{sec:triads}), such as triangles, and in larger functional modules or communities (Sec.~\ref{sec:groups}). We will show how to use these concepts to understand how units in multilayer systems are integrated or segregated, in terms of information flow (Sec.~\ref{sec:intsegration}). Technically speaking, connected components should be described here, but we opted to keep the corresponding section after the description of walks for sake of simplicity.

We will conclude this chapter by discussing existing methods to quantify the correlations between pairs of layers (Sec.~\ref{sec:layercorr}): this class of analytical tools is fundamental to better understand results from versatility and mesoscale analysis, since the existence of layer-layer correlation patterns is reflected in structural measures based on walks and paths.

\subsection{Basic definitions}\label{sec:basicdefs}

\begin{figure}[!t]
\centering\includegraphics[width=0.8\textwidth]{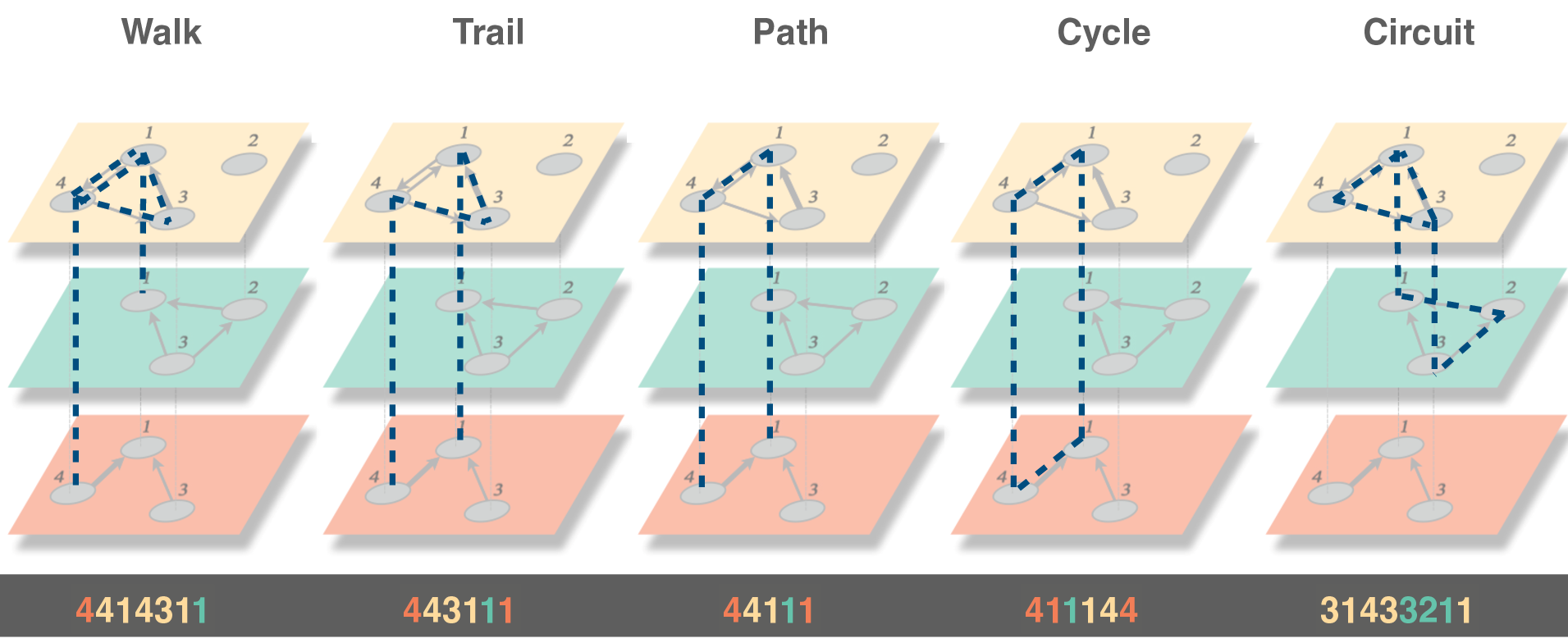}
\caption{\label{fig:chap_gui_multipaths} Illustration of distinct walks possible on a multilayer network. By definition, a walk in a multilayer system is the most general way to move between its layers, nodes and links. To define walks characterized by distinct features, it is sufficient to restrict the number of times nodes or links can be traversed or to put constraints on origin and destination nodes. The illustration shows a sequence of nodes and edges corresponding to each type of walk (multilayer walk, trail, path, cycle and circuit) with a dashed line: note that the sequence of nodes visited by each type of walk is explicitly reported below the corresponding multilayer network. Reproduced with permission from~\citep{de2020illustrations}.}
\end{figure}

In a generic network, a \emph{walk} is defined as a sequence of adjacent nodes and edges visited by a hypothetical walker.  In general, a given edge (but also node) can be traversed more than once, but it is possible to apply some restrictions to a multilayer walk to specify special walks. For instance, one can define a \emph{multilayer trail} as a walk where links can be traversed only one time. A further restriction can be also applied to the identity of origin and destination nodes, and we can define a \emph{multilayer path} as a multilayer trail with the restriction that repeated nodes are not allowed, whereas a \emph{multilayer cycle} is a closed trail where only the origin and destination nodes are repeated. Finally, a closed trail with more than one repeated node is defined as a \emph{multilayer circuit}.  An illustration of different types of walks on a multilayer network is shown in Fig.~\ref{fig:chap_gui_multipaths}. 

The \emph{length of a walk} is the number of edges traversed along the walks. It is possible to calculate the number of walks of length $\ell$ from a node $i$ to any other node $j$. For an unweighted network the element $A^{k}_{l}$ of its adjacency matrix is 1 if there is an edge from $k$ to $l$, and is 0 otherwise. Then, if there is a walk with $\ell=2$ from $i$ to $j$ via $k$,  the product $A^{i}_{k}A^{k}_{j}$ will be 1. By generalizing this argument, we can write the rank--2 tensor encoding information about walk length between any pair of nodes in the network as the $\ell$--th power of the rank--2 adjacency tensor representing the network: 

\begin{eqnarray}
\mathcal{W}^{i}_{j}(\ell) = (A^{i}_{j})^{\ell} = A^{i}_{j_1}A^{j_1}_{j_2}\dots A^{j_{\ell-1}}_{j}
\end{eqnarray}

The same formalism can be used in a weighted network, by defining the weight of a walk as the product of the weights of the traversed links. The entries of $\mathcal{W}^{i}_{j}(\ell)$ will give the sum of weights of the walks of length $\ell$ connecting the corresponding pair of nodes.

An analogous approach is used to calculate the walk length for multilayer networks. If $M^{i\alpha}_{j\beta}$ is the rank--4 adjacency tensor representing the system, then the entries of the $\ell$--th power of this tensor provide the number of multilayer walks of length $\ell$ between a node $i$ in layer $\alpha$ and a node $j$ in layer $\beta$:
\begin{equation}
\mathcal{W}^{i\alpha}_{j\beta}(\ell) = M^{i\alpha}_{j_1\beta_1}M^{j_1\beta_1}_{j_2\beta_2}\dots M^{j_{\ell-1}\beta_{\ell-1}}_{j\beta}.
\end{equation}

The aggregate representation of multilayer networks allows us to aggregate layers to obtain a network where the number of edges or the weight of an edge is the number of different types of edges between a pair of nodes ~\citep{de2013mathematical,cozzo2015structure}.
It is possible to highlight the topological difference between multilayer networks and their aggregated representations by using the aforementioned formalism.
Let $\bar{G}^{i}_{j}=M^{i\alpha}_{j\beta}U^{\beta}_{\alpha}$ be the aggregate network which accounts for inter-layer links: the corresponding rank--2 walk tensor is then given by
\begin{eqnarray}
\bar{\mathcal{W}}^{i}_{j}(\ell) &=& (\bar{G}^{i}_{j})^{\ell} = \bar{G}^{i}_{j_1}\bar{G}^{j_1}_{j_2}\dots \bar{G}^{j_{\ell-1}}_{j}\nonumber\\
&=& M^{i\alpha}_{j_1\beta_1}U^{\beta_1}_{\alpha} M^{j_1\beta_1}_{j_2\beta_2}U^{\beta_2}_{\beta_1}\dots M^{j_{\ell-1}\beta_{\ell-1}}_{j\beta}U^{\beta}_{\beta_{\ell-1}}\nonumber\\
&=& \underbrace{\left(M^{i\alpha}_{j_1\beta_1} M^{j_1\beta_1}_{j_2\beta_2}\dots M^{j_{\ell-1}\beta_{\ell-1}}_{j\beta}\right)}_{\mathcal{W}^{i\alpha}_{j\beta}(\ell)}\underbrace{\left(U^{\beta_1}_{\alpha}U^{\beta_2}_{\beta_1} \dots U^{\beta}_{\beta_{\ell-1}}\right)}_{U^{\alpha}_{\beta} L^{\ell-1}},
\end{eqnarray}
showing that the number of walks with length $\ell$ between two nodes in the aggregate network is not a linear function of the number of walks of length $\ell$ between the same pair of nodes in the multilayer network. 

In a network, a walk that does not intersect itself is named \emph{path} and the \emph{shortest path} is the shortest walk between a given pair of nodes. Shortest paths have an important role in several network phenomena: they allow us to model, for instance, how information is exchanged between two nodes by using the  least number of traversed nodes and links. In undirected networks the length of the shortest path is often used to define a distance between nodes whereas some properties of a \emph{geodesic distance} are, in general, no more satisfied in presence of directed links.

\subsection{Connected components}\label{sec:conncomp}

To identify clusters of nodes\footnote{Note that this notion is different from the one of groups or communities or modules, although the terminology might be sometimes misleading.} that can exchange information in a network, it is useful to analyze the connected components~\citep{newman2018networks,barabasi2016network}. Components are defined as separate parts of the network, i.e., subset of nodes with at least one path between any origin/destination pair belonging to the subset. A network with a single component is \emph{connected} whereas, if there is more than one component, a network is \emph{disconnected}. For instance, isolated nodes count as disconnected components of the system. For a directed network, a component is defined as \emph{weakly connected} if two nodes of the undirected representation of the component are connected by one or more paths. If there is a directed path in both directions between every pair of nodes, the component is defined as \emph{strongly connected}.

For networks with finite size, we define the \emph{largest connected component} (LCC) as the cluster with the maximal subset of nodes. If the size of the network is infinite, such as in the thermodynamic limit, the LCC is usually named \emph{giant connected component} (GCC).

In interdependent networks~\citep{gao2012networks} two systems $A$ and $B$ are interconnected with links and the potentially functional clusters are identified by \emph{mutually connected components}. If we indicate by $\mathcal{A}$ the set of nodes in network $G(A)$ and by $\mathcal{B}$ the corresponding set of nodes in network $G(B)$, they form a mutually connected component if: 

\begin{itemize}
\item each pair of nodes in $\mathcal{A}$ is connected by a path consisting of nodes belonging to $\mathcal{A}$ and links of network $G(A)$;
\item each pair of nodes in $\mathcal{B}$ is connected by a path consisting of nodes belonging to $\mathcal{B}$ and links of network $G(B)$~\citep{buldyrev2010catastrophic}.
\end{itemize}

In multilayer networks we can use the definition of path described in Sec.~\ref{sec:basicdefs} to define a \emph{multilayer connected component} as the subset of nodes connected by a multilayer path~\citep{de2014navigability}. This definition can be used to identify connected components from the aggregate representation of the multilayer network, because it is sufficient that a pair of nodes is connected by a path to be part of the same component.

\begin{figure}[!t]
\centering\includegraphics[width=0.8\textwidth]{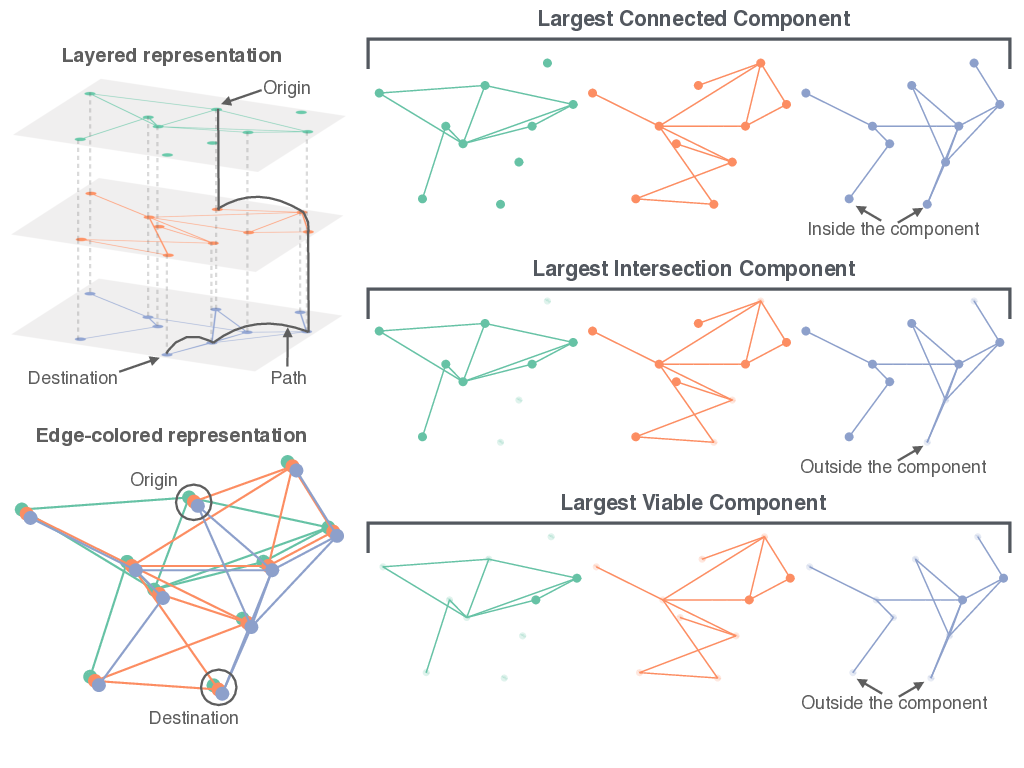}
\caption{\label{fig:chap_gui_multi_component} In many cases of interest, multilayer networks can be represented in different ways, e.g., (top-left) in terms of interconnected layers -- i.e., a multiplex -- or (bottom-left) by collapsing state nodes and explicitly encoding layers as colors of an edge-colored multigraph. To define components, it is important to identify origin and destination nodes and to consider paths connecting them. A multilayer connected component is identified by paths between state nodes. On the right-hand side of the figure, three distinct definitions of multilayer connected components are illustrated: largest connected component (top), largest intersection component (middle) and largest viable component (bottom). See the text for details. Reproduced with permission from~\citep{de2020illustrations}.}
\end{figure}

However, different and complementary information about the structure of a multilayer network can be provided by more restrictive definitions. 
For instance, the \emph{largest intersection component} (LIC) is defined as the largest cluster in which nodes are connected across all layers simultaneously~\citep{baxter2012avalanche} and can be identified by intersecting the LCC of each layer separately. Another alternative is to aggregate the multilayer system with respect to the intersection of edges and then identify the largest connected component of the resulting network~\citep{radicchi2015percolation}. That definition has been recently used to better understand the emergence of continuous or abrupt percolation phase transitions even in systems of finite size.

However, one more definition is possible and useful for applications. The \emph{largest viable component} (LVC) is defined as the maximal subset of \emph{viable} nodes and it consists of nodes that are connected by a path in each layer simultaneously~\citep{baxter2012avalanche}. As a result, all nodes in the LVC are essential to the function of the system and to define its structural core~\citep{azimi2014k,grassberger2015percolation,stella2018multiplex}. The more restrictive condition imposed by the LVC is responsible for a hybrid phase transition which leads to the discontinuous emergence of the giant viable cluster, at variance with ordinary percolation where a continuous phase transition is observed~\citep{baxter2012avalanche}.

Fig.~\ref{fig:chap_gui_multi_component} shows an illustration of multilayer connected components for both interconnected multiplex  and edge-colored multigraph representations. The more restrictive the condition imposed, the smaller the size of the largest connected cluster: in particular, the size of the LVC is equal to or smaller than the size of the LIC, which in turn is equal or smaller than the size of the LCC.

\subsection{Measuring influence: versatility}\label{sec:versatility}

Information exchange with a multilayer network is strongly dependent on the organization of the underlying system into connected components. However, given a (possibly connected) network, it is plausible to wonder \emph{which node(s)} are more important than others for information flow. For practical applications, several additional questions are plausible, and each question encodes an operational definition of \emph{node centrality}.

Revealing the most \textit{central} nodes in complex networks is a key issue in a variety of real-world scenarios~\citep{barabasi2016network}. In epidemiology, for example, finding the most central multilayer nodes helps identifying the pivotal disease spreaders~\citep{de2016physics}, while in cascading failures, by detecting the most central nodes we can recognize the most fragile actors able to trigger the failure of other parts of the network~\citep{buldyrev2010catastrophic}.  

Over the years, a wide spectrum of network centrality measures has been proposed for single-layer networks (for interested readers, we refer to a review from the computational social sciences~\citep{scott2017popularity}). The generalization of such descriptors to the realm of multilayer systems is not always immediate, since nodes peripheral in one layer might be extremely central in another one~\citep{de2013mathematical,kivela2014multilayer,boccaletti2014structure}, and crucial nodes for a given dynamics might not be central across all layers. If, for example, two distinct layers have only one node in common, any exchange of information between those layers will involve the passage through that common node, independently from its centrality in each layer separately: hence, such a node will be highly central for the considered process. \emph{Multilayer versatility}~\citep{de2013mathematical,de2015ranking} is a measure able to capture these features of a node, quantifying how important nodes are for the processes which characterize the definition of specific centrality descriptors -- e.g., in terms of information diffusion or flow through shortest paths -- in a multilayer network.

In 2006, Borgatti and Everett~\citep{borgatti2006graph} approached centrality from a graph-theoretic perspective, claiming that all centralities consider the involvement of a node in the walk structure of the network according to four features: Walk Type, Walk Property, Walk Position and Summary Type. The \textit{Walk Type} dimension concerns the kind of walks considered as well as the kind of constraints on such walks, for example considering only shortest-paths. The \textit{Walk Property} dimension distinguishes between measures that consider the \textit{volume} of walks a node is involved in (e.g., as in betweenness centrality) and measures that regard the \textit{length} of those walks (e.g., as in closeness centrality). The \textit{Walk Position} can be radial, where walks start in a given node (e.g., closeness) or medial, where walks pass through that given node (e.g., betweenness). Finally, the \textit{Summary Type} dimension distinguishes measures according to the chosen summary statistic used to obtain a centrality score vector. 
Based on the possible combinations of these four features -- i.e., Walk Type, Walk Property, Walk Position and Summary Type --  different centrality measures can be obtained. 

Besides identifying the most central nodes, the power of versatility lies also in predicting their role in some emblematic dynamical processes, such as diffusion and congestion~\citep{de2015ranking}. The importance of this aspect is highlighted, for example, when nodes of a multilayer network are ranked by their coverage~\citep{de2014navigability} (see also Sec.~\ref{sec:diffproc}) using the PageRank -- see later in this section for its definition --  obtained from the multilayer model and its aggregated network. PageRank versatility, obtained from the multilayer network, is a better estimator of the evolution of such dynamics, outperforming its single-layer counterpart.

In the following, we present the most important centrality measures in the case of multilayer networks, i.e. Multidegree, K-coreness, Eigenvector and Katz versatility, HITS versatility, Random walk occupation centrality, PageRank versatility, Random walk betweenness and closeness versatility, Betweenness versatility and Interdependence centrality, while we refer the interested reader to other interesting measures which exploiting multiplex features to identify a node's importance~\citep{posfai2019consensus}, recently validated to predict multiplex centrality in the rhesus macaque~\citep{beisner2020multiplex}.

\subsubsection{Multidegree}

Multidegree centrality ($k_{i}$) is the simplest indicator of node importance at the local level, it is obtained by summing up all the links connected to node $i$ across all layers~\citep{de2013mathematical}: 
\begin{eqnarray}
k_{i} = \sum_{\alpha, \beta= 1}^{L} \sum_{j=1}^{N}M_{j\beta}^{i\alpha} = M_{j\beta}^{i\alpha}u_{\alpha}^{\beta}u^{j}.
\end{eqnarray}
In some cases, e.g., when interlayer links are not explicitly considered, it can be more suitable to evaluate the degrees coming from individual layers:

\begin{eqnarray}
k_{i}^{\alpha} = \sum_{j=1}^{N}A_{j}^{i}(\alpha),
\end{eqnarray}
where $ A_{j}^{i}(\alpha) $ represents the adjacency tensor of layer $\alpha$. Other definitions partially related to this one, for the case of non-interconnected multiplex networks, can be found in~\citep{bianconi2013statistical,battiston2014structural}.

\subsubsection{K-coreness}

\begin{figure}[!t]
\centering
\begin{minipage}[c]{0.3\textwidth}
\includegraphics[width=\textwidth]{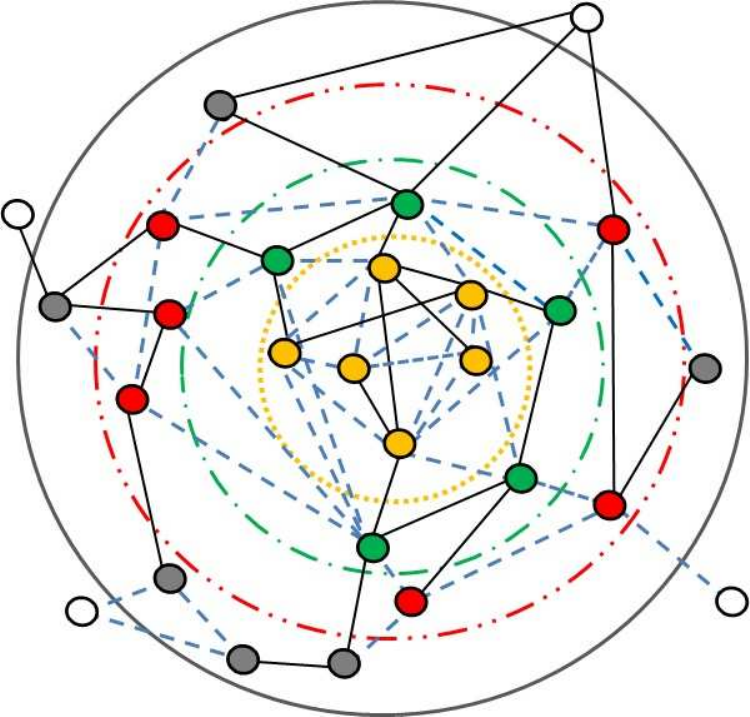}
\end{minipage} 
\quad
\begin{minipage}[c]{0.65\textwidth}
\caption{The ($k_{1},k_{2}$)-core decomposition of a multilayer network characterized by two distinct types of edges. The first type is encoded by solid black links, while the second type is encoded by dashed blue links. The concentric circles help identify the ($k_{1},k_{2}$)-core decomposition: the cores from the most external circle to the innermost are the (1,1)-core, the (1,2)-core, the (2,2)-core, and the (1,3)-core, respectively. Figure from~\citep{azimi2014k}.}
\label{fig:k-coreness}
\end{minipage}
\end{figure}

In single layer network, the $k$-core of a graph is defined as a maximal connected subgraph in which each vertex has at least degree $k$ within that subgraph. The ensemble of all $k$-cores of a graph represents the core decomposition of that specific graph~\citep{seidman1983network}. 
To extend this concept to multilayer networks, we have to take into account the different types of edges, encoding different types of interactions encoded into layers. In this case, the $k$-core is defined as the largest subgraph in which each vertex has at least $k_{\alpha}$ edges of each type, $\alpha = 1,2, . . . ,M$ where $M$ is the total number of layers~\citep{azimi2014k} (see Fig.~\ref{fig:k-coreness}). For more detail on how to efficiently compute the complete core decomposition of a multilayer network, we refer to~\citep{galimberti2020core}. 
	
\subsubsection{Eigenvector Versatility}\label{eigcen1}

Eigenvector centrality is a measure of influence of the single nodes in a network. A particular node has high eigenvector centrality if it is high the centrality of its neighbours~\citep{bonacich1987power}. In the monoplex case, the recursive character of this definition is untangled by the eigenvalue problem:
\begin{eqnarray}
W{\bf x} = \lambda_1 {\bf x},
\end{eqnarray}
where $ \lambda_1$ is the largest eigenvalue of $W$ and the element $x_i$ represents the centrality score of the node $i$ in the network described by the weight matrix W~\citep{de2013mathematical,de2015ranking}. If $W$ is symmetric with positive entries, the Perron-Frobenius theorem grants the existence and uniqueness of this vector.

When complete knowledge of the intra-layer connectivity is available, a natural extension of the definition of eigenvalue centrality for multilayer networks  can be easily obtained as~\citep{de2013mathematical}:
\begin{eqnarray}
\sum_{i,\alpha} M_{j\beta}^{i\alpha}\Theta_{i\alpha} = \lambda_1 \Theta_{j\beta},
\end{eqnarray}
where $\lambda_1$ is the largest eigenvalue of $M$ and $\Theta$ its corresponding eigentensor, whose values represent the centrality of each node in each layer.
The problem of finding the eigenvector centrality consists in
computing $\Theta_{j\beta} = \lambda_1^{-1} \sum_{i,\alpha} M_{j\beta}^{i\alpha}\Theta_{i\alpha} $, which represents the multilayer generalization of Bonacich's eigenvector centrality per node per layer~\citep{de2013mathematical, de2015ranking}. 
By summing up the scores of a node across all the layers, Eigenvector versatility can be condensed across layers: $\theta_{i}= \sum_{\alpha} \Theta_{i\alpha}$.

Note that the summation across layers appears naturally through the tensorial formalism~\citep{de2013mathematical,de2015ranking}. Nevertheless, we can opt for other types of aggregation, based on heuristics specific to the nature of the centrality vector, to summarize the centrality measures computed for all layers into a unique descriptor. For other definitions, we refer the interested reader to~\citep{sola2013eigenvector,halu2013multiplex} and~\citep{taylor2021tunable}, the latter proposing tunable eigenvector-based centralities that can be applied to both temporal and multiplex networks.

\subsubsection{Katz versatility}

In the case of directed networks, nodes with only outgoing edges have by definition eigenvalue centrality equal to zero. This may lead to meaningless results for the eigenvalue centrality, such as in the limiting case of a directed acyclic network, where all nodes have zero centrality~\citep{newman2018networks}. This problem can be overcome by assigning to each node a minimum value of centrality, so that all nodes are taken into account for measuring the influence between neighbours~\citep{katz1953new}. This can be realized~\citep{de2013mathematical} by redefining the eigenvalue problem as:
\begin{eqnarray}
\sum_{i,\alpha} (a M_{j\beta}^{i\alpha}\Phi_{i\alpha} + 1)= \Phi_{j\beta} 
\end{eqnarray}
with $a < 1/\lambda_1$, being $\lambda_1$ the largest eigenvalue of $M$. As for the Eigenvector versatility, the overall Katz centrality of a node is the sum of centrality scores $\Phi_{j\beta}$ across layers~\citep{de2013mathematical,de2015ranking}.

\subsubsection{HITS versatility} 

In single-layer networks, the Hyperlink-Induced Topic Search or HITS centrality (also known as \textit{hubs and authorities}) was originally
introduced for web page rating according to their authority (e.g. their content) and their hub value (e.g. the value of their links to other web-pages)~\citep{kleinberg1999authoritative}.
Again in the case of directed graphs, it can be useful to recognise as important a node that is pointed to by important nodes or, alternatively, a node that points to important nodes. These two behaviours define two different roles a node can play in a directed network: hub and authorities. In the case of multilayer networks, how extensively a node plays the role of a hub or an authority is measured by the HITS versatility, that is defined by two eigenvalue problems~\citep{de2013mathematical,de2015ranking}:

\begin{eqnarray}
\sum_{i,\alpha} (M M^\dagger)_{j\beta}^{i\alpha}   \Gamma_{i\alpha} = \lambda_1 \Gamma_{j\beta} \\
\sum_{i,\alpha} (M^\dagger M)_{j\beta}^{i\alpha}   \Upsilon_{i\alpha} = \lambda_1 \Upsilon_{j\beta},
\end{eqnarray}
where $\lambda_1$ is the largest eigenvalue (which is the same for the two problems), $\Gamma$ represents the authority centrality, and $\Upsilon$, the hub centrality.

\subsubsection{Random walk occupation centrality} \label{rwoc}

A random walk on a network is a stochastic process where a path is defined, starting from a node of origin $X(0)$, with the node $X(t+1)$ chosen at random from among the neighbours of node $X(t)$. For a multilayer network, the probabilities of transition between pairs of nodes are represented as a tensor $T_{j\beta}^{i\alpha}$~\citep{de2013mathematical}, that are often taken as proportional to the edges' weights.
If we let $p_{i\alpha}(t)$ be a time dependent tensor giving the occupation probability of a given node in a given layer at time $t$, the random walk can be modelled as a Markov chain (note that we are using the Einstein convention):
\begin{eqnarray}
p_{j\beta}(t+1) = T_{j\beta}^{i\alpha} p_{i\alpha}(t)
\end{eqnarray}
The steady state for this equation $\Pi_{i\alpha}$ can be obtained as the leading eigentensor in the eigenvalue problem:
\begin{eqnarray}
T_{j\beta}^{i\alpha} \Pi_{i\alpha} =\lambda_1 \Pi_{j\beta}
\end{eqnarray}
The probabilities $\Pi_{i\alpha}$ define the \textit{random walk occupation centrality}~\citep{sole2016random}, a measure that highlights which nodes might experience congestion due to insufficient outflow. 

\subsubsection{PageRank versatility} 

For single-layer networks, PageRank~\citep{page1999pagerank} is a centrality measure originally developed for ranking webpages. It represents the occupation probability of random walkers on the network subjected to teleportation: at any time the walker can walk to a neighbour with a rate $r$ and be teleported to any other node with rate $1-r$. Its extension to a multilayer network can be based on different heuristics, in case of unknown nature of the layer's coupling~\citep{halu2013multiplex}, or on the tensorial formulation~\citep{de2013mathematical,de2015ranking} if the inter-layer connectivity is known. In this second case, the dynamics of the random walker are regulated by a transition tensor defined by:
\begin{eqnarray}
R_{j\beta}^{i\alpha} = rT_{j\beta}^{i\alpha} + \frac{1-r}{NL}u_{j\beta}^{i\alpha},
\end{eqnarray}
where $T_{j\beta}^{i\alpha}$ is the transition tensor of a classical random walk in the absence of teleportation and $u_{j\beta}^{i\alpha}$ is the rank 4 tensor with all components equal to 1. 
If we denote by $\Omega_{i\alpha}$ the eigentensor of $R_{j\beta}^{i\alpha}$, then the PageRank versatility $\omega_i$ of a node is obtained by summing across layers $\sum_{i,\alpha}\Omega_{i\alpha}$. In other words, the Page Rank centrality for multilayer networks is the steady-state solution of the master equation corresponding to the transition tensor $R_{j\beta}^{i\alpha}$. 
 
It is important to remark that, for all nodes without out-going edges, the transition tensor has to be redefined as $R_{j\beta}^{i\alpha} =\frac{1}{NL}u_{j\beta}^{i\alpha}$ to grant the correct normalisation of the transition probabilities. We refer to~\citep{halu2013multiplex,battiston2016efficient,iacovacci2016functional} for existing variants on PageRank definitions in the context of multiplex networks.

 \subsubsection{Random walk betweenness versatility}
 
The random walk betweenness versatility~\citep{sole2016random} measures the importance of nodes in terms of the number of times random walk paths in the network pass by a given node. This quantity is suitable, for instance, for identifying the critical nodes in the random spreading of pieces of information, which are not necessarily following the shortest trajectories~\citep{freeman1991centrality}. To analytically compute the random walk betweenness versatility, it is convenient to take advantage of the concept of absorbing random walks, i.e. walks that will end when a given node $d$ is reached. These walks are defined by the absorbing transition tensor on a particular node $d$:
\begin{eqnarray}
(T_{[d]})^{i\sigma}_{j\beta} = \left\{ 
\begin{array}{l l}
0 & \quad j=d\\
T^{i\sigma}_{j\beta} & \quad j \ne d
\end{array} \right.
\end{eqnarray}

The average number of times a random walk (originating in node $o$ in layer $\sigma$ and destination $d$, independently of the layer) will pass by a node $j$ in layer $\beta$, regardless of the time step, is given by~\citep{sole2016random}:
\begin{eqnarray}
(\tau_{[d]})^{o\sigma}_{j\beta} = \left[(\delta - T_{[d]})^{-1}\right]^{o\alpha}_{j\beta},
\end{eqnarray}
where $\delta^{i\alpha}_{j\beta} = \delta^i_j\delta^\alpha_\beta$ and $\delta$ is the Kronecker delta. The average over all possible starting layers $\sigma$ and the aggregation of the walks that pass through $j$ in the different layers is obtained by
\begin{eqnarray}
(\tau_{[d]})^{o}_{j} = \frac{1}{L} (\tau_{[d]})^{o\sigma}_{j\beta} u^{\beta}u_{\sigma}.
\end{eqnarray}
Finally, the overall betweenness versatility is given by averaging over all possible origins and destinations:
\begin{eqnarray}
\tau_{j}  = \frac{1}{N(N-1)}\sum_{d=1}^{N}  (\tau_{[d]})^{o}_{j}u_{o}
\end{eqnarray}

\subsubsection{Random walk closeness versatility}

Random walk closeness centrality~\citep{sole2016random} of a node $i$ is defined as the inverse of the average number of steps required by a random walker, starting from any other node in the multilayer network, to reach $i$ for the first time.
As for the case of betweenness, by using the concept of absorbing random walks one can find that for walks originating in node $o$ in layer $\sigma$ the probability of visiting node $j$ in layer $\beta$ after $t$ time steps is given by $(T_{[d]})^{o\alpha}_{j\beta}$ to the power of $t$. 
Considering the walk starting from node $o$ in layer $\sigma$, each tensor encoding the mean first passage time to node $d$ is given by~\citep{sole2016random}:
\begin{eqnarray}
(H_{[d]})^{o\sigma}= \left[ (\delta - T_{[d]})^{-1}\right]^{o\sigma}_{j\beta} u^{j\beta}.
\end{eqnarray}
The average mean first passage time $h_{[d]}$ to node $d$ is obtained
by averaging $(H_{[d]})^{o\sigma}$ over all possible starting nodes and layers as:
\begin{eqnarray}
h_{[d]} = \frac{1}{(N-1)L} u_{o\sigma} (H_{[d]})^{o\sigma} + \frac{1}{N}\Pi^{-1}_{[d]},
\end{eqnarray}
where $\delta^{i\alpha}_{j\beta} = \delta^i_j\delta^\alpha_\beta$ and $\Pi_{[d]}$ is the random walk occupation centrality. The last term $\frac{1}{N}\Pi^{-1}_{[d]}$ is included to account for the average return times, usually excluded when using absorbing random walks. Finally, the random walk closeness centrality of node $d$ is simply obtained as the inverse of the distance $\xi_d= 1/h_{[d]}$.

\subsubsection{Betweenness centrality}

If a metric distance can be defined in a multilayer topology, it is possible to extend the classical definitions of node betweenness centrality, edge betweenness centrality and closeness centrality to obtain their multilayer counterparts~\citep{morris2012transport, magnani2013combinatorial}. 
Given the existence of layers, it is possible to select a subset of layers $\Omega$ and consider only the paths belonging to that subset, defining the \emph{cross betweenness centrality}~\citep{buccafurri2013measuring} of the node $i$:
\begin{eqnarray}
CBC(i,\alpha,\Omega) = \sum_{o \ne i,d \ne i, \beta \in \Omega, \beta \ne \beta'} \frac{\sigma_{o\beta}^{d\beta'}(i,\alpha)}{\sigma_{o\beta}^{d\beta'}},
\end{eqnarray}
which counts the fraction of inter-layer shortest paths, having their destination in $\Omega$, that pass through node $i$ of layer $\alpha$.
Taking advantage of this definition, the multiplex betweenness centrality (BC) can be decomposed into the following contributions:
\begin{eqnarray}
BC(i,\alpha) = CBC(i,\alpha,\Omega)+CBC(i.\alpha,\bar{\Omega}) + IBC(i,\alpha),
\end{eqnarray}
where $\bar{\Omega}$ indicates the layers not belonging to $\Omega$ and $IBC(i,\alpha)$ is called \emph{internal betweenness centrality} and represents the contribution at the BC of the paths that never leave the layer $\alpha$:
\begin{eqnarray}
IBC(i,\alpha) = \sum_{o \ne i,d \ne i, \alpha=\beta=\beta'} \frac{\sigma_{o\beta}^{d\beta'}(i,\alpha)}{\sigma_{o\beta}^{d\beta'}}.
\end{eqnarray}

Other definitions are also available and define betweenness versatility as another natural extension of the monoplex betweenness centrality~\citep{sole2014centrality,SoleRibalta2016congestion}.

\subsubsection{Interdependence centrality}

In general, the presence of more than one layer within a system increases the number of available paths with respect to the case where only one single layer is present. The richness of multilayer paths allows for the possibility that single-layer shortest paths are longer than multilayer ones. To quantify the value added by multiplexity to potential communication routes, it is possible to define the reachability of each physical node $i$ in terms of interdependence centrality~\citep{nicosia2013growing}, defined as:
\begin{eqnarray}
\lambda_i = \frac{1}{N-1} \sum_{j\ne i}\frac{\psi_{ij}}{\sigma_{ij}}, 
\end{eqnarray}
where $\psi_{ij}$ is the number of shortest paths between nodes $i$ to $j$ that pass by more than one layer and $\sigma_{ij}$ is the total number of shortest paths from $i$ to $j$. In both cases, the number of shortest paths can be greater than 1 if there are multiple paths having the same, minimal, length. Note that these quantities are not tensors or matrices: they can be better understood in terms of matrix entries, since they encode pairwise information about physical nodes. The global interdependence is computed by summing up over the interdependence centrality of all nodes: $\lambda = \langle \lambda_i \rangle$.

In some systems, such as human transportation, the flows between different pairs of nodes might be strongly dis-homogeneous. In such cases, the global interdependence can be redefined to take into account the real loads of the network by weighting the aforementioned sums by $T_{ij}$, which is an origin-destination matrix whose entries are normalised such that $\sum\limits_{i,j} T_{ij} = 1$. The weighted interdependence, also known as \emph{coupling}~\citep{morris2012transport}, is therefore given by

\begin{eqnarray}
\lambda' =  \sum_{i,j\ne i}T_{ij}\frac{\psi_{ij}}{\sigma_{ij}}.
\end{eqnarray}

\subsection{Mesoscale Organization}\label{sec:mesoscale}

\subsubsection{Triadic Closures}\label{sec:triads}

\begin{figure}[!t]
\centering\includegraphics[width=\textwidth]{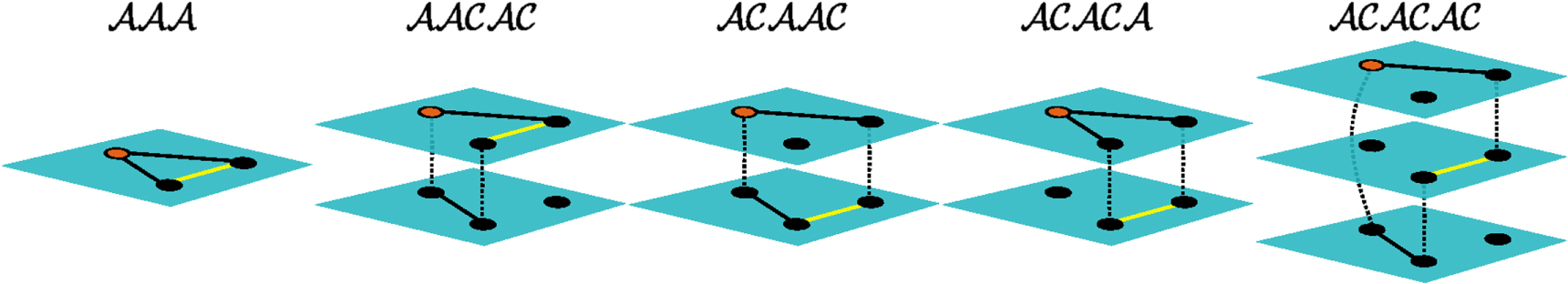}
\caption{\label{fig:cozzo_triadic} An elementary cycle on a multilayer network can include edges from up to three different layers, identifying a triad of physical nodes. The starting point of the cycle is highlighted in orange, intra-layer edges are dotted curves and intra-layer edges are solid lines. To simplify the reading of the cycle, the second intra-layer step is represented by a yellow line. Above each cycle it is indicated the tensorial form of the cycle in terms of the tensors $\mathcal{A}$ and $\mathcal{C}$ described in the text. Figure from~\citep{cozzo2015structure}.}
\end{figure}

The clustering coefficient, a measure of the cohesiveness of triads, is an important topological property of a node and a whole network. Nevertheless, it can also be defined as the number of 3-cycles that start and end at the same node~\citep{cozzo2015structure}, relating this network descriptor to how information flows and localizes. As exemplified in Fig.~\ref{fig:cozzo_triadic} for an interconnected multiplex network, a 3-cycle can go through different layers, but is characterized only by 3 intra-layer steps. These type of walks, in which after or before each intra-layer step the walker can choose between continuing on the same layer or changing to some other adjacent layer, can be explicitly decomposed into combinations of intra- and inter-layer steps by using adequately defined tensors:
\begin{equation}
\mathcal{A}_{i\alpha}^{j\beta} =\left\{ 
\begin{array}{l l}
M_{i\alpha}^{j\beta} & \quad \alpha=\beta\\
0 & \quad \alpha\ne \beta
\end{array} \right.
\end{equation}
encoding intra-layer steps, and
\begin{equation}
\mathcal{C}_{i\alpha}^{j\beta} =\left\{ 
\begin{array}{l l}
M_{i\alpha}^{j\beta} & \quad i=j\\
0 & \quad i\ne j
\end{array} \right.
\end{equation}
encoding inter-layer steps. Since changing layer is optional, we represent layer's choice as
\begin{equation}
\mathcal{\hat C}_{i\alpha}^{j\beta} = b I_\alpha^\beta + g \mathcal{C}_{i\alpha}^{j\beta},
\end{equation}
where $I$ is the identity matrix, $b$ represents the weight of staying in the layer $i$ and $g$ the weight of stepping from layer $\alpha$ to layer $\beta$.

With this definition, two types of cycles can be defined, and therefore two different number of cycles starting in $i$ can be obtained. In the first case, in which two consecutive inter-layer steps are forbidden, this number is $t_{(w),i\alpha} = 2\left((\mathcal{A}\mathcal{\hat C})^3\right)_{i\alpha}^{i\alpha}$,  corresponding to the examples shown in Fig.~\ref{fig:cozzo_triadic}), while in the second case, if those steps are permitted, this number is $ t_{(sw),i\alpha} = \left(( \mathcal{\hat C} \mathcal{A} \mathcal{\hat C})^3\right)_{i\alpha}^{i\alpha}$. 

To compute the multiplex clustering a normalisation factor $d_{(*),i\alpha}$ is also required, where with the label $_{(*)}$ we indicate both types of cycles, $_{(w)}$ and $_{(sw)}$. The corresponding definitions for both factors $d$ can be obtained from the $t$ numbers by replacing the tensor $\mathcal{A}$ in the second intra-layer step with the tensor:
\begin{equation}
F_{i\alpha}^{j\beta} =\left\{ 
\begin{array}{l l}
1 & \quad i \ne j\ \text{and} \  \alpha = \beta\\
0 & \quad \text{elsewhere}
\end{array} \right.
\end{equation}
For example, for the $_{(w)}$ type of cycles we have $d_{(w),i\alpha} = 2 \mathcal{A}\mathcal{\hat C} F \mathcal{\hat C} \mathcal{A}\mathcal{\hat C}$.
The local multiplex clustering coefficient for the state node $(i,\alpha)$ can be then calculated as:
\begin{equation}
C_{(*),i\alpha}= \frac{t_{(*),i\alpha} }{d_{(*),i\alpha} },
\end{equation}
while the global multiplex clustering coefficient, the scalar value representative of triadic closure in the whole system, is given by
\begin{equation}
C_{(*)} = \frac{\sum\limits_{i\alpha}t_{(*),i\alpha} }{\sum\limits_{i\alpha}d_{(*),i\alpha} }
\label{eq:multiplex_clustering_coeff}
\end{equation}

\begin{figure}[!t]
\centering\includegraphics[width=0.7\textwidth]{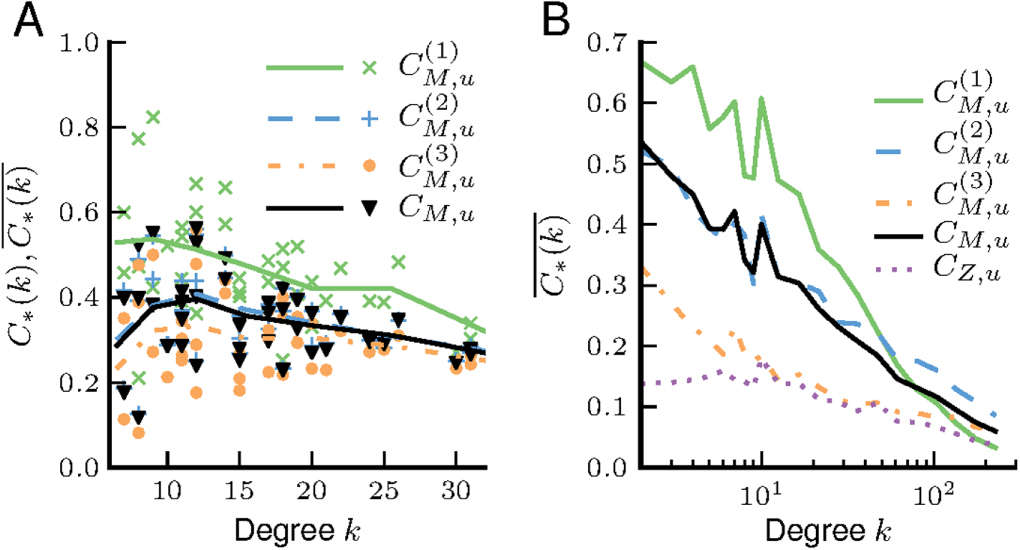}
\caption{\label{fig:cozzo_clustering}
Relation between local clustering coefficient, an indicator of triadic closure, and node's unweighted degrees for $m$-layers cycles $C^{(m)}_{M,u}$ as defined in Eq.~(\ref{eq:local_clustering_limited}), and for all cycles but considering the multilayer network with its unweighted $C_{M,u}$ or weighted $C_{Z,u}$ representation. Two empirical systems are considered: a) Kapferer tailor-shop social network; b) the airport network obtained from openflights.org. The curves interpolate the local clustering coefficient for each type of cycle. Figure from~\citep{cozzo2015structure}.}
\end{figure}

We can decompose Eq.(~\ref{eq:multiplex_clustering_coeff}) into its contributions coming from cycles traversing a different number $m$ of layers as
\begin{equation}
C^{(m)}_{(*)} = \frac{\sum\limits_{i\alpha}t_{(*),i\alpha}^{(m)}} {\sum\limits_{i\alpha}d_{(*),i\alpha}^{(m)}},
\label{eq:local_clustering_limited}
\end{equation}
where $t_{(*),i\alpha}^{(m)}$ and $d_{(*),i\alpha}^{(m)}$ are restricted to cycles involving exactly $m$ layers, with $m=1,2,3$. This perspective can be useful to appreciate the differences observed for cycles of a different nature, evident in Fig.~\ref{fig:cozzo_clustering}, where local clustering coefficients are drawn as a function of the nodes' degrees. In several empirical multiplex systems~\citep{cozzo2015structure}, it has been observed that $C < C^{(1)}$ and $C^{(1)} > C^{(2)} > C^{(3)}$ as a general pattern which provides a kind of hierarchy of triadic closure across layers.

\begin{figure}[!t]
\centering\includegraphics[width=0.55\textwidth]{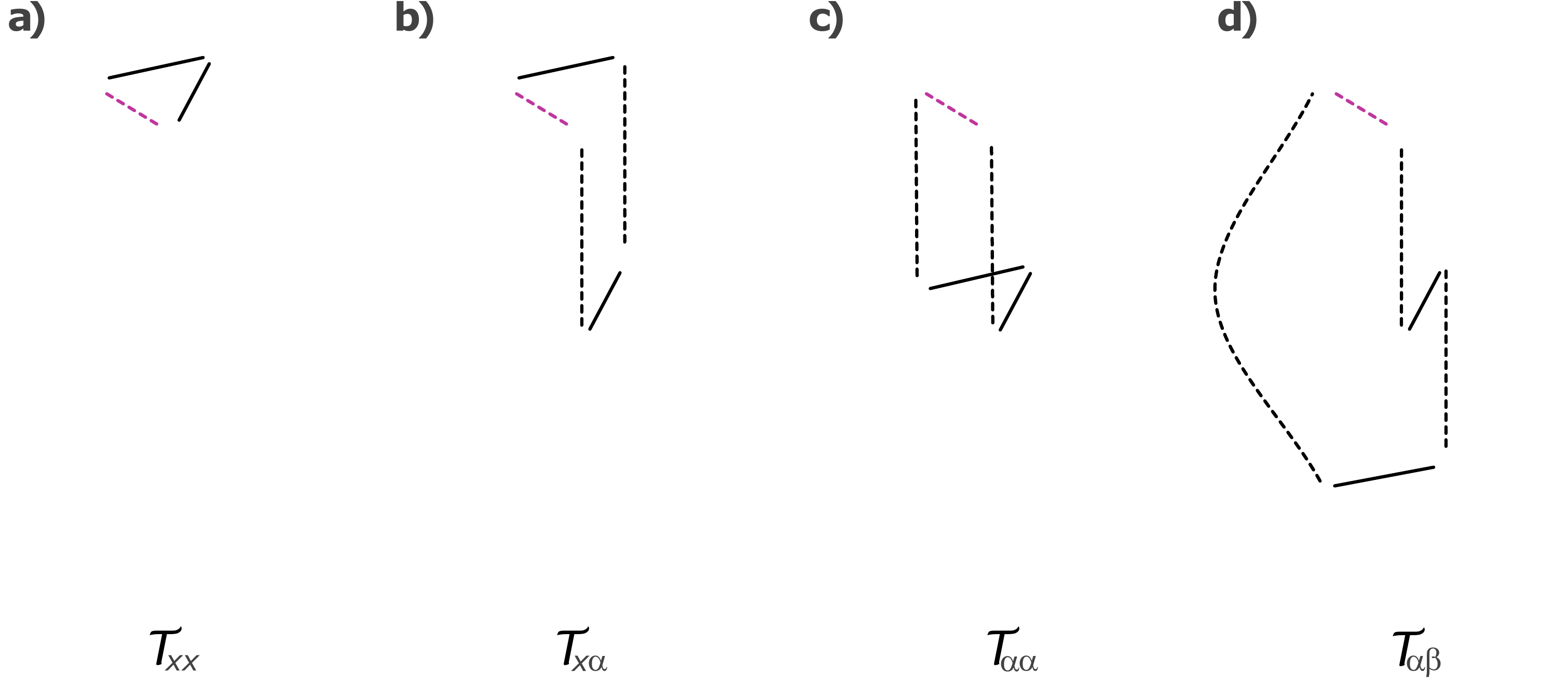}
\caption{\label{fig:aleta_clos_1}Four possible types of triadic closures on a multilayer network that can be used for predicting the existence of missing links. (a) $\mathcal{T}_{xx}$, if the two links belong to the same layer $x$ where the cycle is closed. (b) $\mathcal{T}_{x\alpha}$, if one link belongs to $x$ and another to a different layer $\alpha$. (c) $\mathcal{T}_{\alpha\alpha}$ if both existing links belong to the same layer, different from $x$. (d) $\mathcal{T}_{\alpha\beta}$ if the two existing links belong to two separate layers, both different from $x$. Figure from~\citep{aleta2020link}.}
\end{figure}

It is worth noticing that triadic closure can also be used to predict future or missing links on a multilayer network, as recently shown in~\citep{aleta2020link}. As can be seen from Fig.~\ref{fig:aleta_clos_1}, depending on the location of the links there are four possible triadic relations which allow us to to predict a missing link by closing a cycle of two links, $u$ on layer $\alpha$ and $v$ on layer $\beta$, with a third link on layer $x$. This observation allows for a multilayer generalization of the Adamic-Adar method~\citep{adamic2003friends}, which predicts links on the basis of a score based on the number of common neighbors weighted by their degree.

In fact, in multilayer networks, the neighbors of a node can belong to different layers. An appropriate ``Multilayer Adamic-Adar'' (MAA) score counts the common neighbors closing the triads of each of the aforementioned types, weighting each contribution by the logarithm of the degree as usually done to calculate the Adamic-Adar score:
\begin{equation}
\label{eq:MAA_aleta}
MAA(u,v)  =  \sum_{\alpha, \beta}    \sum_{w\in\mathcal{T}_{\alpha\beta}}  \frac{\eta_{x\alpha} \eta_{x\beta}}{\sqrt{\langle k \rangle_\alpha \langle k \rangle_\beta}} 		 \frac{1}{\sqrt{\ln(k_w^\alpha ) \ln(k_w^\beta)}}
\end{equation}
where $\langle k \rangle_\alpha$ indicates the average degree of nodes in layer $\alpha$, $k_w^\alpha$ the degree of node $w$ in layer $\alpha$ and $\eta_{x\alpha}$ are free coefficients allowing to control the relative weight of each type of triadic closure in the link's total score.

\begin{figure}[!ht]
\centering\includegraphics[width=\textwidth]{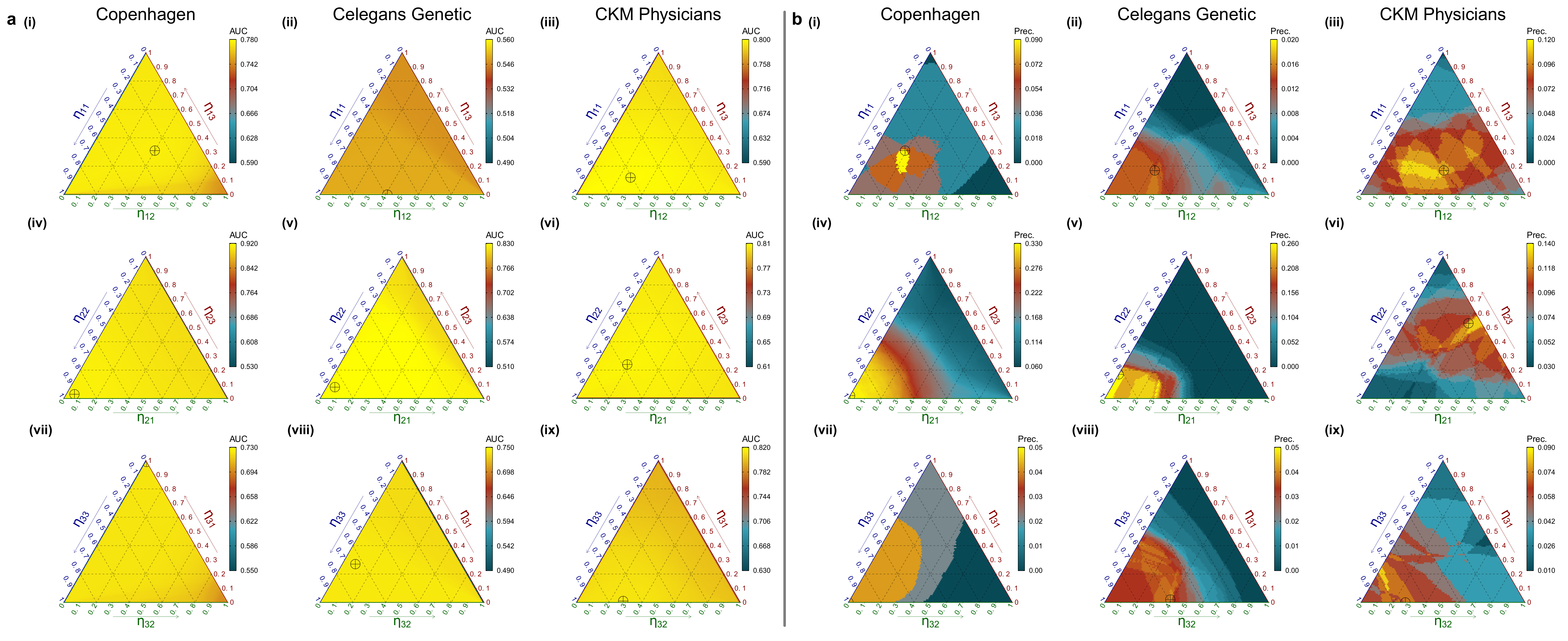}
\caption{\label{fig:aleta_clos_2}How areas under the curve (AUC) (a) and precision (b) change for different values of the coefficients $\eta_{x\alpha}$, for three real-world multiplex networks. For sake of simplicity, in each network only three layers are considered. In both panels, each column corresponds to a different data set, from left to right: a social network of students in Copenhagen, the \emph{C. elegans} genetic network, and the CKM social network of physicians. Each row corresponds to a prediction in a different layer. The cross marker indicates the location of the maximum value for each plot, corresponding to the combination of coefficients optimizing the indicator. Figure from~\citep{aleta2020link}.}
\end{figure}

Fig.~\ref{fig:aleta_clos_2} shows the AUC (panel a) and the precision (panel b) of MAA for three empirical systems, as a function of the coefficients $\eta_{x\alpha}$. The results show that single-layer link prediction can be largely improved by exploiting information encoded in other layers.  

\subsubsection{Communities and modules}\label{sec:groups}

Complex systems are characterized by the mesoscale organization of their units into groups, also known as \emph{modules} or \emph{communities}~\citep{fortunato2010community,newman2012communities,fortunato2016community}. However, the definition of what a group exactly corresponds to is an open problem. Here, we will briefly describe the major advances in this direction, considering four methods widely used in the literature, namely: multilayer modularity maximization, multilayer tensor factorization, multilayer Bayesian inference and multilayer description length minimization through the map equation. A variety of methods is available, including the analysis of intermittent communities in time-varying networks~\citep{aslak2018constrained}, but they are beyond the scope of this work and deserve a dedicated review~\citep{holme2012temporal,holme2015modern}.

\paragraph{Multilayer modularity maximization}

\begin{figure}[!ht]
\centering
\begin{minipage}[c]{0.4\textwidth}
\includegraphics[width=\textwidth]{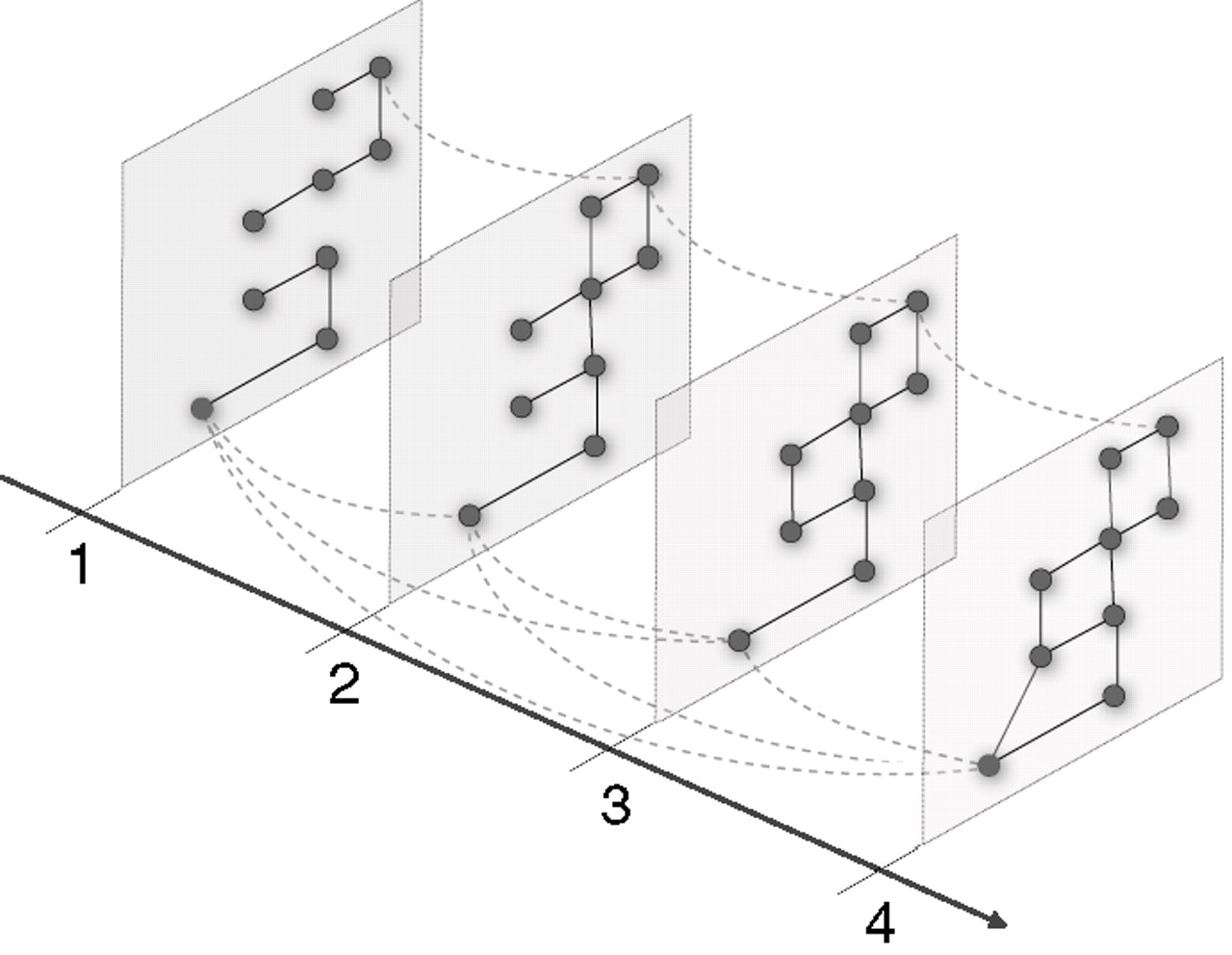}
\end{minipage}
\quad
\begin{minipage}[c]{0.5\textwidth}
\caption{\label{fig:mod_mucha1} Illustration of a \emph{multislice} network representing, for instance, a time-varying network where each temporal snapshot is encoded by a layer and inter-layer connectivity encodes temporal relationships among state nodes. Figure from~\citep{mucha2010community}.}
\end{minipage}
\end{figure}

The simplest -- and also one of the most popular ones -- definition concerns the density of links within a group with respect to inter-group density of links: for a module, we expect units to be mostly connected with other units inside the module and poorly connected outside. This approach is based on the calculation of a function named \emph{modularity} which, for a given partitioning of the units into groups, quantifies the deviation of the number of links from what is expected by chance according to the corresponding configuration model~\citep{newman2006modularity}. This function is therefore calculated for all possible partitions, and the partition with maximum modularity is chosen as the most representative mesoscale organization of the underlying network, providing a hint about its structural and functional organization. For this reason, identifying such an organization is of paramount importance for applications in many disciplines, from social sciences to biology~\citep{girvan2002community,guimera2005functional}.

\begin{figure}[!t]
\centering
\begin{minipage}[c]{0.45\textwidth}
\includegraphics[width=\textwidth]{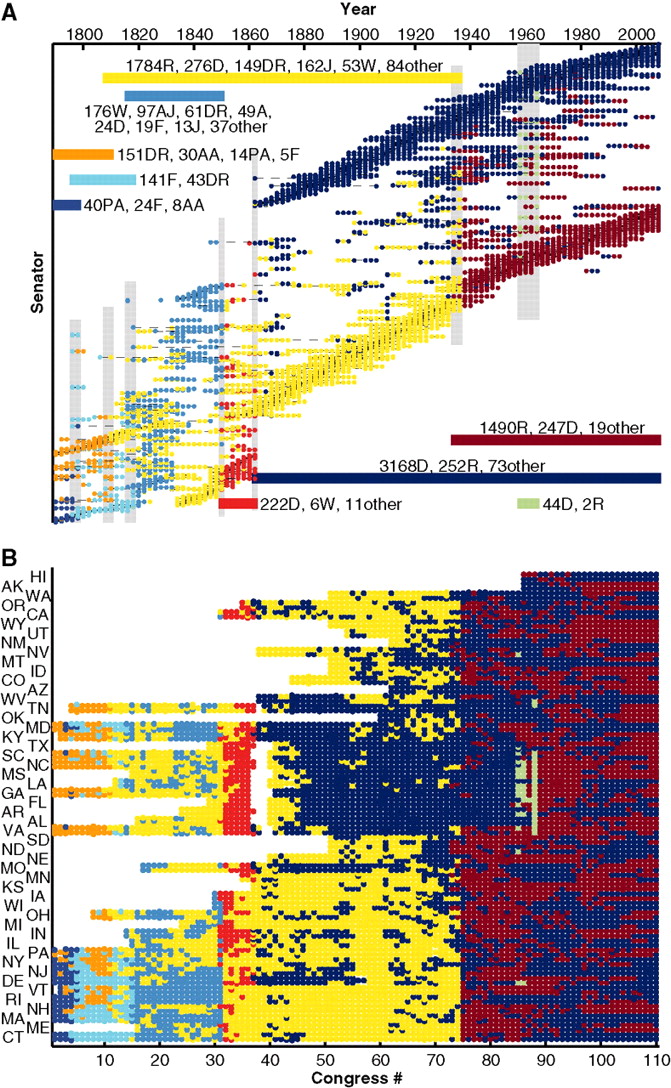}
\end{minipage}
\quad
\begin{minipage}[c]{0.5\textwidth}
\caption{\label{fig:mod_mucha2} Evolution over time of political groups, identified by multilayer modularity maximization, in the U.S. Senate across more than two centuries (110 layers encoding the number of 2-year Congresses from 1789 to 2008). Intra-layer links are given by roll call vote similarities and inter-layer connectivity is weighted by $\omega=0.5$. (A) Results from senators' assignments. (B) Results from state affiliations. Note that less sophisticated analytical techniques, applied for instance on aggregated representation of this system, would not identify the same patterns. Analysis and Figure from~\citep{mucha2010community}.}
\end{minipage}
\end{figure}

The definition of the multilayer generalization of modularity was given by~\citep{mucha2010community} a decade ago, become a standard for applications~\citep{carchiolo2011communities,amelio2019modularity}. It assumes that the system can be represented by a three-dimensional tensor, encoding within-layer connectivity as for edge-colored multigraphs and time-varying networks\footnote{Where each snapshot is encoded by a layer.} (see Fig.~\ref{fig:mod_mucha1}) and by another three-dimensional tensor encoding inter-layer connectivity. To allow for a multi-resolution analysis\footnote{We refer the interested reader to \citep{fortunato2007resolution,arenas2008analysis,Taylor2017superresolution} for more information about the resolution and super-resolution problems in community detection. Note that recent studies have shown that by aggregating layers by summation, for instance, it is also possible to enhance the detectability of community structures~\citep{Taylor2016detectability}.}, a resolution parameter $\gamma$ is introduced, and inter-layer connectivity is weighted by another tunable parameter $\omega$. Using the tensorial notation, considering the more general systems that can be represented by the multilayer adjacency tensor $M^{i\alpha}_{j\beta}$, and indicating by $P^{i\alpha}_{j\beta}$ the tensor encoding the connectivity values obtained from a null model (e.g., the configuration model), one can define the rank-4 tensor 
\begin{eqnarray}
B^{i\alpha}_{j\beta}=M^{i\alpha}_{j\beta} - P^{i\alpha}_{j\beta},
\end{eqnarray}
which is then used to define the modularity function
\begin{eqnarray}
Q \propto S^{a}_{i\alpha}B^{i\alpha}_{j\beta}S^{j\beta}_{a}
\end{eqnarray}
to be maximized while varying the membership tensor $S^{i\alpha}_{a}$, with entries equal to 1 if node $i$ in layer $\alpha$ belongs to the community $a$, and zero otherwise~\citep{de2013mathematical}. This approach can be suitably used to identify groups in time-varying systems, like the U.S. Senate (Fig.~\ref{fig:mod_mucha2}). We refer to~\citep{bazzi2016community} for an enhanced version of this method and to the recent work demonstrating that, under certain conditions, modularity maximization corresponds to maximizing the posterior probability of community assignments under suitably chosen stochastic block models~\citep{pamfil2019relating}. For further information about the latter, see also the subsection on multilayer Bayesian inference later in this text.

\begin{figure}[!t]
\centering\includegraphics[width=0.7\textwidth]{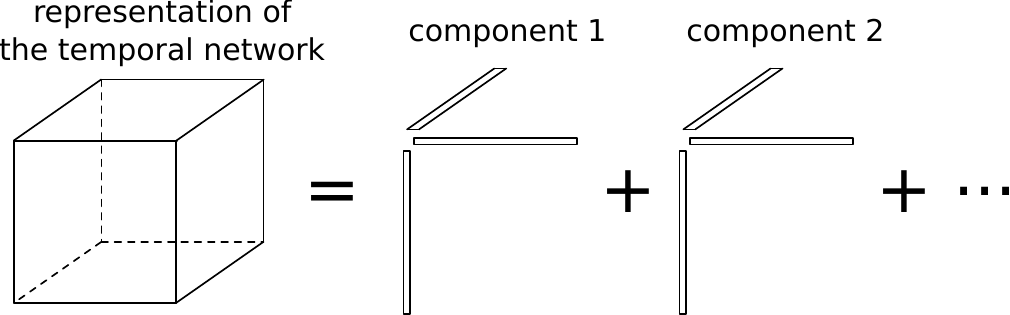}
\caption{\label{fig:gauvain_1} Illustrating the Kruskal decomposition of a rank-3 tensor representing a time-varying network. The tensor is factorized and decomposed into the sum of the outer product of rank-1 tensors. Figure from~\citep{gauvin2014detecting}.}
\end{figure}

\paragraph{Multilayer tensor factorization}

At variance with modularity maximization, this approach is based on decomposing the three-dimensional tensor $T_{ij\alpha}$, encoding the inter-layer connectivity of a time-varying network where the order of tensors follows the arrow of time, by means of non-negative factorization. This procedure maps a tensor into the sum of outer products of rank-1 tensors as schematically shown in Fig.~\ref{fig:gauvain_1}.

\begin{figure}[!ht]
\centering\includegraphics[width=0.7\textwidth]{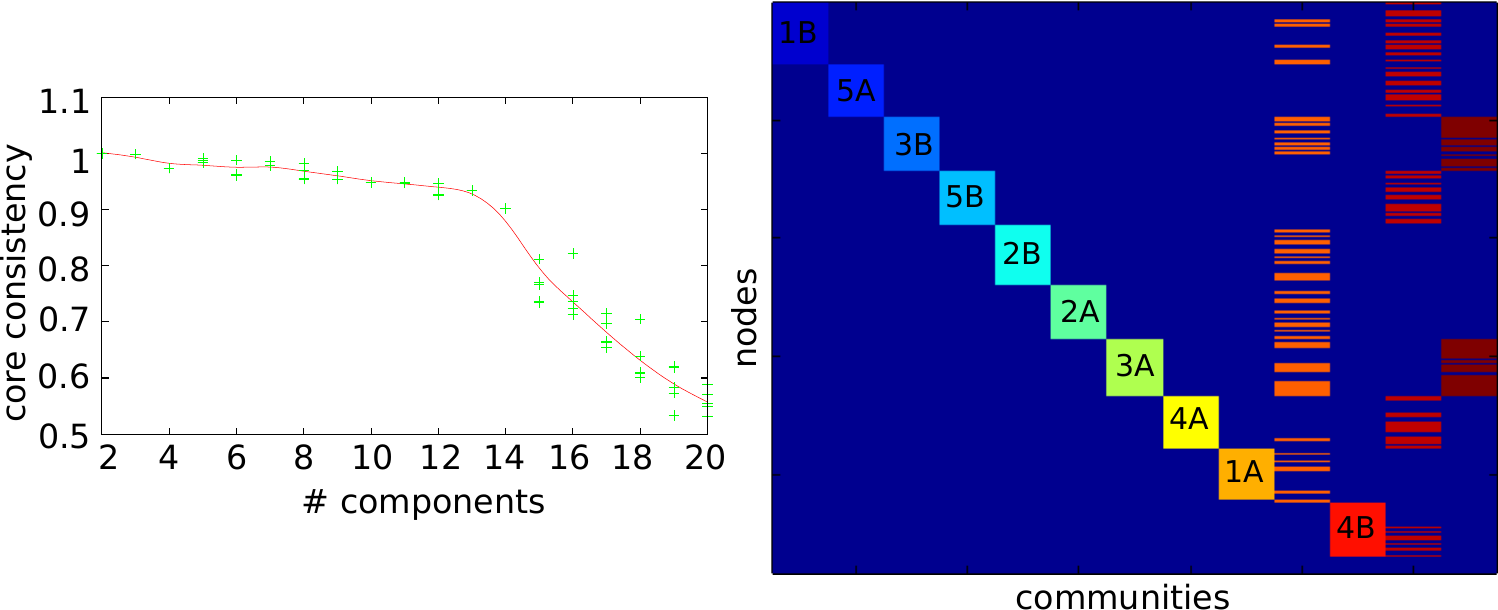}
\caption{\label{fig:gauvain_2} Non-negative tensor factorization of temporal snapshots (the layers) describing close-range interactions (the links) of students and teachers (the nodes), divided into classes. The panels show the distribution of membership weights encoded by the first output of the procedure. Figure from~\citep{gauvin2014detecting}.}
\end{figure}

The result of this procedure is a coarse-grained representation of the system in terms of two assignments: nodes to groups, giving the membership weight of units, and time intervals to groups, giving the activity level of each group at different temporal snapshots. When applied to a contact network, for instance, the result of this approach is shown in Fig.~\ref{fig:gauvain_2}. 

\paragraph{Multilayer Bayesian inference}

\begin{figure}[!t]
\centering
\begin{minipage}[c]{0.5\textwidth}
\includegraphics[width=\textwidth]{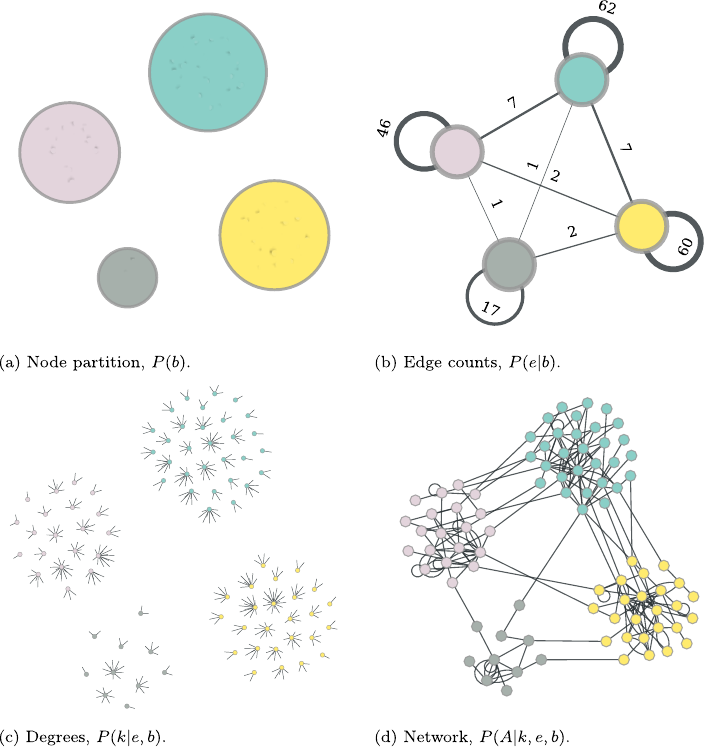}
\end{minipage}
\quad
\begin{minipage}[c]{0.45\textwidth}
\caption{\label{fig:peixoto_0} Summarizing a non-parametric generative process in the case of a monoplex network, according to degree-corrected stochastic block modeling (DCSBM): (a) sampling the partition of units; (b) counting edges between and within groups; (c) imposing the observed degree sequence; (d) sampling the overall network by accounting for (a--c). Figure from~\citep{peixoto2017nonparametric}.}
\end{minipage}
\end{figure}

A different approach is to consider the problem of recovering the best partition of the nodes, where one assumes that there is an underlying model encoding some mechanisms at work to generate the observed system. This class of methods, originally developed for analysis of social networks~\citep{holland1983stochastic,snijders1997estimation,nowicki2001estimation}, assumes that groups can be encoded by blocks: probability to have links within and between blocks, as well as the number of blocks, are the parameters of the model to be calculated. This model is known as the Stochastic Block Model (SBM), and it finds extensive applications in machine learning~\citep{airoldi2008mixed,goldenberg2010survey,qin2013regularized,anandkumar2014tensor}. More recently, a variation based on Bayesian inference and statistical physics has been proposed for applications to multilayer networks~\citep{peixoto2015inferring,valles2016multilayer}. A thorough discussion of the mathematical framework behind this powerful approach is beyond the scope of this section, therefore we refer the interested reader to an accurate and recent work fully devoted to this purpose~\citep{peixoto2019chapter}.

\begin{figure}[!t]
\centering
\begin{minipage}[c]{0.5\textwidth}
\includegraphics[width=\textwidth]{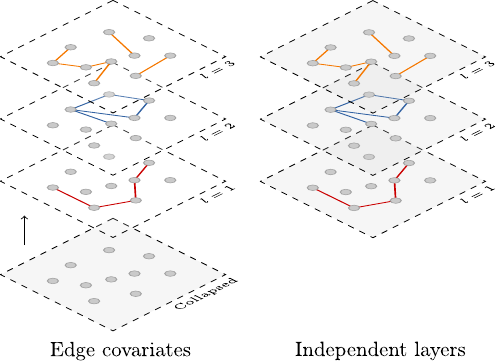}
\end{minipage}
\quad
\begin{minipage}[c]{0.45\textwidth}
\caption{\label{fig:peixoto_1} Generative models of multilayer networks. Left: generate the aggregated (collapsed) representation of the system and then create layers whose aggregation leads to the same representation. Right: each layer is generated independently from the others. Figure from~\citep{peixoto2015inferring}.}
\end{minipage}
\end{figure}

Here, we outline the basic idea behind this method~\citep{peixoto2015inferring}, which is schematically summarized in Fig.~\ref{fig:peixoto_0} for the case of a monoplex. The input is considered an edge-colored multigraph or a time-varying network that can be represented by an array of matrices, similar to the ones we have described in the previous sections. Let us indicate this object with $M_{ij}^{\alpha}$, and its aggregation obtained by entry-wise summation across layers as $A_{ij}=M_{ij}^{\alpha}u_{\alpha}$. Let $\boldsymbol{\Theta}$ indicate the set of all parameters of the model to fit. The Bayes theorem in this case reads
\begin{eqnarray}
P(\boldsymbol{\Theta} | M_{ij}^{\alpha} )=\frac{P(M_{ij}^{\alpha}|\boldsymbol{\Theta})P(\boldsymbol{\Theta})}{P(M_{ij}^{\alpha})},
\end{eqnarray}
where $P(\boldsymbol{\Theta})$ is the prior probability on the parameters, $P(M_{ij}^{\alpha})$ is a normalization factor and $P(M_{ij}^{\alpha}|\boldsymbol{\Theta})$ is the likelihood of observing the multiplex system encoded by $M_{ij}^{\alpha}u_{\alpha}$ given the parameters. Here, $P(\boldsymbol{\Theta} | M_{ij}^{\alpha} )$ is the posterior likelihood: the higher its value the more likely our model describes the data. By noticing that: 

\begin{itemize}
\item $P(\boldsymbol{\Theta})=e^{-\mathcal{L}(\boldsymbol{\Theta})}$, where $\mathcal{L}(\boldsymbol{\Theta})$ is the microcanonical entropy of the parameter ensemble; 
\item $P(M_{ij}^{\alpha}|\boldsymbol{\Theta}) = e^{-\mathcal{S}(M_{ij}^{\alpha})}$, where $\mathcal{S}(M_{ij}^{\alpha})$ is the microcanonical entropy of the model ensemble; 
\item the term $P(M_{ij}^{\alpha})$ can be neglected since it does not depend on parameters and only acts as a constant; 
\end{itemize}
it is straightforward to show\footnote{Hint: take the natural logarithm of both sides of the Bayes formula, discard the constant term and solve by $\Sigma$.} that minimizing the function
\begin{eqnarray}
\Sigma=\mathcal{S}(M_{ij}^{\alpha}) + \mathcal{L}(\boldsymbol{\Theta}),
\end{eqnarray}
known as the \emph{description length}, is equivalent to maximizing the posterior probability. Note that $\Sigma$ has a very nice interpretation in terms of information theory: it is the number of bits required to describe the data summed to the number of bits required to describe the model.

\begin{figure}[!t]
\centering\includegraphics[width=0.55\textwidth]{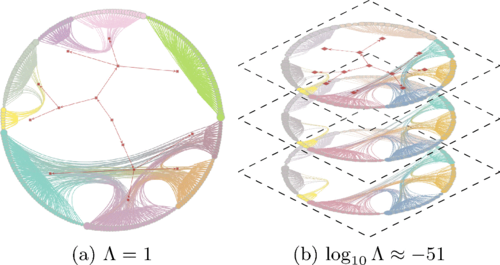}
\caption{\label{fig:peixoto_2} A non-interconnected multiplex social network of physicians is reproduced by means of two distinct generative models. The layout is specially designed to highlight the presence of hierarchies, indicated by red markers and links, of nodes -- the circular dots arranged on a circle, linked by observed edges drawn by means of bundling to better put in evidence inter-group connections. (a) the inferred degree-corrected SBM is directly applied to the aggregated network; (b) inference is performed while accounting for edge covariates, by assigning each layer to a type of social interaction. Below each panel, the value of the posterior odds ratio is shown. Figure from~\citep{peixoto2015inferring}.}
\end{figure}

The explicit definition of the prior probability $P(\boldsymbol{\Theta})$ is often a subtle issue involving the choice of a generative process for the parameters. This is a problem-specific task which depends on the \textit{a priori} knowledge and assumptions about the data. In a more general setting, and in those cases in which the prior knowledge is missing, it is desirable to choose \textit{uninformative} priors. In \citep{peixoto2015inferring} the author proposes a nested uninformative prior which minimizes the influence on the posterior.

\begin{figure}[!t]
\centering\includegraphics[width=\textwidth]{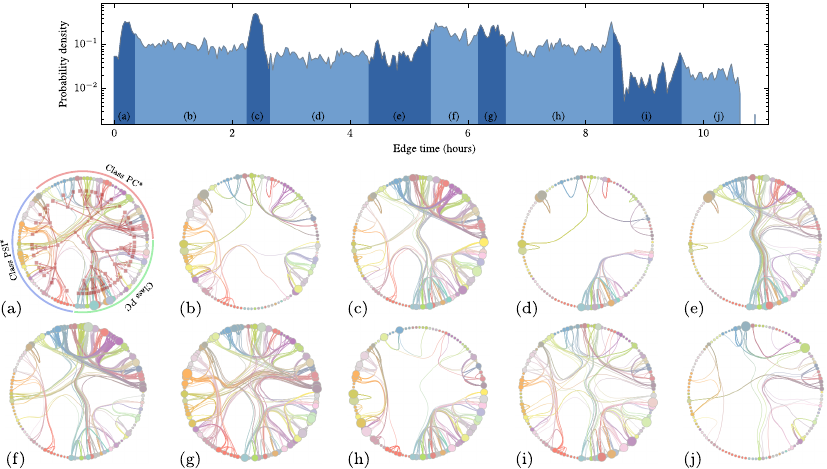}
\caption{\label{fig:peixoto_3} A proximity network of high-school students, varying over time, is considered. The probability of existence of an edge is shown, across time, in the top panel: the colored areas, labeled from (a) to (j), are the ones obtained from the inference process as the best partition into layers. For each area, the corresponding partitioning is shown in the bottom panels by using the hierarchical layout. Figure from~\citep{peixoto2015inferring}.}
\end{figure}

One can choose different models to generate the observed network and, specifically, plausible options are to either generate layers conditional to the aggregate representation of the system or to generate each layer independently from the others (Fig.~\ref{fig:peixoto_1}), as well as other generative models -- such as the M-DCSBM -- specifically designed for multilayer systems~\citep{bazzi2020framework}. The reader could be concerned about how to choose the best model: in fact, the framework also allows for model selection and one should take the model which minimizes the description length. When one needs a more refined approach, where a model or its alternative should be chosen, non-parametrically, with some degree of confidence, it is possible to calculate posterior odds ratio as
\begin{eqnarray}
\Lambda = \frac{ P(\boldsymbol{\Theta}_{a} | M_{ij}^{\alpha}, \mathcal{H}_{a}) P(\mathcal{H}_{a}) }{ P(\boldsymbol{\Theta}_{b} | M_{ij}^{\alpha}, \mathcal{H}_{b}) P(\mathcal{H}_{b}) }=e^{-\Delta \Sigma} \frac{ P(\mathcal{H}_{a}) }{ P(\mathcal{H}_{b}) },
\end{eqnarray}
where each model class is encoded into the hypothesis $\mathcal{H}_{p}$ ($p=a,b$), $P(\mathcal{H}_{p})$ is the prior belief for that hypothesis and $\Delta \Sigma=\Sigma_{a}-\Sigma_{b}$ is the difference between the description length of each hypothesis. The operational prescription is that if $\Lambda<1$ then $\mathcal{H}_{b}$ should be preferred to $\mathcal{H}_{a}$, whereas when $\Lambda\approx 1$ both models are equally plausible. More generally, if $\Lambda\in[1/3,1]$ the difference between the two hypotheses can be considered negligible. An application of this procedure is shown in Fig.~\ref{fig:peixoto_2} in the case of an empirical social network.

Finally, this procedure can be used to identify the best division of time-varying edges into layers, as shown in the case of a real-world contact network in Fig.~\ref{fig:peixoto_3}.

We refer the interested reader to a more recent work~\citep{pamfilinference}, where an SBM-like model -- not relying on sampling of edges in different layers independently -- is proposed to account for edge correlations, and to define a new measure of layer-layer correlation which incorporates similarity between connectivity patterns in different layers. It is used for link prediction in multilayer networks.

\paragraph{Multilayer description length minimization through the map equation}

\begin{figure}[!t]
\centering\includegraphics[width=\textwidth]{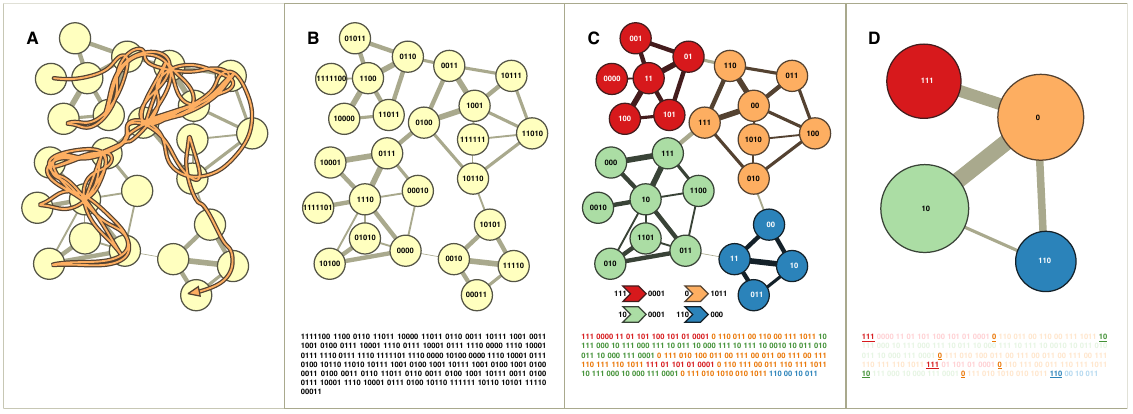}
\caption{\label{fig:chap_infomap} Detection of modules based on the Infomap algorithm. (A) The solid line represents a trajectory of random walks on the top of a classical network. (B) By encoding each node with a codeword, e.g. using a Huffman scheme, one can efficiently describe the trajectories in terms of strings by appending the codeword of each node visited by the walker. The trajectory shown in (A) is encoded into a string of 314 bits. (C) Given a partitioning of the system, the persistence of the trajectory in a certain module is encoded by a certain number of bits: the lower this number the more persistent the trajectory in that module. This regularity can be used to coarse-grain the network into functional modules as in (D). Figure from~\citep{rosvall2008maps}.}
\end{figure}

Instead of relying only on topological information -- such as the abundance of edges or the presence of blocks -- to identify groups, one can observe how information flow is trapped within modules, the rationale being that information is more likely to flow within a group than between groups~\citep{pons2006computing,lambiotte2014random}. Another approach might be to map the potential regularities of the flow into strings and to look for the partitioning of the system that minimizes the corresponding description length~\citep{rosvall2007information,rosvall2008maps,esquivel2011compression}, as schematically described in Fig.~\ref{fig:chap_infomap}.

\begin{figure}[!t]
\centering\includegraphics[width=0.8\textwidth]{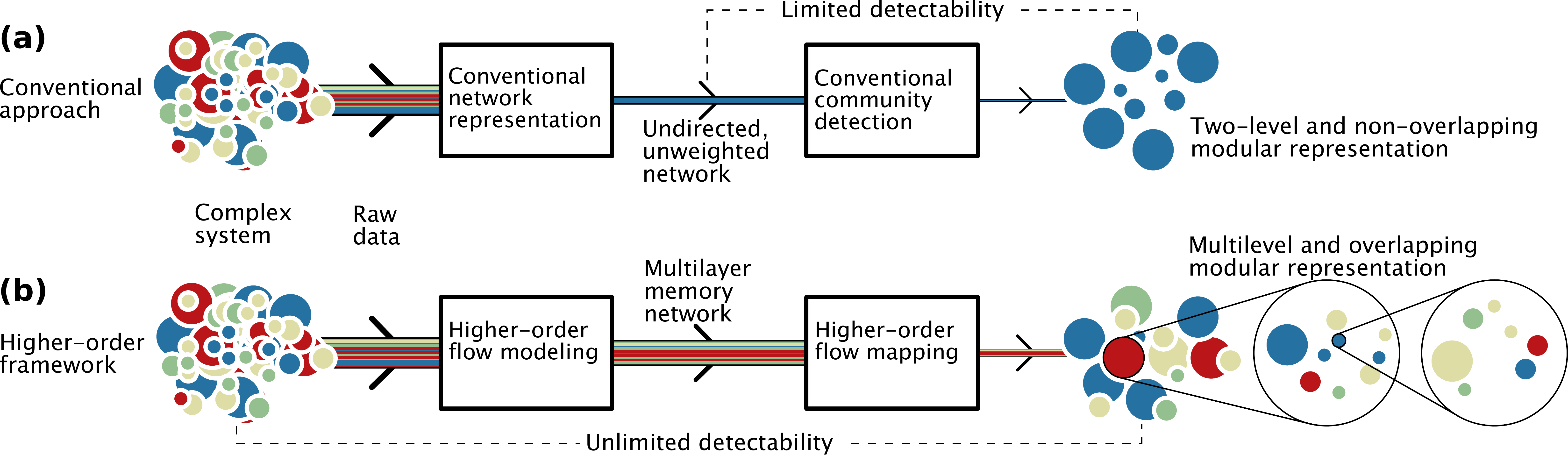}
\caption{\label{fig:chap_highorder} A complex system is analyzed using a conventional approach (a) and the higher-order framework (b) provided by the map equation. The latter allows us to account for the richness of the data, such as weights and directionality, without being affected by limited detectability (e.g., due to the resolution problem mentioned for modularity maximization) and, remarkably, is suitable to detect multilevel and overlapping modular information within the same elegant mathematical framework. Figure from~\citep{edler2017mapping}.}
\end{figure}

The generalization to multilayer networks of this method, known as Multiplex InfoMap, was introduced in~\citep{de2015identifying} and, more recently, it was better formalized under the perspective of higher-order modeling~\citep{rosvall2014memory,salnikov2016using,edler2017mapping,lambiotte2019networks}, as schematically illustrated in  Fig.~\ref{fig:chap_highorder}.

\begin{figure}[!t]
\centering
\begin{minipage}[c]{0.5\textwidth}
\includegraphics[width=\textwidth]{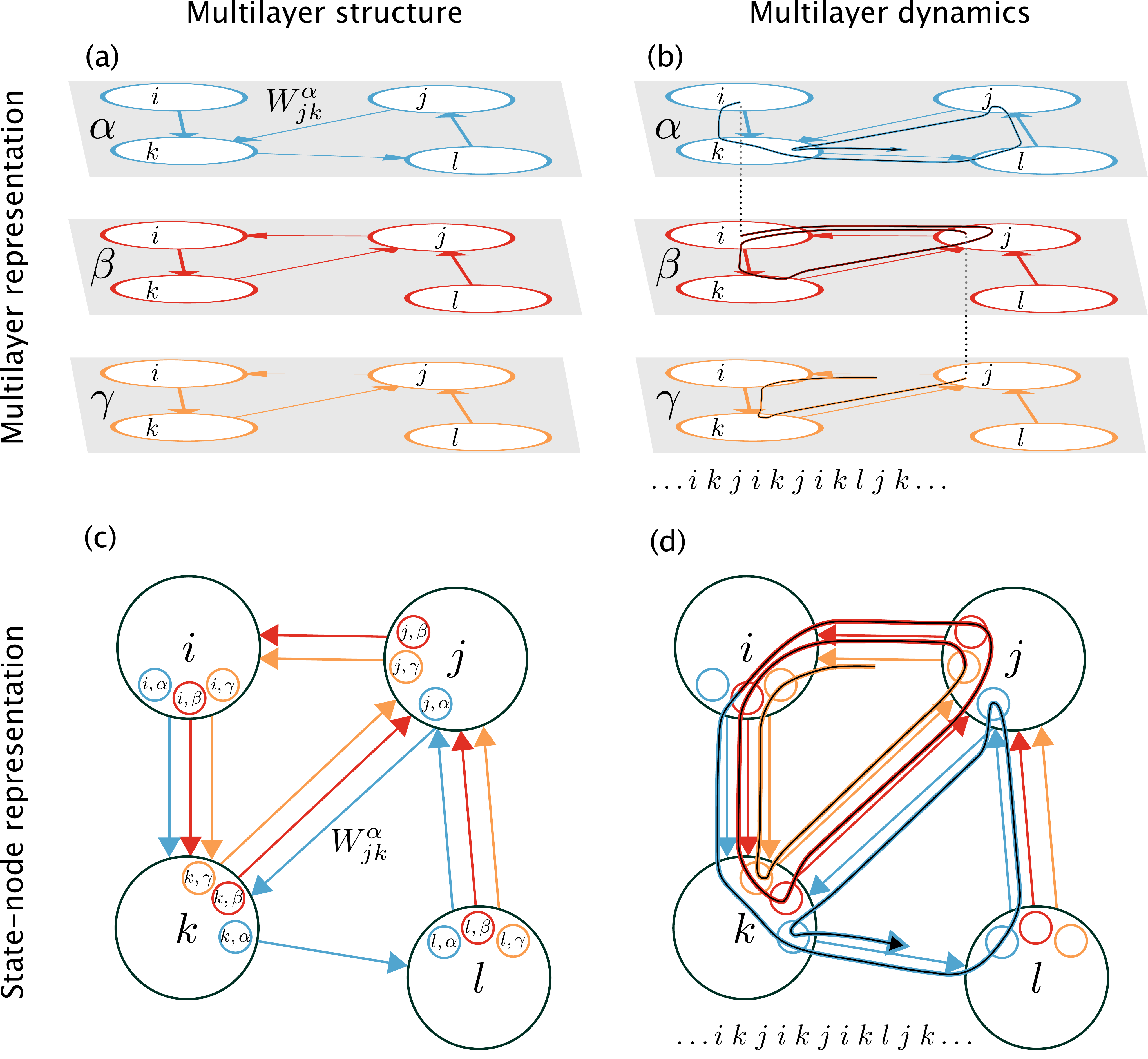}
\end{minipage}
\quad
\begin{minipage}[c]{0.45\textwidth}
\caption{\label{fig:chap_muxinfomap} Toy multilayer system with 4 physical nodes and 3 layers, represented in terms of structure (a-c) and dynamics (b-d) with a layered perspective (a-b) and with respect to state-node representation (c-d), the latter highlighting the existence of memory (encoded by small colored nodes) which alters the information flow (solid, colored, lines). Figure from~\citep{de2015identifying}.}
\end{minipage}
\end{figure}

Multiplex InfoMap is based on the same principles of its single-layer counterpart, with some important differences. First, the monoplex InfoMap is based on Markovian dynamics, whereas this feature is kept only for within-layer dynamics. Second, coding only captures the visits to physical nodes, not the ones to state nodes, an effective non-Markovian dynamics across layers. The remainder of the method is the same: compressing persistent multilayer trajectories\footnote{The curious reader might wonder about the governing equations of the corresponding random walk: this will be discussed in the chapter devoted to dynamics and, specifically, in Sec.~\ref{sec:diffproc}.} to identify modules according to the map equation
\begin{eqnarray}
\Sigma(\mathcal{M})=q_{in}\mathcal{S}(\mathcal{Q}) + \sum\limits_{\ell=1}^{m} p_{\ell}\mathcal{S}(\mathcal{P}_{\ell})
\end{eqnarray}
which encodes information flows within and across layers, as shown in Fig.~\ref{fig:chap_muxinfomap}. This equation deserves a careful explanation of its terms. 

Here, $\mathcal{M}$ is a partitioning of the system -- i.e., a model for the observed modules -- and $\Sigma$ is its description length, in bits. Indicating with $q_{\ell,in}$ and $q_{\ell,out}$ the transition rates at which a random walker enters and exits, respectively, a module $\ell$, with $q_{in}=\sum_\ell q_{\ell,in}$ the sum of those rates and with $\mathcal{S}(\mathcal{Q})$ the information entropy of the normalized probability distribution of the transition rates ($Q=\{q_{\ell,in}/q_{in}\}$), the first term in the right-hand side of the map equation captures the number of bits required to describe the dynamics. Indicating with $p_{l \in \ell}$ the visit rates of the physical nodes for the module codebook $\ell$, with $p_{\ell}$ the sum of those rates and with $\mathcal{S}(\mathcal{P}_{\ell})$ the information entropy of the corresponding normalized probability distribution ($\mathcal{P}_{\ell}=\{p_{l \in \ell}/p_{\ell}\}$), the second term of the map equation captures the number of bits required for coding. An application of Multiplex InfoMap to a toy system is schematically summarized in Fig.~\ref{fig:chap_muxinfomap}.

The interested reader can play interactively with an online version of this method, which is publicly available\footnote{\url{http://www.mapequation.org/apps/sparse-memory-network/index.html}.}.

\subsubsection{Integrated and segregated systems}\label{sec:intsegration}

\begin{figure}[!t]
\centering
\begin{minipage}[c]{0.45\textwidth}
\includegraphics[width=\textwidth]{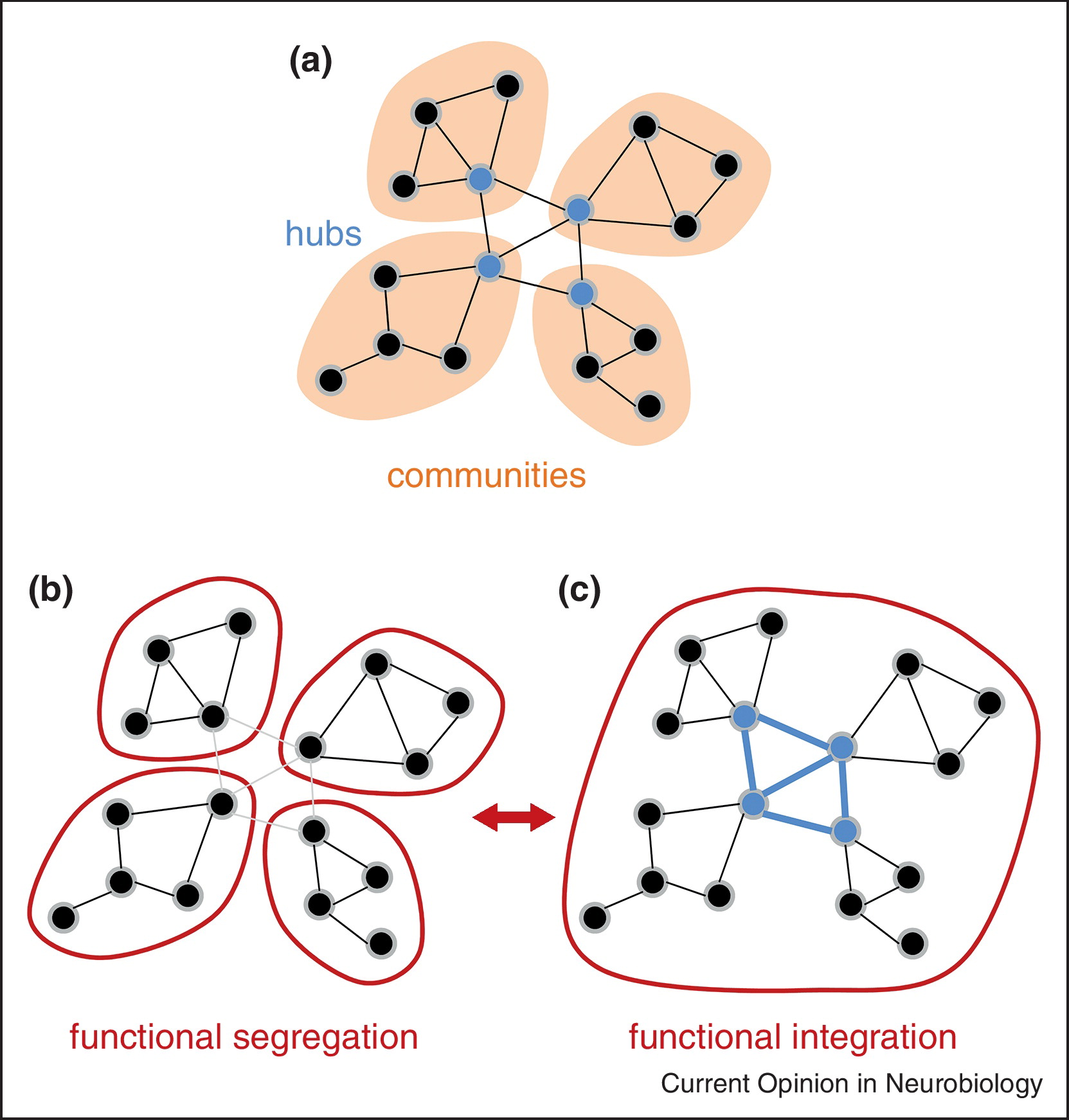}
\end{minipage}
\quad
\begin{minipage}[c]{0.5\textwidth}
\caption{\label{fig:sporns_intsegregation}Segregated communities and integrating hubs. (a) Network with four communities interconnected by four central hubs. (b) Functional segregation due to community structure is estimated by within-community coupling. (c) Functional integration favors information flow across the whole network through a subset of interconnecting hubs and is measured by the hubs or the paths crossing the network. Figure from \citep{sporns2013network}.}
\end{minipage}
\end{figure}

To understand how a network operates, it is essential to understand if information flows in such a way that, from a functional perspective, the system is either integrated, i.e., operating like a whole, or segregated, i.e., operating like independent modules or groups of units (see Fig.~\ref{fig:sporns_intsegregation}).

How to correctly study integration and segregation of a complex network is a still debated topic, and several proxies have been proposed across a broad spectrum of disciplines. The analysis stem almost independently from sociology and neuroscience, in both cases within the frameworks of classical single-layer networks~\citep{latora2001,latora2003economic,achard2007efficiency,rubinov2010complex,centola2015social,louf2016patterns,yamamoto2018impact,gallotti2019disentangling,bertagnolli2021quantifying}.

In multilayer systems, where multiple relationships co-exist simultaneously, the evaluation of integration must be accounted for by more complex topological models. In fact, it is critical to choose which kind of paths are to be evaluated, e.g., by adding a cost to inter-layer links used in paths that allow us to switch between two or more layers. The literature on this topic is rather poor and an agreement on the most suitable methodology to adopt is far from being reached. In the following, we describe two recent attempts in this direction.

\begin{figure}[!ht]
\centering\includegraphics[width=0.9\textwidth]{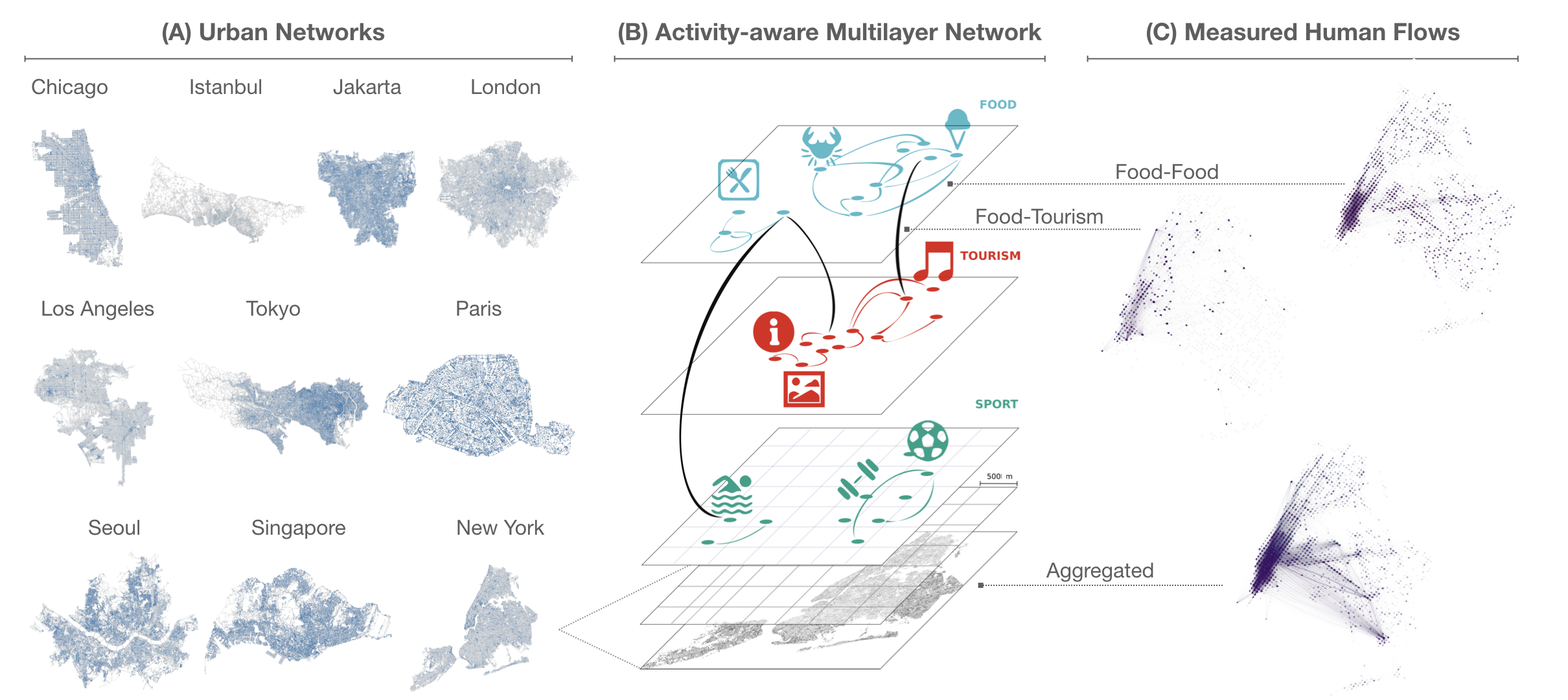}
\caption{\label{fig:foursquare_1}Multilayer modeling of structure and function of urban systems. (A) Urban  backbone of 10 megacities described by their street networks. (B) Urban functional networks described by the flows between cells of size 500m$\times$500m. The Foursquare data used for this analysis allow us to disentangle mobility flows into different layers where subsequent check-ins have been made between activities of the same type (e.g., leisure, eating, shopping, sport, etc.). (C) Measured flows for New York City represented as the weighted edges, which may be very different across distinct activity layers. Figure from~\citep{gallotti2019disentangling}.}
\end{figure}

The first application comes from the analysis of socio-technical systems, and in particular movement flows. One important approach in this direction is communicability~\citep{akbarzadeh2018communicability}, for which an explicit multilayer definition is available~\citep{Estrada2014Communicability}. However, the simultaneous analysis of integration and segregation can provide a more comprehensive perspective of a city's structural form and functional behavior. In a city, distinct layers can represent different activities (e.g., leisure, eating, shopping, sport, etc.; see Fig.~\ref{fig:foursquare_1}) that can be performed in the same area: two areas are connected within or across layers if there is a human flow between them in either direction. The analysis of empirical flow networks of this type measured from 10 megacities around the world would reveal, for instance, which activity was most influential for the urban system, from an integration-segregation perspective~\citep{gallotti2019disentangling}. Using modularity as a proxy for integration and normalised global efficiency as a proxy for integration, allowed scholars to unravel the functional organization of each megacity and to discover the presence of distinct ``cities within a city''. By comparing the deviations measured after aggregating all layers into a single-layer network and after aggregating while excluding one layer at a time, it can be shown that  ``transportation'' is the activity whose disruptions would remarkably hinder the urban function: in fact, the integration granted by long range connections would be sensibly reduced, while segregation due to the interrupted connections characterised by large flows would increase. Conversely, the same analysis shows that activities such as restoration, leisure and going to shopping malls poorly contribute to a city's integration~\citep{gallotti2019disentangling}. 

The fact that this result aligns with our expectations of macroscopic and easy-to-interpret systems confirms that the methodologies discussed in this section are suitable for characterising a multilayer network in terms of integration and segregation. In the specific case of layered urban systems, this type of analysis is potentially relevant to support policy makers with quantitative information on which activities can be temporary limited or promoted to achieve a desired amount of human flows integrated across the city.

The second application concerns the human brain. In fact, a possible way to analyze its functional connectivity is to map relationships between distinct regions of interest across distinct frequency bands~\citep{de2016mapping,de2017multilayer,DeDomenico2018}, each one encoding a distinct layer, and we can wonder whether these layers are integrated or segregated~\citep{tewarie2016integrating}. However, since the metric used to this aim is edge overlap, according to the aforementioned prescriptions this application is not entirely compatible with the study of integration and segregation and would be better understood as an analysis of layer-layer correlations. It will be interesting to see more studies on this topic in the future, for instance, based on multilayer modularity and multilayer generalization of global efficiency.

\subsection{Layer-layer correlations}\label{sec:layercorr}

Single layer networks usually exhibit correlations between entities acting in the system. Specifically, these correlations occur between some specific nodes' properties. In the case of assortativity, for instance, there is a preference for network's nodes to attach to others that are similar in some way (e.g. in degree), while it is the opposite in the case of disassortativity. Expanding the concept of \emph{correlation} to multilayer networks has to take into account the increased degrees of freedom introduced by the multilayer structure, where nodes' involvement across layers exhibits nontrivial and more variegated patterns than those observed in the single layer~\citep{nicosia2015measuring}. 
In fact, one can study the degree-degree correlations of each layer in the network but, in the case of a multilayer network, it is far more intriguing, for example, to explore how a given property of a node at a certain layer is correlated to the same or other properties of the same node at another layer.  
In the following we present the main correlation measures designed for multilayer networks, divided into three main sub-sections: interlayer degree correlation, overlap and degree of multiplexity and pairwise multiplexity, although other interesting measures -- e.g., based on SBM-like models used to infer edge correlations and for link prediction~\citep{pamfilinference}, on estimating a joint probability distribution describing edge existence over all layers to quantify correlations through conditional mutual information~\citep{wu2020correlated} or on a set-theoretic approach to quantify if a layer correlates with a second layer directly or via the indirect mediation with a third layer~\citep{lacasa2021beyond}-- are available.

\subsubsection{Inter-layer degree correlation}

This kind of correlation points out if high-degree nodes in one layer maintain this property in other layers.
We can measure the \textit{inter-layer assortativity} by studying the conditional degree distribution $P(k^{\alpha'}|k^{\alpha})$\citep{nicosia2013growing}, where $k_i^\alpha $ is the degree of node $i$ at layer $\alpha$, evaluating the average degree $\bar k^{\alpha'}$ at layer $\alpha'$ of nodes having degree $k^\alpha$ at layer $\alpha$:
\begin{eqnarray}
\bar k^{\alpha'}(k^\alpha) = \sum_{k_{\alpha'}}k^{\alpha'} P(k^{\alpha'}|k^{\alpha});
\end{eqnarray}
if there is no correlation between the layers $\alpha$ and $\alpha'$ we expect $\bar k^{\alpha'}(k^\alpha) = \langle k^{\alpha'}\rangle$, 
i.e. the average degree $\bar k^{\alpha'}$  does not depend on $k^{\alpha}$. If, instead, $\bar k^{\alpha'}(k^\alpha)$ is an increasing function in $k^\alpha$ the degrees have an assortative correlation, while they have a dissortative correlation if the function is decreasing.
The \textit{inter-layer assortativity} could also be defined using the Pearson, Spearman or Kendall correlation
between the degrees of the same node in the different layers~\citep{boccaletti2014structure}.

\subsubsection{Overlap and degree of multiplexity}

In this case correlation is evaluated in terms of node connectivity patterns in different layers. Specifically, we refer to the correlations among the links of the different intra-layer networks. In fact, the internal connectivity in different layers of the multiplex can be in certain cases correlated. To clarify with an example, you can be a friend of the same person in two different social networks, and thus the link between you and your friend exists in both layers of an imaginary multilayer network representing your social connections. Many different measures have been proposed to quantify this topological similarity. For the unweighted case, a first, associated to a couple of layers, is called \textit{global overlap}~\citep{bianconi2013statistical}, and is defined as the total number of pairs of nodes connected at the same time by a link in both layers, or, in other words, the total number of links that are in common between layer $\alpha$ and layer $\alpha'$:
\begin{eqnarray}
O^{\alpha,\alpha'} = \sum_{i<j} M_{j\alpha}^{i\alpha}M_{j\alpha'}^{i\alpha'}
\end{eqnarray}
(for other variants on \textit{overlapping} please refer to~\citep{battiston2014structural}).
A second quantity, associated this time to the whole multiplex, is the \textit{degree of multiplexity}~\citep{kivela2014multilayer}, defined as the fraction of node pairs that have multiple edge types between them. This quantity is obtained by dividing the number of node pairs that have multiple edge types between them by the total number of adjacent node pairs~\citep{kivela2014multilayer}. For the specific case of weighted multilayer networks we refer to~\citep{menichetti2014weighted} and to \citep{boccaletti2014structure}.

In particular, here we present two main weighted measures of multiplex networks, i.e. \textit{multistrength} and the \textit{inverse multi-participation ratio}~\citep{menichetti2014weighted}.  The \textit{multistrenght} for a node $i$ in layer $\alpha$, indicated by $s_{i, \alpha}^{\bar{m}}$, is obtained by summing the weights of a certain type of multilink $\bar{m}$ -- i.e., the set of links connecting a given pair of nodes in the different layers of the multiplex -- incident to a single node:
\begin{eqnarray}
s_{i, \alpha}^{\bar{m}}=\sum_{j=1}^{N} a_{i j}^{\alpha} A_{i j}^{\bar{m}}.
\end{eqnarray}
Given that for every layer $\alpha$ and node $i$ the non trivial multistrengths must include multilinks $\bar{M}$ with $\bar{M}_{\alpha}=1$, the number of non trivial multistrengths $\alpha$ $s_{i, \alpha}^{\bar{m}}$ is given by $2^{M-1}$. It follows that the number of multistrengths that can be obtained for each node in a multiplex network of $M$ layers is $M2^{M-1}$.
The \textit{inverse multiparticipation ratio} for a layer $\alpha$ $Y_{i, \alpha}^{\vec{m}}$ is used to measure the heterogeneity of the weights of multilinks incident upon a single node:
\begin{eqnarray}
Y_{i, \alpha}^{\vec{m}}=\sum_{j=1}^{N}\left(\frac{a_{i j}^{\alpha} A_{i j}^{\bar{m}}}{\sum_{r} a_{i r}^{\alpha} A_{i r}^{\bar{m}}}\right)^{2}.
\end{eqnarray}
Within the context of a weighted multiplex network, the importance of considering the interacting layers in the analysis of complex system has been proven by ~\citep{menichetti2014weighted} which evaluated the additional amount of information provided by weighted properties of multilinks over the one contained in single layers, further strengthening the evidence that the analysis of multilayer networks cannot be confined to the partial analysis of single layers.

\begin{figure}[!t]
\centering\includegraphics[width=\textwidth]{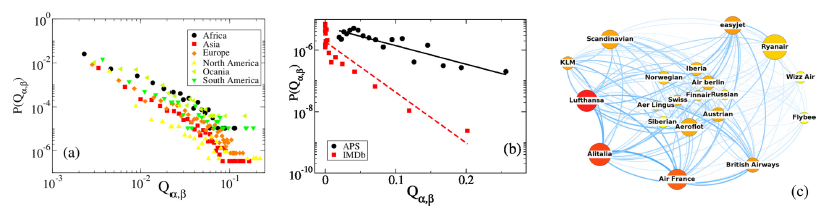}
\caption{\label{fig:p_multiplexity} (a) Airline, (b) APS and IMDb networks. (c) Top 20 airlines in Europe by number of covered airports. Figure from~\citep{nicosia2015measuring}.}
\end{figure}

\subsubsection{Pairwise multiplexity}

In this last case, correlation is provided in terms of \textit{correlated activity patterns} in the multilayer. 
A node $i$ is said to be active at a layer $\alpha$ if $k_{i}^{\alpha}>0$. For each node $i$, we can associate a node activity
vector \mbox{$\mathbf{b_{i}} = \{b_{i}^{[1]},b_{i}^{[2]}, ...,b_{i}^{[L]} \}$} where $b_{i}^{[\alpha]}= 1$ if node $i$ has at least one edge at layer $\alpha$ and is 0 otherwise. The node activity $B_{i}$ of the node $i$ corresponds to the total number of layers where $i$ is active~\citep{nicosia2015measuring}: 
\begin{eqnarray}
B_{i}=\sum_{\alpha}b_{i}^{[\alpha]}.
\end{eqnarray}
By definition, $0\leq B_{i} \leq L$. Analogously, the layer activity of $\alpha$ is given by the number of active nodes in layer $\alpha$~\citep{nicosia2015measuring}:
\begin{eqnarray}
N^{[\alpha]}=\sum_{i}b_{i}^{[\alpha]}.
\end{eqnarray}
By definition, $0\leq N^{[\alpha]} \leq N$. We can now define the \textit{layer pairwise multiplexity} $Q_{\alpha, \beta}$, which is a measure of correlation between the layers, as~\citep{nicosia2015measuring}: 
\begin{eqnarray}
Q_{\alpha, \beta}=\frac{1}{N} \sum_{i} b_{i}^{[\alpha]} b_{i}^{[\beta]},
\end{eqnarray}
that corresponds to the fraction of nodes that are active in both layer $\alpha$ and $\beta$. Examples of the distribution of the pairwise multiplexity for a continental airports networks, for the papers published in the journals of the American Physical Society (APS) and for the movies in the Internet Movie Database (IMDb) are reported in Fig.~\ref{fig:p_multiplexity}. Similarly, we can define the \textit{node pairwise multiplexity}, measuring the correlation of activities between two nodes, as the fraction of layers in which both node $i$ and node $j$ are active~\citep{criado2012mathematical,boccaletti2014structure}.

\begin{figure}[!h]
\centering
\begin{minipage}[c]{.45\textwidth}
\includegraphics[width=\textwidth]{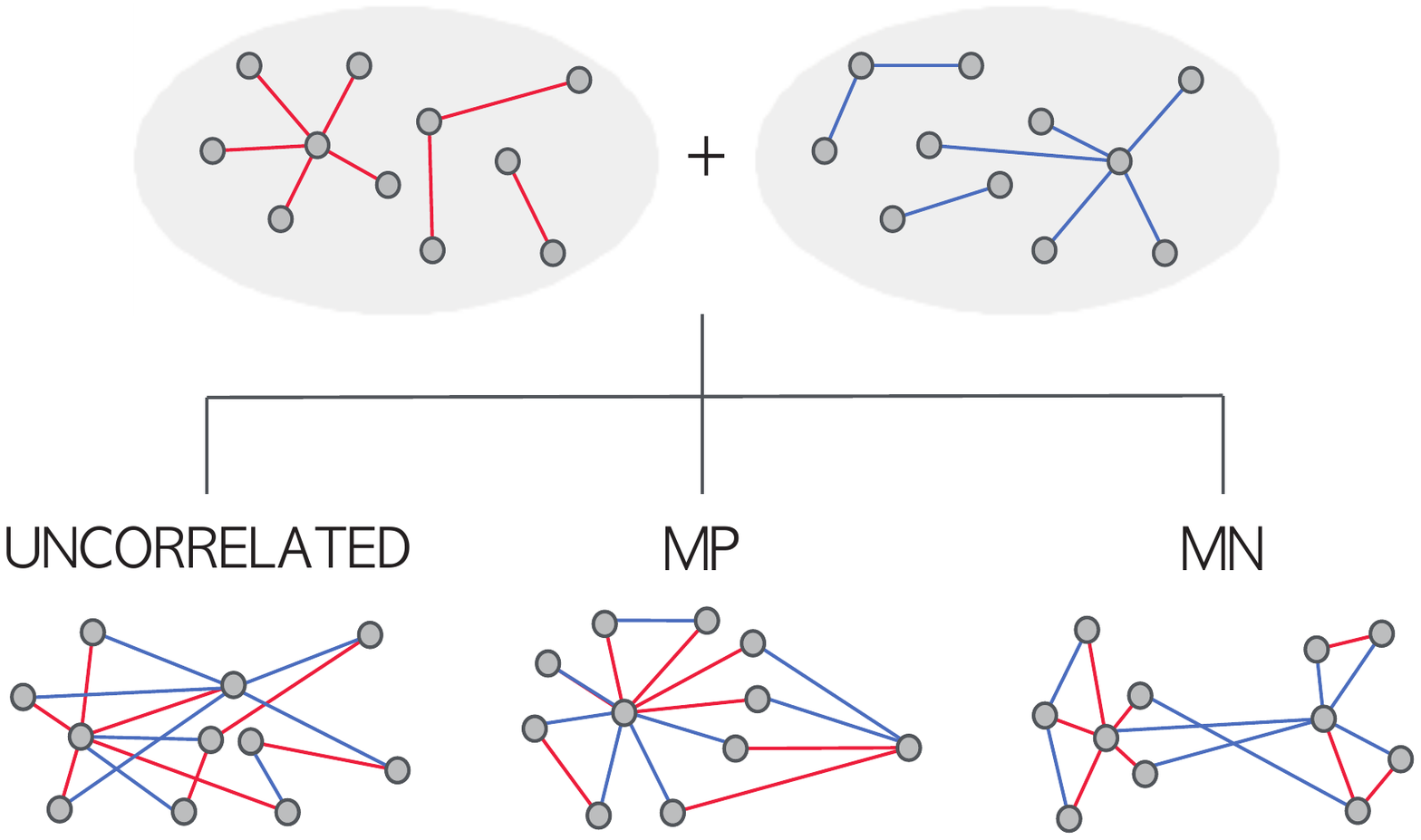}
\end{minipage}
\quad
\begin{minipage}[c]{.5\textwidth}
\caption{\label{fig:duplex_c}Different types of duplex couplings: uncorrelated, maximally positive (MP) and maximally negative (MN) correlated multiplexity. Figure from~\citep{lee2012correlated}.}
\end{minipage}
\end{figure}

\begin{figure}[!ht]
\centering\includegraphics[width=0.8\textwidth]{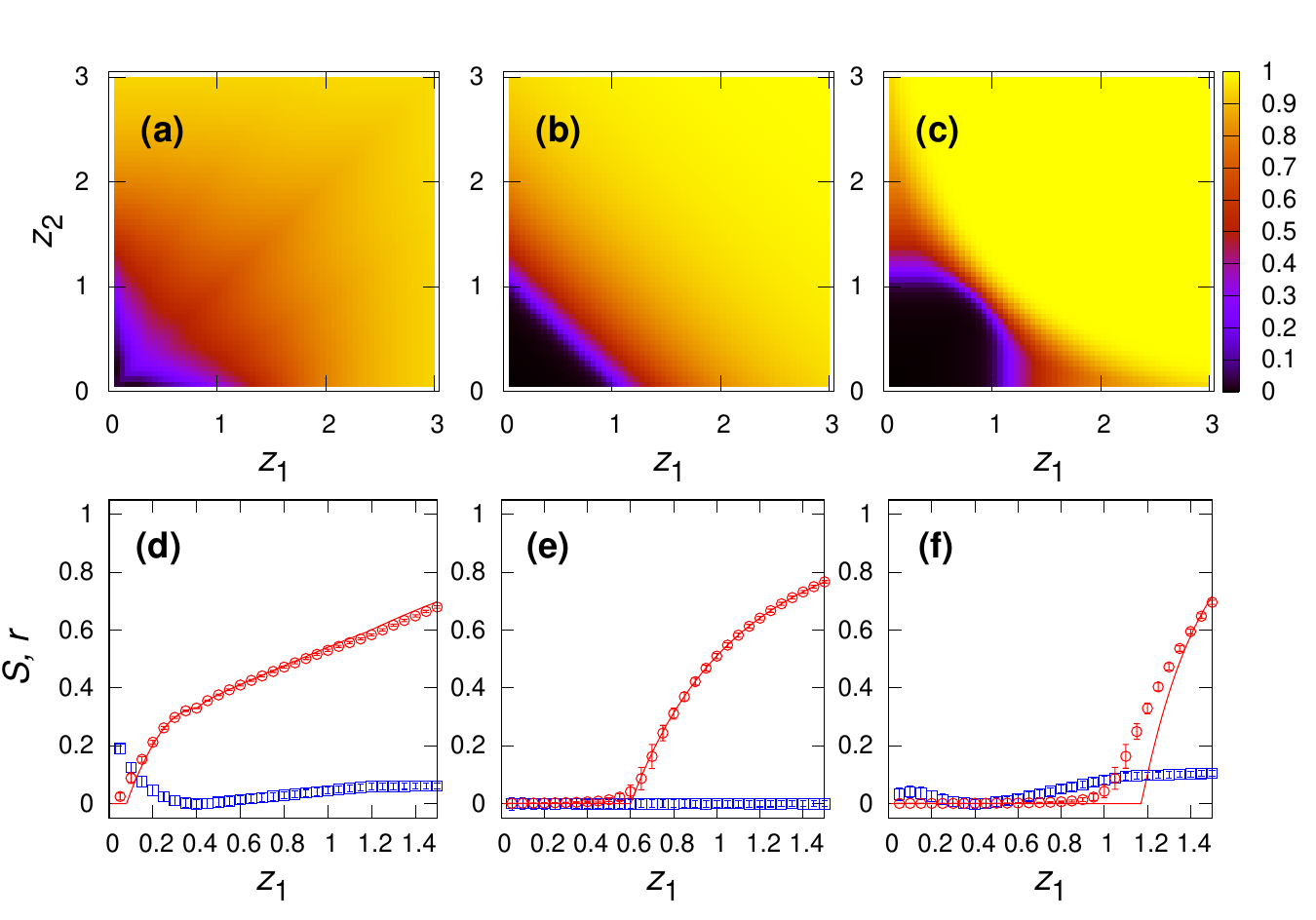}
\caption{\label{fig:effect_on_LCC}Size of the giant component of a ER duplex network for (a) maximally positive correlated multiplexity, (b) uncorrelated and (c) maximally negative correlated multiplexity. (d--f) Size of the giant component (red) for $z_2 = 0.4$ and assortativity coefficient $r$ (blue). Figure from~\citep{lee2012correlated}.}
\end{figure}

We conclude this section showing how the study of the correlation is fundamental since it alters the critical properties of the network.
In fact, the pattern of correlated multiplexity (Fig.~\ref{fig:duplex_c}) is crucial since it affects the multiplex system's connectivity as shown in Fig.~\ref{fig:effect_on_LCC} where the different choices of correlations have an impact on the largest connected component.

In this chapter, we have described the most widely used measures for the topological analysis of multilayer systems, although in some cases we have used some dynamical process (e.g., random walks, continuous-time diffusion or shortest-path routing) to define some network descriptors.
We refer to~\citep{brodka2018quantifying} for a taxonomy and an experimental evaluation of the approaches to compare different layers in multiplex networks. The next chapter will be entirely dedicated to introduce dynamical processes on (and of) multilayer networks.

%%%%%%%%%%%%%%%%%%%%%%%%%%%%%
%%%%%%%%%%%%%%%%%%%%%%%%%%%%%
%%%%%%%%%%%%%%%%%%%%%%%%%%%%%
\section{Multilayer dynamics}
In the previous chapter we discussed many structural descriptors of a multilayer network. However, some definitions adopted for the structural analysis are, in fact, based on some notion of dynamical process on the top of the network. In this chapter, the reader will have the opportunity to shed light on a spectrum of important network-driven phenomena, such as diffusive processes (Sec.~\ref{sec:diffproc}), synchronization dynamics (Sec.~\ref{sec:synchro}), cooperation dynamics (Sec.~\ref{sec:coopdyn}), intertwined/interdependent processes (Sec.~\ref{sec:interdepproc}), percolation (Sec.~\ref{sect:perc}), cascade failures (Sec.~\ref{sec:cascades}). Note that we refer to~\citep{nicosia2013growing,Santoro2017pareto} and~\citep{kivela2014multilayer,boccaletti2014structure} for further information about the dynamics of multiplex networks, like growth processes.

\begin{figure}[!t]
\centering\includegraphics[width=0.6\textwidth]{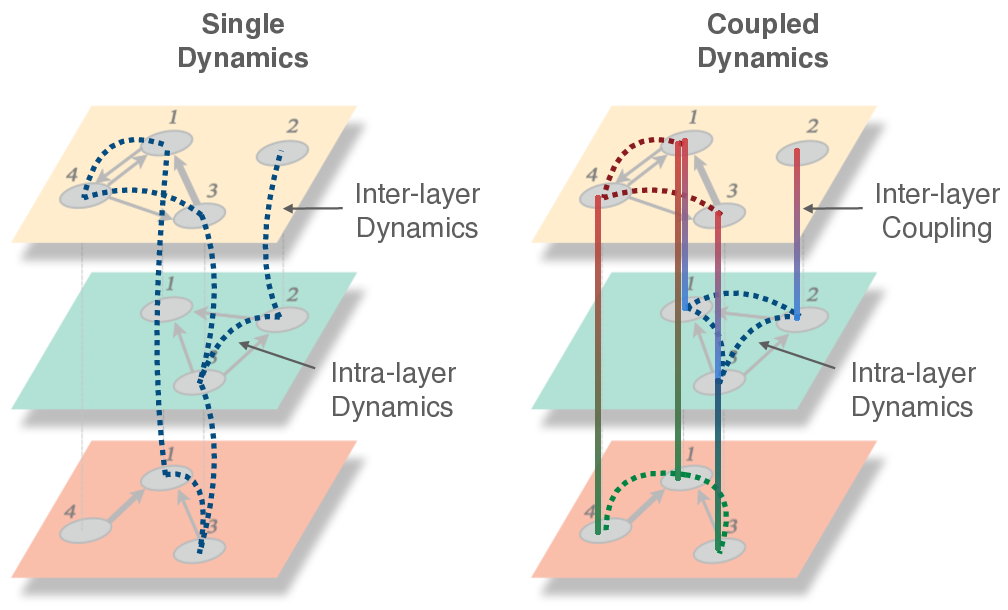}
\caption{\label{fig:chap_2dynams} Two classes of dynamical processes on the top of a multilayer network. Left: single dynamics, e.g. a random walk or the diffusion of some quantity, is defined for all layers. Right: coupled dynamics, where a distinct process is defined in each layer and interdependency is specified by means of coupled governing equations. Reproduced with permission from~\citep{de2020illustrations}.}
\end{figure}

In the following we will distinguish two main classes of dynamics~\citep{de2016physics} on the top of multilayer systems (see Fig.~\ref{fig:chap_2dynams}):
\begin{itemize}
\item \textbf{Type--I}: \emph{single dynamics} is defined and it is the same for all layers. Here we can find  discrete~\citep{de2014navigability,sole2016random,Lacasa2018,Valdeolivas2018} and continuous-time~\citep{gomez2013diffusion,sole2013spectral,Requejo2016,Brechtel2018,Tejedor2018diffusion} diffusion, pattern formation~\citep{asllani2014turing,asllani2015turing,kouvaris2015pattern,busiello2018homogeneous}, communicable information~\citep{Estrada2014Communicability}, synchronization~\citep{Singh2015,Skardal2015,Zhang2015,Saa2018,Leyva2018}, epidemic spreading of one disease~\citep{Wang2013,sahneh2013generalized,buono2014epidemics,valdano2015analytical,bianconi2017epidemic}, adoption dynamics, diffusion of innovation and other contagion processes~\citep{Yagan2012contagion,Hu2014viral,ramezanian2015diffusion,traag2016complex,Wang2017} and opinion dynamics~\citep{Jang2015opinion,Diakonova2016irreducibility,artime2017joint,amato2017opinion,Antonopoulos2018}. In this class we find also system control~\citep{yuan2014exact,Posfai2016}, congestion dynamics~\citep{tan2014traffic,SoleRibalta2016congestion,Chodrow2016demand}, cooperation processes~\citep{gomez2012evolution,matamalas2015strategical,battiston2017determinants}, percolation~\citep{leicht2009percolation, lee2015towards, kryven2019bond} and cascade failures~\citep{buldyrev2010catastrophic, valdez2020cascading}.

\item \textbf{Type--II}: \emph{coupled dynamics}, where multiple dynamical processes are defined on each layer and their interdependency is operationally encoded into adequately coupled equations, for instance by means of inter-layer connectivity when present. In this class we can find processes which combine epidemics spreading with human behavior~\citep{Funk2009, wu2012impact, lima2013cellular,Granell2013,massaro2014epidemic,lima2015disease,funk2015,wang2015coupled,azimi2016cooperative,VelasquezRojas2017interacting}, simple and complex contagion~(\citep{Czaplicka2016}), evolutionary game dynamics with social influence~\citep{amato2017interplay}, cooperative with competitive epidemics spreading~\citep{Dickison2012epidemics,Cozzo2013social,Sanz2014,DeArruda2017,danziger2019dynamic}, transport with synchronization~\citep{nicosia2017collective}, interdependent human flows~\citep{SorianoPanos2018framework} and other interdependent processes~\citep{GomezGardenes2015competition}.
\end{itemize}

\subsection{Diffusive processes}\label{sec:diffproc}

Diffusive processes, either discrete or continuous in time, provide reasonable models for a wide spectrum of phenomena, from random searches to population dynamics, from system exploration and navigability to information diffusion. In the following we will consider two emblematic classes of models: random walks and continuous-time diffusion.

\paragraph{Random walks on single-layer networks}

Random walks on networks~\citep{noh2004random,masuda2017random} are discrete processes governed by a transition probability from $i$ to $j$ that, at each time step, is given by $T_{ij} = \frac{W_{ij}}{s_i}$, where $s_i$ is the outgoing strength of node $i$ or, equivalently, $s_i=\sum\limits_j W_{ij}$. Let us introduce the matrix diagonal $\mathbf{D}$ whose entries are defined by $D_{ii}=s_{i}$ and the vector $\mathbf{p}(t)\in \mathbb{R}^{N}$, whose $i$-th entry gives the probability to find the random walker at time $t$ on node $i$: the master equation governing the random walk dynamics can be written in compact form as
\begin{eqnarray}
\label{eq:DTRW-master}
\mathbf{p}(t+1) = \mathbf{p}(t)\mathbf{T},
\end{eqnarray}
where $\mathbf{T}=\mathbf{D}^{-1}\mathbf{W}$ is known as the transition matrix. We have just used the standard notation but, for instance, the same equation can be written using our tensorial formalism as
\begin{eqnarray}
\label{eq:DTRW-master2}
p_j(t+1) = p_i(t)T^{i}_{j},
\end{eqnarray}
where Einstein convention (sum over $i$) is used and $i,j$ indicate covariant and contravariant indices, not the entries of the corresponding rank-1 and rank-2 tensors.

\begin{figure}[!ht]
\centering
\begin{minipage}[c]{0.35\textwidth}
\includegraphics[width=\textwidth]{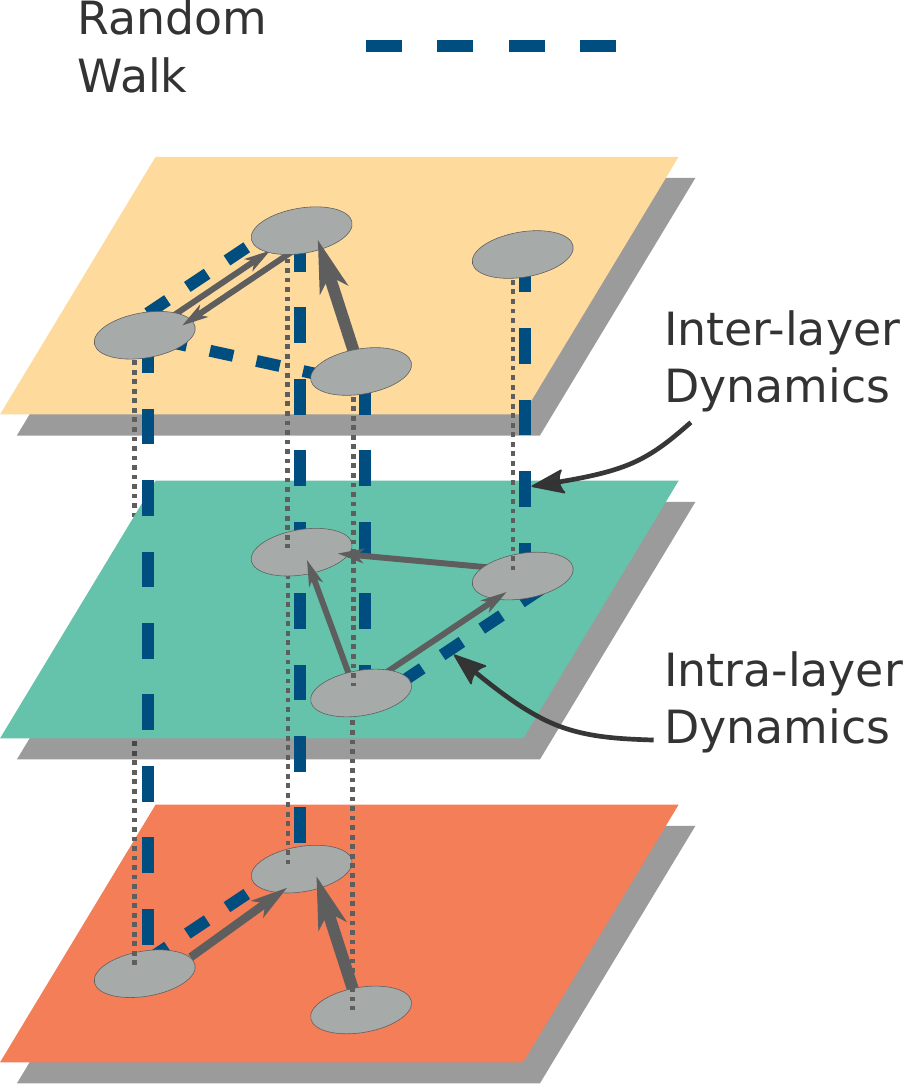} \hfill
\end{minipage} 
~
\begin{minipage}[c]{0.6\textwidth}
\caption{\label{fig:chap_rw_multi} A random walker visits the nodes of a multilayer network: starting from one state node in a specific layer, it is allowed to i) stay in the same state node, ii) jump to a neighbor within the same layer, iii) switch to another state node on another layer or iv) jump elsewhere while subjected to the transition tensor $T^{i\alpha}_{j\beta}$. The walk is defined by the ordered sequence of visited intra- and inter-layer connections. Reproduced with permission from~\citep{de2020illustrations}.}
\end{minipage}
\end{figure}

The reader familiar with the Markov chains formalism will immediately recognize the continuous-time Markov chain on $N$ states (i.e., on the set of nodes), with independent exponential holding times with parameter $\lambda=1$ and jumping probabilities given by the transition probabilities of the discrete-time random walk, and which can be equivalently defined by the forward equation
\begin{equation}\label{eq:CTRW-ode}
  \dot{\mathbf{p}}(t) = -\mathbf{p}(t) \tilde{\mathbf{L}}
\end{equation}
where $\tilde{\mathbf{L}} = \mathbf{I}-\mathbf{D}^{-1} \mathbf{W}$ is known as the random walk normalized Laplacian and $-\tilde{\mathbf{L}}$ is the generator of the continuous-time process with an initial condition $\mathbf{p}(0) = \mathbf{p}_0$. 
Equivalently, in tensorial notation we can rewrite the equation and initial condition as 
\begin{equation}\label{eq:CTRW-ode-tensor}
\begin{cases}
  & \dot{{p}}_{j}(t) = - \sum\limits_{i=1}^N \tilde{{L}}^{i}_{j} {p}_{i}(t) \\
  & {p}_{i}(0) = q_i
  \end{cases}
\end{equation}
with $\tilde{{L}}^{i}_{j} = \delta^{i}_{j} - T^{i}_{j}$ indicating the component $(i,j)$ of the random walk normalized Laplacian tensor, $\delta^{i}_{j}$ the Kronecker delta and $T^{i}_{j}$ the component $(i,j)$ of the transition probability tensor. The solution of this equation is given by
\begin{eqnarray}
\mathbf{p}(t)=\mathbf{p}_0 e^{-t\tilde{\mathbf{L}}},
\end{eqnarray}
where the intervening exponential matrix is known as the \emph{propagator of the dynamics}. Note that the transition matrix can encode other rules governing the jumps of a random walker from one node to another: by changing the \emph{flavor} of the walk one can explore a broad spectrum of stochastic processes, as we will see in the next sections, where we will generalize these dynamics to multilayer networks, where a walker can jump between nodes but can also switch between layers. In general, the transition matrix is replaced by a rank-4 transition tensor $T^{i\alpha}_{j\beta}$ which governs the probability that a walker on node $i$ in layer $\alpha$ will move to node $j$ in layer $\beta$. An illustration of a random walk on the top of such a system is shown in Fig.~\ref{fig:chap_rw_multi}, while a schematic representation of the supra-adjacency matrices corresponding to the multilayer systems considered in the reminder of this section, is shown in Fig.~\ref{fig:mux_rw_multilayers}, and is useful to get some visual insights about the structure of the corresponding supra-transition matrices.

\paragraph{Random Walks on edge-colored multigraphs}

On this class of systems (note that $M^{i\alpha}_{j\beta}=0$ for all $\alpha\neq \beta$), it is possible to define random walk dynamics in two distinct ways. On one hand, we can allow the walker to jump through edges with different colors, which are considered as multiple edges, regardless of their colors, where a node's degree is the sum of its state nodes' degree~\citep{battiston2016efficient}. Operationally, this approach is equivalent to performing a classical random walk on the top of the aggregate representation of the system: in general this could not be desirable, because it is difficult to quantify the effects of neglected information (i.e., connectivity patterns per layer) on the dynamics.

On the other hand, we can allow walkers to explore each layers independently and then integrate the transition matrices governing their dynamics~\citep{de2020EnhanceTransport} as
\begin{equation}\label{eq:tmatrix-ec}
\langle T^i_{j} \rangle = \sum\limits_{\alpha = 1}^L m_{i, \alpha} T_{j \alpha}^{i \alpha}, \qquad \sum\limits_{\alpha=1}^L m_{i, \alpha} = 1 \quad \forall i \in V,
\end{equation}
where $m_{i, \alpha}\geq 0$ are weights enabling us to tune the importance of each layer, relative to node $i$ in the final dynamics of the system.
If $m_{i, \alpha} = \frac{1}{L}$ for all $i$ and $\alpha$, then $\langle T^i_{j} \rangle$ is simply the average transition matrix over all layers. 
However, if we choose $m_{i, \alpha} = \frac{1}{\mu_i} \mathbf{1}_{\{s_{i}(\alpha) \neq 0\}}$, with $s_{i}(\alpha)$ being the out-strength of vertex $i$ in layer $\alpha$ and $\mu_i = \sum\limits_{\alpha=1}^L \mathbf{1}_{\{s_{i}(\alpha) \neq 0\}}$ the multiplicity of node $i$, i.e., the number of layers in which $i$ is not isolated, then we can discard the effect of $i$ being isolated in one or more layers. 
This approach preserves the diversity of information coming from layers' connectivity patterns and allows for more flexibility. 
In fact, in Equation~\eqref{eq:tmatrix-ec}, we use a finite mixture of probability mass functions with weights $m_{i, \alpha}\geq 0$ such that $\sum\limits_{\alpha=1}^L m_{i, \alpha} = 1$, but if additional information from the data is available then the transition matrices in Eq.~(\ref{eq:tmatrix-ec}) can be combined by means of weights encoding the relative importance given to each layer, thus providing an operational way to prioritize the available information from specific layers. Here, for the sake of simplicity, we are just considering that all layers are equally important and adopting the weights $m_{i, \alpha} = \frac{1}{\mu_i} \mathbf{1}_{\{s_{i}(\alpha) \neq 0\}}$.

\paragraph{Random walks on multilayer networks}

In the following, we indicate by $D(i; \alpha, \beta)$ the inter-layer connections between replicas across nodes, i.e., the entries of $M_{j\beta}^{i \alpha}$ for $i=j$. In this framework, the outgoing strength of node $i$ in layer $\alpha$ is given by $s_i(\alpha) = \sum\limits_{j=1}^{N} M^{i \alpha}_{j \alpha}$, whereas the multilayer strength -- discarding the inter-layer edges -- is given by $s_{i} = \sum\limits_{\alpha} s_{i}(\alpha) = \sum\limits_{\alpha=1}^{L} \sum\limits_{j=1}^{N} M_{j \alpha}^{i \alpha}$. The inter-layer strengths are similarly given by $S_{i}(\alpha) = \sum\limits_j \sum\limits_{\beta \neq \alpha} M^{i \alpha}_{j \beta}$ and, consequently, $(s_{i}(\alpha) + S_{i}(\alpha))_{i=1, \alpha=1}^{N, L}$ is the out-strength supra-vector with $NL$ components obtained as the row sums of the supra-adjacency matrix. The reader will notice that the same results can be simply obtained by using the tensorial formalism and Einstein convention as $s^{i\alpha}=M_{j \beta}^{i \alpha}u^{j}u^{\beta}$.

Usually, for the sake of simplicity, isolated nodes and components, as well as more complex patterns of inter-layer connectivity, are discarded from the analysis, e.g., focusing on the largest connected component and considering only nodes which exist in all layers, since a more general framework has a non-negligible impact on the calculation of transition probabilities. In general, one can overcome these limitations, typical of real-world systems, by adding state nodes as isolated and adequately accounting for them in calculations. In the following, we consider three cases:
\begin{enumerate}
\item $S_i(\alpha) = 0$, i.e., inter-layer connectivity is missing for node $i$;
\item $s_i(\alpha) = 0$, i.e., intra-layer connectivity is missing for node $i$ in layer $\alpha$ (i.e., the node is isolated);
\item $s_i = 0$, i.e., the node is isolated everywhere.
\end{enumerate}
In the last case one can simply exclude the node from the analysis, since it is not part of the network, in practice. Therefore we can safely assume that $s_i>0$ for all nodes. Concerning the cases 1 and 2, they require special attention only if $S_i(\alpha) + s_i(\alpha) = 0$ for some $\alpha$, since the state node $(i,\alpha)$ is an absorbing state, i.e., the walker reaching that node will stay there with probability equal to 1. To avoid absorbing states of this type, one option is to add a \emph{teleportation} dynamics, i.e., a uniform but small probability equal to $\frac{1}{NL}$ to jump elsewhere in the system even if outgoing connections are missing. This approach has been successfully used in Google's PageRank to rank web pages~\citep{brin1998anatomy}, as well as to identify functional modules in monoplex~\citep{rosvall2008maps} and multilayer networks~\citep{de2015identifying}. Here, we use a similar approach, with the effect that the occupation probability of the state node $(i,\alpha)$ is effectively reduced. Once the transition tensor is well defined, as we will see later in this section for some specific processes, one can define the master equation in discrete time as
\begin{align*}
  p_{j \beta}(t + 1) =
  & \underbrace{T_{j \beta}^{j \beta} p_{j \beta}(t)}_{\text{stay}} +
  \underbrace{\sum_{\substack{\alpha = 1 \\ \alpha \neq \beta }}^L T_{j \beta}^{j \alpha} p_{j \alpha}(t)}_{\text{switch}} + 
  \underbrace{\sum_{\substack{i=1 \\ i \neq j}}^{N} T_{j \beta}^{i \beta} p_{i \beta}(t)}_{\text{jump}} +
  \underbrace{\sum_{\substack{\alpha=1 \\ \alpha \neq \beta}}^{L} \sum_{\substack{i=1 \\ i \neq j}}^{N} T_{j \beta}^{i \alpha} p_{i \alpha}(t)}_{\text{switch and jump}}
\end{align*}
where $p_{j \beta}(t)$ indicates the probability of finding a random walker in node $j$ of layer $\beta$ at time $t$, and the contributions of jumps and switches are made explicit. Note that using the tensorial formalism, this expression would be compressed into the elegant master equation
\begin{eqnarray}\label{eq:multilayer_random_walk}
  p_{j \beta}(t + 1) = p^{i \alpha}(t) T_{i \alpha}^{j \beta},
\end{eqnarray}
with the continuous-time version described by the forward equation
\begin{equation}\label{eq:CTRW-ml-ode-tensor}
  \dot{p}_{j \beta}(t) = -\tilde{L}_{j \beta}^{i \alpha} p_{i \alpha}(t),
\end{equation}
where $\tilde{L}_{j \beta}^{i \alpha}=\delta_{j \beta}^{i \alpha}-T_{j \beta}^{i \alpha}$ is the random walk normalized Laplacian tensor.

It is important to remark that a random walk dynamics is affected by both structure and transition mechanisms: i.e., by fixing the topology one can always define distinct transition rules encoding different stochastic movements between nodes and exploration strategies, which we have previously named flavors. 

As mentioned before, in PageRank random walks (PRRW) a teleportation (or non-local jumping) parameter tunes the ability of the random walker to escape from absorbing states and to reach nodes which are not in the neighborhood of the current node~\citep{brin1998anatomy,de2015ranking}. The random walk introduced in the previous section for monoplex networks is known as a classical random walk (CRW)~\citep{de2013mathematical,de2014navigability}. There are several walks that can be defined on multilayer networks, which generalize their monoplex counterparts or are specific to multilayer systems. For instance, one can define multilayer diffusive random walks (DRW) -- generalizing the one defined in~\citep{de2014navigability} --, maximal-entropy random walks (MERW)~\citep{de2014navigability} -- generalizing monoplex walks which localize around topological defects~\citep{burda2009localization}, and a physical random walk with relaxation (PrRW)~\citep{de2015identifying}. Transition probabilities for these processes are reported in Tab.~\ref{tab:rws}.

\begin{tiny}
\begin{table*}[!htb]
\begin{tabular}{l|ccccr}
& CRW & PRRW & DRW & MERW & PrRW \\ \hline
$T_{j \beta}^{j \beta}$  & $\frac{M_{j \beta}^{j \beta}}{\sigma_j(\beta)}$   & $r \frac{M_{j \beta}^{j \beta}}{\sigma_j(\beta)} + \frac{1 - r}{NL}$   & $\frac{s_{\max }+M_{j \beta}^{j \beta} - \sigma_j(\beta)}{s_{\max }}$ & $\frac{M_{j \beta}^{j \beta}}{\lambda_{\max }}$ & $(1-r) \frac{M_{j \beta}^{j \beta}}{s_j(\beta)} + r \frac{M_{j \beta}^{j \beta}}{s_j}$ \\
$T_{j \beta}^{j \alpha}$ & $\frac{M_{j \beta}^{j \alpha}}{\sigma_j(\alpha)}$ & $r \frac{M_{j \beta}^{j \alpha}}{\sigma_j(\alpha)} + \frac{1 - r}{NL}$ & $\frac{M_{j \beta}^{j \alpha}}{s_{\max }}$ & $\frac{M_{j \beta}^{j \alpha}}{\lambda_{\max }} \frac{V_{j\beta}}{V_{j\alpha}}$ & $r \frac{M_{j \beta}^{j \beta}}{s_j}$ \\
$T_{j \beta}^{i \beta}$  & $\frac{M_{j \beta}^{i \beta}}{\sigma_i(\beta)}$   & $r \frac{M_{j \beta}^{i \beta}}{\sigma_i(\beta)} + \frac{1 - r}{NL}$   & $\frac{M_{j \beta}^{i \beta}}{s_{\max }}$ & $\frac{M_{j \beta}^{i \beta}}{\lambda_{\max }} \frac{V_{j \beta}}{V_{i \beta}}$   & $(1-r) \frac{M_{j \beta}^{i \beta}}{s_{i}(\beta)} + r \frac{M_{j \beta}^{i \beta}}{s_i}$ \\
$T_{j \beta}^{i \alpha}$ & $\frac{M_{j \beta}^{i \alpha}}{\sigma_i(\alpha)}$ & $r \frac{M_{j \beta}^{i \alpha}}{\sigma_i(\alpha)} + \frac{1 - r}{NL}$ & $\frac{M_{j \beta}^{i \alpha}}{s_{\max }}$ & $\frac{M_{j \beta}^{i \alpha}}{\lambda_{\max }} \frac{V_{j \beta}}{V_{i \alpha}}$ & $r \frac{M_{j \beta}^{i \beta}}{s_i}$ \\
\hline
\end{tabular}
\caption{Entries of the transition tensor for distinct random walks mentioned in the text, namely classical (CRW), PageRank (PRRW), diffusive (DRW),  maximal-entropy (MERW), and physical with relaxation (PrRW) random walks. To keep the notation simple, we use $\sigma_j(\beta)=s_j(\beta) + S_j(\beta)$ and $s_{\max }=\max\limits_{i, \alpha}\left\{\sigma_i(\alpha) \right\}$ for the outgoing strengths (see the text for details). Note that the teleportation parameter is usually denoted by $\alpha$: to avoid confusion, here we indicate it by $r$. Finally, $\lambda_{\max }$ is the largest eigenvalue of the multilayer adjacency tensor (see Sec.~\ref{eigcen1}). It is worth remarking that for CRW, DRW and MERW, these transition rules generalize to any multilayer system the ones introduced in~\citep{de2014navigability} for multiplex networks; PrRW is defined as in~\citep{de2015identifying}; PRRW generalizes the walk introduced in~\citep{de2015ranking}. Reprinted table with permission from~\citep{bertagnolli2020diffusion}. Copyright  (2021) by the American Physical Society.}
\label{tab:rws}
\end{table*}
\end{tiny}

We conclude this section with a short discussion about the physical random walk, introduced in~\citep{de2014navigability} to describe those dynamics where the state nodes have a ``common memory'' such that information reaching one state node is \emph{instantaneously} diffused to all state nodes of the corresponding physical node. This mechanism encodes a variety of processes in the real world. For instance, in a network of digital interactions among individuals, such as social media platforms, we can get a rumor in a particular platform through our intra-layer connections, but at the same time that rumor will be known to our alter egos in the other social media platforms. In this scenario, inter-layer connectivity does not carry physical meaning and, consequently, is ignored. The physical random walk with relaxation (PrRW)~\citep{de2015identifying} can be seen as a variant of PRW, where the knowledge of inter-layer connectivity is not required and is therefore discarded: from Tab.~\ref{tab:rws} it is easy to identify that its transition probabilities balance intra- and inter-links by means of probability weights $1-r$ and $r$, respectively.

It is worth remarking here that several other types of walks can be defined, including ones describing quantum processes. See~\citep{biamonte2019complex} for a review of this topic. From a mathematical perspective, one can write a master equation similar in shape to the one of a classical random walk:
\begin{eqnarray}
|\dot{\psi}\rangle = -i \mathbf{L}_{Q} |\psi\rangle,
\end{eqnarray}
where $i$ is the imaginary unit and $\mathbf{L}_{Q}=\mathbf{D}^{-1/2}\mathbf{L}\mathbf{D}^{-1/2}$, $\mathbf{L}$ being the combinatorial Laplacian matrix. The evolution equation for this type of continuous-time quantum walk is time-reversibile and, unlike classical random walks, quantum ones: i) do not admit a stationary distribution and ii) are deterministic in their dynamic nature and stochastic with respect to measurements~\citep{bottcher2020walks}. In the future, it will be of interest to study the emergence of potentially unseen physical properties related to quantum walks on the top of multilayer networks.

\paragraph{Continuous-time diffusion}

Let us assume we have some quantity, e.g., water, free to flow in a network of nodes through pipes, which can be represented by links between pairs of nodes. Given an initial distribution, what is the level of this quantity in each node at a given time $t$? This problem can be casted into continuous-time diffusion on a network. 

In the following, let us indicate by $x_{i}(t)$ the state vector, in tensorial formalism, carrying information about the quantity in each node of a monoplex network at time $t$. Let $x_{i}(0)$ be the initial state vector: its evolution over time can be modeled by the diffusion equation
\begin{eqnarray}
	\dot{x}_{j}(t)=\mathcal{D}\left[W^{i}_{j}x_{i}(t)-W^{i}_{k}u_{i}e^{k}(j)x_{j}(t)\right]\,,
\end{eqnarray}
where $\mathcal{D}$ is a diffusion constant, $u_{i}$ is the vector of ones and $e^{k}(j)$ is a canonical vector (see Sec.~\ref{sec:representation}). Since $s_{k}=W^{i}_{k}u_{i}$ is the outgoing strength and $s_{k}e^{k}(j)x_{j}(t)=s_{k}e^{k}(j)\delta^{i}_{j}x_{i}(t)$, the diffusion equation can be written in more compact form as:
\begin{eqnarray}
\label{eq:diff-lapl-chapdyn}
	\frac{dx_{j}(t)}{dt}=-\mathcal{D}L^{i}_{j}x_{i}(t)\,,
\end{eqnarray}
where $L^{i}_{j}=W^{l}_{k}u_{l}e^{k}(j)\delta^{i}_{j}-W^{i}_{j}$ is the combinatorial Laplacian tensor~\citep{chung1997}. It is worth remarking here that this tensor differs from the normalized random walk Laplacian introduced in the previous section, although in some cases -- as for a classical random walk -- they are related by the simple relationship $\tilde{L}_{j}^{i}=(D^{-1})^{i}_{k}L^{k}_{j}$. The solution of Eq.~(\ref{eq:diff-lapl-chapdyn}) is given by $x_{j}(t)=x_{i}(0)e^{-\mathcal{D}L^{i}_{j}t}$, similarly to what we have seen for continuous-time random walks. However, at variance with random walks, it can be shown that in the stationary regime, the solution takes the form $x_{j}(\infty)\propto u_{j}$, i.e., one will find exactly the same fraction of the quantity in each node, uniformly distributed.

We might wonder how fast diffusion happens. Since the Laplacian matrix is semi-positive definite, it is possible to show that it can be decomposed as
\begin{eqnarray}
L^{i}_{j} = Q^{i}_{h}\Lambda^{h}_{k} (Q^{-1})^{k}_{j},
\end{eqnarray}
where $Q^{i}_{h}$ is a matrix whose columns are the eigenvectors of the Laplacian and $\Lambda^{h}_{k}$ is a diagonal matrix whose entries are the Laplacian's eigenvalues. This algebraic feature allows us to prove that $e^{-\mathcal{D}L^{i}_{j}t}=Q^{i}_{h}(e^{-\mathcal{D}\Lambda t})^{h}_{k} (Q^{-1})^{k}_{j}$, highlighting that the exponential decay of $x_{j}(t)$ is dominated by the smallest positive eigenvalue, which usually is indicated by $\Lambda_2$. Therefore, the diffusion temporal scale is given by $\tau \approx 1 / \Lambda_2$.

\begin{figure}[!th]
\centering\includegraphics[width=\textwidth]{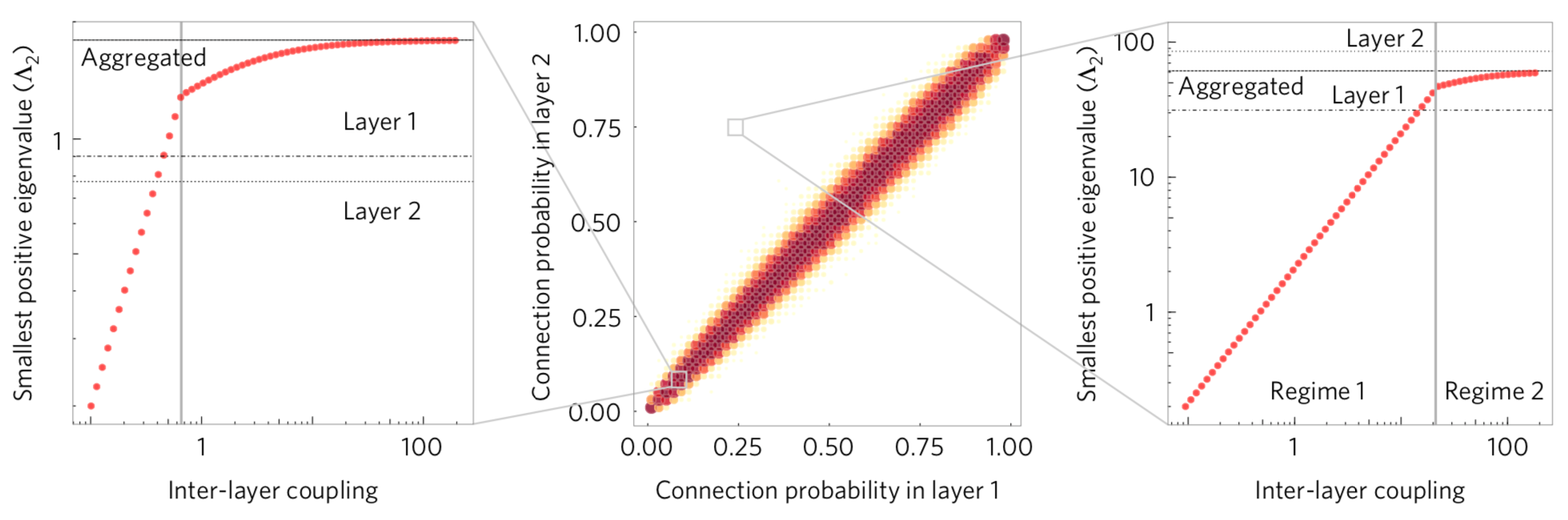}
\caption{\label{fig:superdiff} Continuous-time diffusion on a duplex network, i.e., a multiplex consisting of two layers, obtained from two Erd\H{o}s--R\'{e}nyi networks with independently wiring probabilities $p_1,p_2 \in [0,1]$, used to connect two pairs of nodes within the same layer. Diffusion speed is quantified by the second smallest eigenvalue of the Laplacian tensor, $\Lambda_2$: depending on the wiring probabilities, diffusion in the duplex can be faster (left-hand side panel) or slower (right-hand side panel) than in each layer separately. The condition for enhanced diffusion to happen is given by $\Lambda_2^{\text{multiplex}}\geq \max\{\Lambda_2^{\text{layer 1}}, \Lambda_2^{\text{layer 2}}\}$: a sharp change in the behavior of the characteristic temporal scale can be observed for varying weight of the inter-layer connections (left and right panels), with a clear transition between two distinct regimes above a certain critical value of inter-layer coupling. The middle panel shows when the enhanced diffusion condition holds (encoded by colors), while varying the wiring probabilities. Figure from~\citep{de2016physics}.}
\end{figure}

In the case of interconnected multilayer networks, the diffusion equation has been generalized by means of the supra-adjacency matrix~\citep{gomez2013diffusion} and the tensorial formulation~\citep{de2013mathematical}. In this new setup, a quantity can diffuse through inter-layer connections as well. If we indicate by $X_{i\alpha}(t)$ the rank-2 state tensor at time $t$, then the multilayer diffusion equation can be written as
\begin{eqnarray}
	\frac{dX_{j\beta}(t)}{dt}=M^{i\alpha}_{j\beta}X_{i\alpha}(t)-M^{i\alpha}_{k\gamma}U_{i\alpha}E^{k\gamma}(i\beta)X_{i\beta}(t)\,,
\end{eqnarray}
where $U_{i\alpha}=u_{i}u_{\alpha}$ and $E^{k\gamma}(i\beta)=e^{k}(i)e^{\gamma}(\beta)$. If we define the multilayer combinatorial Laplacian tensor as
\begin{equation}
	L^{i\alpha}_{j\beta}=M^{l\epsilon}_{k\gamma}U_{l\epsilon}E^{k\gamma}(j\beta)\delta^{i\alpha}_{j\beta}-M^{i\alpha}_{j\beta},
\end{equation}
the diffusion equation can be written more compactly as
\begin{eqnarray}
\label{eq:diff-multilapl}
	\frac{dX_{j\beta}(t)}{dt}=-L^{i\alpha}_{j\beta}X_{i\alpha}(t)\,,
\end{eqnarray}
whose solution is given by $X_{j\beta}(t)=X_{i\alpha}(0)e^{-L^{i\alpha}_{j\beta}t}$, a clear generalization of the result obtained in the case of single-layer networks. Also in this case, the second smallest eigenvalue $\Lambda_{2}$ -- calculated from the supra-adjacency matrix representation -- governs the speed of diffusion~\citep{de2013mathematical,sole2013spectral,gomez2013diffusion}, leading to interesting phenomena (see Fig.~\ref{fig:superdiff} for details).

\paragraph{Topological transition with diffusive processes}

\begin{figure}[!t]
\centering\includegraphics[width=\textwidth]{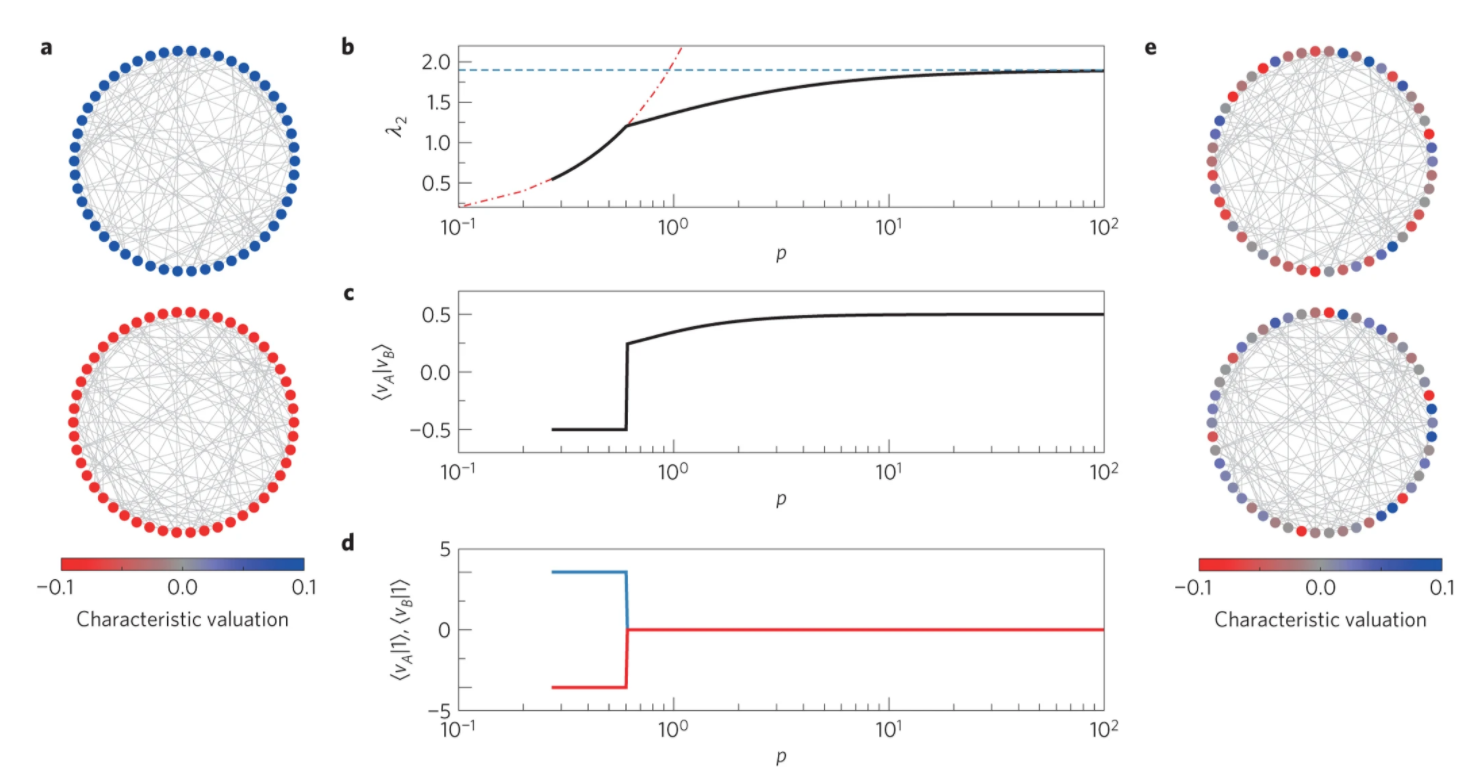}
\caption{\label{fig:chap_abrupt_transition} Algebraic phase transition. In (a) and (e), the intra-layer connectivity of a duplex (i.e., a multiplex consisting of 2 layers) is shown, where the color of each node refers to the value of the corresponding component in the Fiedler vector, i.e., the eigenvector associated with the second smallest eigenvalue $\lambda_2$. In (a) $p = p^* = 0.602$ and in (e) $p = 0.603$, In the middle column is displayed the eigenvalue $\lambda_2$ (b), the inner product $\langle v_A | v_B \rangle$ (c), and, in (d), inner products of $|v_A \rangle$ and $|v_B \rangle$ with the unit vector, i.e., the sum of their elements (see the text for details). Figure reproduced from~\citep{radicchi2013abrupt}.}
\end{figure}

An important question concerns the emergence of such effects in spreading processes, but not only. One can adapt the dynamical rules of models defined in monolayers to encompass the fact that the topology is more complex and this, as we will see, can lead to substantially different behaviors than those observed in isolated networks. Another point of view, though, is to ask why and when a certain model could effectively behave as in an isolated network even though it runs in an a layered structure, or in the other way around, why and when the multilayer dimension dominates, thus disregarding any role of the individual networks of the layers. In some special cases, such as interconnected multiplex networks with identical coupling, it is possible to show the existence of two distinct regimes as a function of the inter-layer coupling strength~\citep{radicchi2013abrupt}, highlighting how the multilayer structure can influence the outcome of several physical processes. By considering a duplex network -- i.e., a multiplex with 2 layers, $A$ and $B$ -- and by using spectral properties of multilayer systems, it was found an abrupt transition between the two aforementioned regimes as a function of $p$, the coupling strength. The two regimes are inferred by analyzing the behavior of eigenvectors and eigenvalues of the supra-Laplacian matrix and are clearly distinguishable, separated by a critical point $p^*$. For $p \leq p^*$, the second smallest eigenvalue\footnote{Note that here we are using $\lambda_2$ instead of $\Lambda_2$, as in the previous section, to keep the same notation of the original paper by~\citep{radicchi2013abrupt}.} $\lambda_2$, which is associated with a plethora of network properties, is independent of the structure of the layers and hence the dynamical processes can be studied separately, while in the regime $p > p^*$, $\lambda_2$ tends to a value independent of $p$, i.e., depending only on the details of the intra-layer networks.

The eigenvector $|v\rangle$ associated with $\lambda_2(p)$ can be split into $|v_A \rangle$ and $|v_B \rangle$, which correspond to the elements of $|v\rangle$ associated with nodes of networks $A$ and $B$, respectively.
It can be proved~\citep{radicchi2013abrupt} that for, $p \leq p^*$, $|v_A \rangle = - |v_B \rangle = 0$ holds, where $|v_A \rangle = \pm \frac{1}{\sqrt{2N}} |1\rangle$, while for $p > p^*$ we have $\langle v_A | 1 \rangle = \langle v_B | 1 \rangle = 0 $. Physically, this means that in the subcritical regime the layers are structurally independent whereas in the supercritical regime the interlayer connection dominates, imposing the same sign in the eigenvector for nodes across networks and alternating the sign for nodes in the same layer. Therefore, the algebraic phase transition can be visualized in several ways. Panels in Fig.~\ref{fig:chap_abrupt_transition} show the behavior of the eigenvectors in the subcritical and supercritical regions. Here, $\lambda_2(p)$ displays a singular point $p^*$ at which the first derivative is not continuous, a sign of an abrupt transition. For $p \leq p^*$, $\lambda_2$ grows as $2p$, while in the other regime it tends to the value that would take for a weighted superposition of the two layers $A$ and $B$, whose Laplacian is $(1/2)(\mathcal{L}_A + \mathcal{L}_B)$.  This abruptness can be observed more directly in the middle and lower panels, which show the behavior of $\langle v_A | v_B \rangle$, $\langle v_A | 1 \rangle$ and $\langle v_B | 1 \rangle$ as a function of $p$.

\subsection{Synchronization processes} \label{sec:synchro}

Synchronization is an emergent phenomenon of a population of dynamically interacting units that, usually with a second-order phase transition~\citep{nelson1977recent,stanley1999scaling,dorogovtsev2008critical}, start operating in a collective, coherent way. Synchronization phenomena may be found in biology, sociology, and ecology and include birds flocking, fireflies flashing, people singing, and neurons spiking, just to mention a few examples.
Once a fully synchronized state is reached, the (linear) stability of such a state is tested by studying the effects of a small perturbation of the system state. This approach was introduced in~\citep{Pecora1998a} for simple network configurations and has been widely used and extended to complex network topologies. Here we briefly introduce the framework of Master Stability Function for a network of oscillators and then extend the formalism to multilayer networks. For a more exhaustive description see~\citep{Boccaletti2018,Arenas2008}.

It is worth remarking that, in the following, we will use the more traditional vector notation where $\mathbf{x}(t)$ indicates the state of the system at time $t$. In some cases, we will also use the Kronecker product operator $\otimes$, as is usual in equations governing synchronization dynamics. Our choice is to help the reader link these concepts to the original studies which introduced them. Nevertheless, the whole section could consider the tensorial formalism, where the system state is indicated by the rank-1 tensor $x_{\ell}(t)$ and where products such as $\mathbf{A}\otimes \mathbf{B}$ are indicated as $A^{\alpha}_{\beta}B^{\gamma}_{\delta}$.

Let us consider a network of $N$ identical oscillators in an $m$-dimensional space, where, in the absence of any interaction, the dynamics of each node $i$ is described by:
\begin{equation}
\dot{\textbf{x}_i} = \textbf{F}(\textbf{x}_i), \qquad i=1,2,...,N; \textbf{x}_i\in \mathbb{R}^{m}.
\label{syn_eq_1}
\end{equation}

We introduce an interaction between oscillators due to the fact that they are coupled in an unweighted network specified by the adjacency matrix $\textbf{A} = \{ A_{ij}\}$ and  we define the output function $\textbf{H}(\textbf{x})$ as the function that governs the interaction between nodes. We also assume that the coupling between the oscillators is diffusive, that is the effect that node $j$ has on node $i$ is proportional to the difference between $\textbf{H}(\textbf{x}_j)$ and $\textbf{H}(\textbf{x}_i)$. 
Then, the evolution of the state of node $i$ is given by:
\begin{equation}
\dot{\textbf{x}_i} = \textbf{F}(\textbf{x}_i) + \sigma \sum_{i=1}^{N} A_{ij} [\textbf{H}(\textbf{x}_j) - \textbf{H}(\textbf{x}_i) ] =
\textbf{F}(\textbf{x}_i) - \sigma \sum_{j=1}^{N} L_{ij} \textbf{H}(\textbf{x}_j),
\label{syn_eq_2}
\end{equation}
where $\textbf{L}$ is the Laplacian matrix and $\sigma$ is the coupling strength . 

\paragraph{Stability of synchronized states} 

The MSF approach to test the stability of a fully synchronized state, described in Appendix~\ref{sec:MSF appendix}, was extended to multilayer complex systems~\citep{DelGenio2016}. In this work, it is considered a network with $M$ different layers, each layer representing a different kind of interaction between nodes. Eq.~(\ref{syn_eq_2}) is thus extended to describe the dynamics of the whole system:
\begin{equation}
\dot{\textbf{x}_i}  =
\textbf{F}(\textbf{x}_i) - \sum_{\alpha=1}^{M}\sigma_\alpha \sum_{j=1}^{N} L^{(\alpha)}_{ij} \textbf{H}^{(\alpha)}(\textbf{x}_j), 
\label{syn_eq_7}
\end{equation}
where $\alpha$ is the index accounting for layers. We obtain the $M$-parameter equation describing the time evolution of perturbation error:
\begin{equation}
 \dot{\boldsymbol{\xi}_i} = \left[J\textbf{F}(\textbf{s})  - \sum_{\alpha=1}^{M} \sigma_\alpha \lambda^{(\alpha)}_i J\textbf{H}_\alpha(\textbf{s})\right] \boldsymbol{\xi}_i,
\label{syn_eq_8}
\end{equation}
where $\boldsymbol{\xi}_i$ is the eigenmode associated with the eigenvalue $ \lambda_i$ of $\textbf{L}$.
As in the case of a dynamics evolving on top of a single layer, in a multilayer system the stability of the synchronized state is completely specified by the sign of the maximum conditional Lyapunov exponent $\Lambda_{max}$. In particular, it is also found that stability of the complete synchronization state may be reached even if each layer, taken individually, is unstable: a very interesting feature for practical application.

It is worth noting that, together with complete synchronization, a network may exhibit other forms of synchronization where clusters of nodes have a synchronized dynamics but different clusters evolve on distinct time evolutions. This type of synchronization is called clustered synchronization (CS) and it has been well studied in terms of cluster formation, stability and role of network symmetries~\citep{Nicosia2013,Pecora2014,Sorrentino2016}. CS has been also studied in multiplex~\citep{Jalan2016} and, more recently, in multilayer~\citep{DellaRossa2020} networks, and cluster stability has been tested as a function of intra- and inter-layer symmetries. 
To describe the evolution of the perturbation error for complex synchronization patterns, such as in cluster synchronization, it is useful to rewrite Eq.~(\ref{syn_eq_3}) with a more compact formalism. To this end, we write the state vector as a vector of vectors,  $\textbf{X} = (\textbf{x}_1; \textbf{x}_2; ...; \textbf{x}_N)$, and the variational equation assumes the form:
\begin{equation}
 \delta \dot{\textbf{X}}= \left[ \textbf{I}_N \otimes J\textbf{F}(\textbf{s}) - \sigma \textbf{L} \otimes J\textbf{H}(\textbf{s}) \right]\delta \textbf{X},
\label{syn_eq_9}
\end{equation}
where $\textbf{I}_N$ is the identity matrix and $\otimes$ is the Kronecker product. Eq.~(\ref{syn_eq_9}) can be decoupled into N independent equations by diagonalizing $\textbf{L}$. However, to deal with cluster synchronization and multilayer interactions we have to further generalize Eq.~(\ref{syn_eq_9}). The variational equations for complex synchronization patterns on generalized networks has the form~\citep{zhang2020}: 
\begin{equation}
 \delta \dot{\textbf{X}} = \left[ \sum_{l=1}^{L} \textbf{D}^{(l)} \otimes J\textbf{F}(\textbf{s}^l)  - \sum_{l=1}^{L} \sum_{\alpha=1}^{M} \sigma_{\alpha} \textbf{L}^{(\alpha)} \textbf{D}^{(l)} \otimes J\textbf{H}^{(\alpha)}(\textbf{s}^l) \right] \delta \textbf{X}
\label{syn_eq_10}
\end{equation}
where the identity matrix has been replaced by the diagonal matrix $\textbf{D}^{(l)}$, whose generic element $D_{ii}^{(l)}=1$ if node $i$ belongs to $l$th dynamical cluster and $D_{ii}^{(l)}=0$ otherwise.

In a recent paper~\citep{zhang2020}, it was established that to optimally decouple Eq.~(\ref{syn_eq_10}), the matrices encoding the synchronization pattern and the interaction pattern, that is $\{\textbf{D}^{(l)} \}$ and $\{\textbf{L}^{(\alpha)} \}$, should be simultaneously diagonalized. In this work, an algorithm was also developed to find the finest simultaneous block diagonalization.

\paragraph{Synchronization in a network of phase oscillators}

One of the first approaches to describe  phase synchronization in an ensemble of oscillators on multiplex networks was done in 2015~\citep{Gambuzza2015}, to investigate the synchronization of indirectly coupled units using a system composed by two layers, where the top layer was made of disconnected oscillators and the bottom one, modeling the medium, consisted of oscillators coupled according to a given topology and with a characteristic natural frequency. Each node of the multiplex was modeled as a Stuart-Landau (SL) oscillator, that is an oscillator with amplitude as well as phase dynamics. Therefore, the Kuramoto model (KM, see Appendix~\ref{sec:Kuramoto}) can be retrieved as a limiting case when the amplitude dynamics vanishes. The Kuramoto order parameter  can be generalized to the multiplex framework as:
\begin{equation}
r^{\alpha\beta}_{ij}=\big| \big \langle  e^{i[\theta^\alpha_i (t)-\theta^\beta_j (t)]}\big\rangle_t \big|,
\label{syn_eq_15}
\end{equation}
while intra- and inter-layer coherence, respectively, can be also defined by
\begin{equation}
r^\alpha = \frac{1}{N(N-1)}\sum^N_{i,j=1}r^{\alpha\alpha}_{ij},
\label{syn_eq_16}
\end{equation} 
and
\begin{equation}
r^{\alpha\beta} = \frac{1}{N}\sum^N_{j=1}r^{\alpha\beta}_{jj}.
\label{syn_eq_17}
\end{equation} 

\begin{figure}[!ht]
\centering
\includegraphics[width=.9\textwidth]{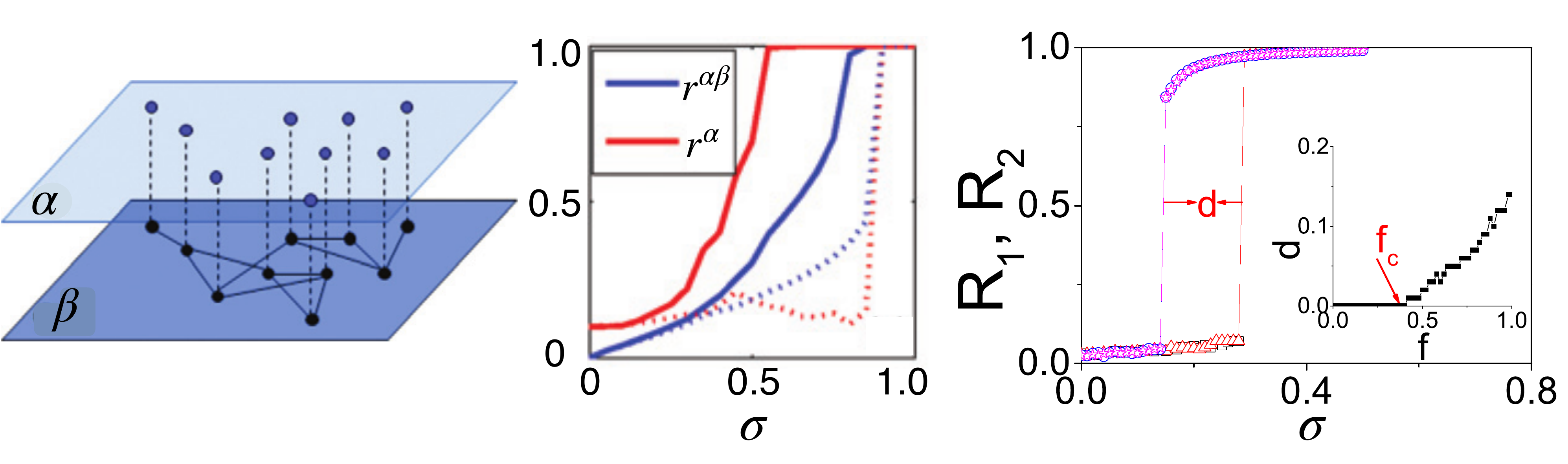}

\caption{Left: Two-layer multiplex network with one-to-one coupling between the layers. In the top layer ($\alpha$) the nodes only interact with those in the bottom one whereas in the bottom layer ($\beta$) the nodes also interact with other members of the same layer. Middle: Kuramoto order parameters ($r^\alpha$) and ($r^{\alpha\beta}$) vs. coupling coefficients $\lambda=\lambda_{\alpha\beta}=\lambda_\beta/5$. Continuous lines refer to a multilayer network of Stuart-Landau
oscillators with $a = 1$, whereas the dashed ones to purely phase oscillators ($a \xrightarrow{}\infty$). Right: Synchronization transitions in two-layer networks  with a fraction of the nodes adaptively controlled by a local order parameter $f = 1$. Squares and circles (triangles and stars) refer to the values of $R1$ ($R2$), and the insets show the corresponding dependence of the width of hysteretic loop $d$ on $f$. Left and middle panels readapted from~\citep{Gambuzza2015}, right panel from~\citep{Zhang2015}}
\label{fig:dyn_synchro}
\end{figure}

By studying a population of $N$ disconnected oscillators, indirectly coupled through an inhomogeneous medium, authors have shown the onset of intra-layer synchronization without inter-layer coherence, i.e. a state in which the nodes of a layer are synchronized between them without being synchronized with those of the other layer (see Fig.~\ref{fig:dyn_synchro}, left panel). Synchronization of units that are not connected requires the presence of an amplitude dynamics as the regime of intra-layer synchronization is not observed in purely phase oscillators, such as those in the KM.

As previously mentioned, not only were continuous second-order transitions observed in an ensemble of networked phase oscillators, but examples of an abrupt first-order transition, named explosive synchronization (ES), were also observed~\citep{Gomez-Gardenes2011,Leyva2013}. It was first observed in a network of oscillators presenting a positive correlations between natural frequencies and the degree of the nodes. Moreover, ES was also studied in systems where a local order parameter for the $i$-th oscillator is defined~\citep{Zhang2015}:
\begin{equation}
r_i(t)e^{i\Phi(t)}= \frac{1}{k_i}\sum_{j=1}^{k_i}\sin(\theta_j),
\label{18}
\end{equation}
and where the phase dynamics is expressed as:
\begin{equation}
 \dot{\theta_i}=\omega_i + \sigma \alpha_i \sum_{j=1}^{N} A_{ij}\sin(\theta_j-\theta_i).
\label{syn_eq_19}
\end{equation}
The overall amount of phase coherence in the network is measured by means of the global order parameter $R$:
\begin{equation}
 Re^{i\Psi}=\frac{1}{N}\sum_{j=1}^{N}e^{i\theta_j},
\label{syn_eq_20}
\end{equation}
where $0 \leq R \leq 1$ and $\Psi$ denotes the average phase. 

Explosive synchronization onset has been reported in a system of two interdependent networks (see Fig.~\ref{fig:dyn_synchro}, right panel), with the same size and where nodes on the two layers are coupled in a one-to-one correspondence, so that a group of oscillators in the first layer is controlled by the local order parameters of the corresponding nodes on the second layer, and vice-versa~\citep{Zhang2015}. Nevertheless, it was shown that ES is a property of a generic multilayer network as long as some microscopic suppressive rule can prevent the formation of the giant synchronization cluster which characterizes second-order transitions.

\subsection{Game dynamics: cooperation processes}\label{sec:coopdyn}

Human cooperation may be intended as a collective behavior that emerges as the result of the interactions among individuals. In the past few years cooperation has been studied in social sciences with methods of statistical physics~\citep{Perc2017}, in particular Monte Carlo methods and the theory of collective behavior of interacting particles near phase-transition points. That approach has proven very valuable for understanding cooperation and its spatio-temporal dynamics. 
The mathematical framework used to study human cooperation is usually evolutionary game theory, which quantitatively describes social interactions using example games and formalizes the concept of the social dilemma, intended as the conflictual choice that an individual has between doing what is best for society or doing what is best for themselves. 

\begin{figure}[!ht]
\centering
\includegraphics[width=0.6\textwidth]{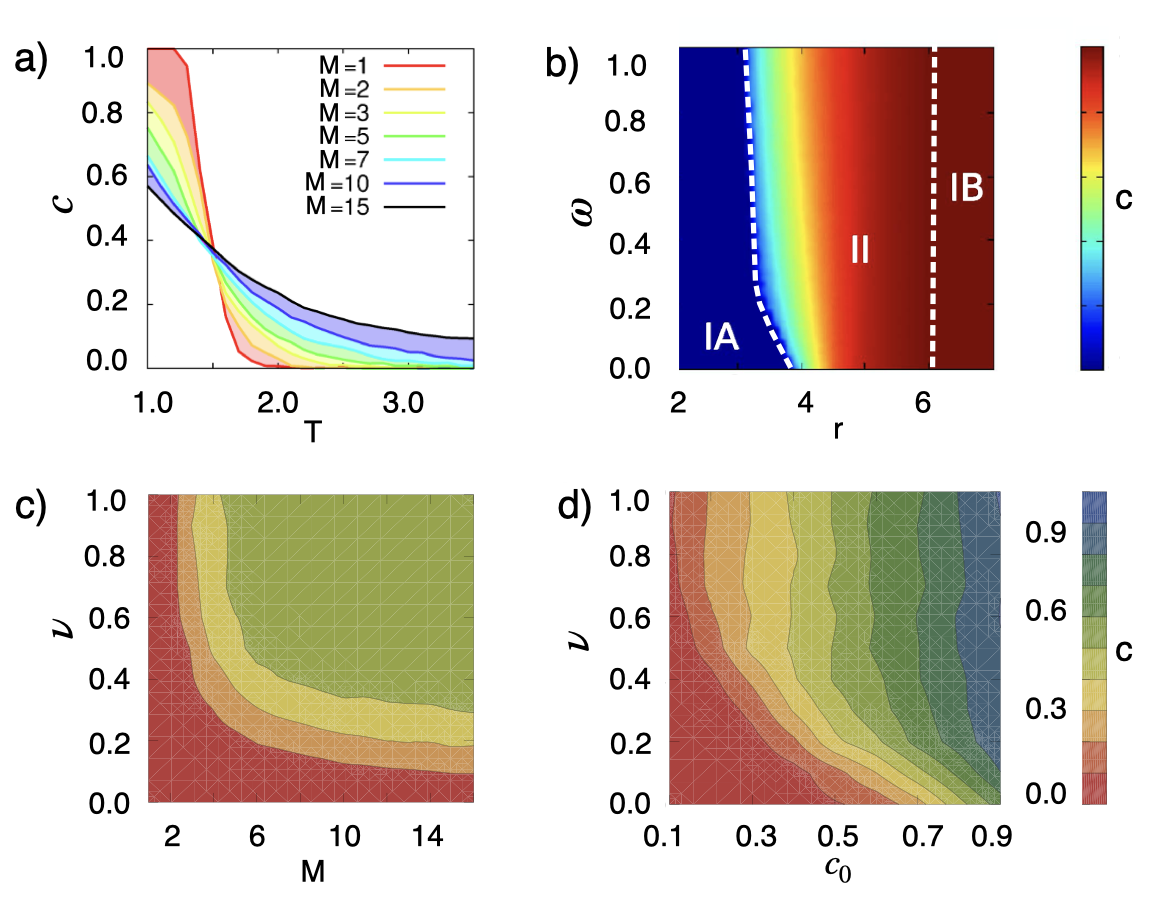}
\caption{a) Mean final cooperation as function of the temptation to defect $T$, for several multiplex networks with different numbers of layers $M$.  b) Multiplex with $M=2$ layers, where cooperation is studied as a function of synergy factor $r$ and edge overlap $\omega$. Dashed lines separate regions with either full defection $c = 0$ (IA) or full cooperation $c = 1$ (IB), from the region of continuously evolving coexistence of cooperators and defectors $0 < c < 1$ (II). Panel a) readapted  from ~\citep{gomez2012evolution} and panel b) readapted from~\citep{battiston2017determinants}.
c) Mean final cooperation (color coded) for the Prisoner's Dilemma as a function $M$ and the strength of degree correlations $\nu$.
d) Mean final cooperation as a function of the initial density of cooperators $c_0$ and the strength of correlations $\nu$. Panels c) and d) readapted from~\citep{Kleineberg2018}.}
\label{fig:dyn_game}
\end{figure}

Evolution of cooperation strongly depends on the population interaction network: individuals who have the same strategy are more likely to interact \citep{Nowak1992,Nowak2010}, effectively creating resilient cooperative clusters in a structured population, a phenomenon named network reciprocity. When the edges that determine the interaction among individuals were fixed on a time scale, it was demonstrated~\citep{Santos2006} that in heterogeneous populations -- modeled by networks with degree distributions exhibiting a power-law behavior -- the sustainability of cooperation is simpler to achieve than in homogeneously structured populations. Here we address the problem of how human cooperation emerges in multiplex networks, with different interaction layers that can account for different kinds of social ties an individual may be involved in. 

In particular, we consider a Prisoner's Dilemma game implemented in a set of $M$ interdependent networks, characterized by an average degree $\langle k\rangle$, each of them containing the same number $N$ of nodes. Each individual is represented by one node in each of the layers and we define a set of adjacency matrices $\{A^\alpha\}$ so that $A_{ij}^\alpha=1$ when two nodes are connected in layer $\alpha$ and  $A_{ij}^\alpha=0$ otherwise.

At each time step, for each of the $k_i^\alpha$ games played, each individual $i$ facing a cooperator collects a payoff $\pi_{ij}=R$ or $\pi_{ij}=T$  when playing as a cooperator or as a defector, respectively. Conversely, if $i$ faces a defector, $i$ collects a payoff $\pi_{ij}=S$  or $\pi_{ij}=P$  playing as a cooperator or as a defector, respectively. For game parameters it holds that $S<P<R<T$. The aggregated payoff of node $i$ is given by the sum of the payoffs  $\pi_{i}$ over all layers. Furthermore, after each round of the game, individuals update their strategies with a rule that can use the global knowledge about the benefits of the neighbors or be random~\citep{Gomez-Gardenes2011b}. 

Finally, it worth noting that cooperation is also studied in public good games, where the Prisoner's Dilemma is played in overlapping groups of individuals: cooperators contribute with a ``cost'' $d$ to the public good, while defectors do not contribute and the total amount in the common pool of each group is multiplied by a factor $r$ and distributed equally among all members of the group~\citep{Santos2008a}.

Previous research \citep{gomez2012evolution} demonstrated that the resilience of cooperative behavior can be enhanced by the multiplex structure through the simultaneous formation of correlated clusters of cooperators across different layers, i.e. through multiplex network reciprocity. However, it was also proven \citep{battiston2017determinants} that, to gain benefits from multiplex structure, a high topological overlap is needed or, in other words, individuals have to be similarly linked across different layers  (see Fig.~\ref{fig:dyn_game}, panels a) and b)). Other studies~\citep{Kleineberg2018} investigated the role of degree correlation $\nu$ between nodes in different layers for the emergence of cooperation. They found that in the absence of degree correlations, increasing the number of layers only leads to mild changes. However, if degree correlations are present we observe a mean final cooperation of $c=0.5$, and this value is nearly independent of the game payoff parameters. This mechanism is called \emph{topological enslavement} and can be understood by considering that, if degree correlations are strong, hubs dominate the game dynamics, since they have the potential to earn higher payoffs (because they play more games) and they are more likely to be selected by other nodes as imitation candidates. Furthermore, topological enslavement implies that the outcome of the evolution of the system is determined by the initial conditions (see Fig.~\ref{fig:dyn_game}, panels c) and d)).

\begin{figure}[!th]
\centering
\includegraphics[width=0.55\textwidth]{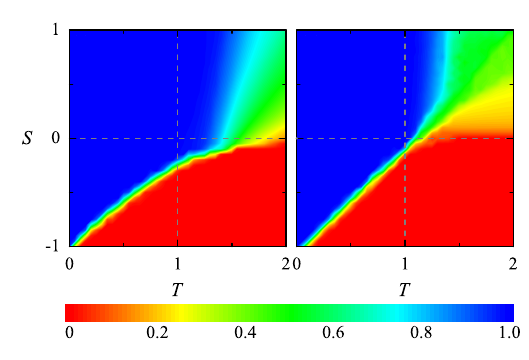}
\caption{Fraction of cooperators as a function of the payoff of a cooperator when playing with a  defector ($S$) and the payoff of a defector when facing a cooperator ($T$). The interaction layer is subject to disassortative mixing  (assortative coefficient  $A_I<0$) while the updating network is subject to assortative mixing (disassortative coefficient $A_U>0$). Results show the disruptive effect of symmetry breaking for the evolution of cooperation. Parameter values are $A_I=-0.1$ and $A_U=0.1$ (Left) and $A_I=-0.2$ and $A_U=0.2$ (Right). Figure from~\citep{wang_pre14a}.}
\label{fig:symbreak}
\end{figure}

Finally, as anticipated in Sec.~\ref{sec:layercorr}, layer-layer correlations might have a deep impact on the dynamics on the top of multilayer systems. This is the case for game dynamics, where cooperation might be hampered, rather than enhanced, by specific correlation patterns combining assortative and disassortative degree mixing across layers~\citep{wang_pre14a,duh_njp19} (see Fig.~\ref{fig:symbreak} for details).

\subsection{Interdependent processes}\label{sec:interdepproc}

The second class of dynamical processes on multilayer networks is the one of ``interdependent'' or ``coupled'' dynamics, consisting of systems characterized by \textit{different} processes on each layer. The inter-layer interactions couple such processes and are responsible for emerging phenomena which could not be detected in a single-layer network framework (see Tab.~\ref{tab:table_app2}). 

As in the case of the structure, it is possible to introduce a unifying framework in terms of a dynamical SNXI decomposition, outlined in Eq.~(\ref{eq:sexi-decomp-structure}), to describe dynamics on multilayer networks. Let $x^{[l]}_{i \alpha}$  (where $l \in {1, 2,..., C }$) denote the $l$-th component of a $C-dimensional$ vector $\mathbf{x}_{i\alpha}$
that represents the state of node $i$ in layer $\alpha$. The most general (and possibly nonlinear) dynamics governing the evolution of each state is given by the systems of equations
\begin{align}\label{eq:sexi_dyn}
\dot{\mathbf{x}}_{i \alpha}(t) &=F_{i \alpha}(\mathbf{X}(t))=\sum_{\beta=1}^{L} \sum_{j=1}^{N} f_{i \alpha}^{j \beta}(\mathbf{X}(t)) \\ \nonumber
&=\underbrace{\sum_{\beta=1}^{L} \sum_{j=1}^{N} f_{i \alpha}^{j \beta}(\mathbf{X}(t)) \delta_{\alpha}^{\beta} \delta_{i}^{j}+\sum_{\beta=1}^{L} \sum_{j=1}^{N} f_{i \alpha}^{j \beta}(\mathbf{X}(t)) \delta_{\alpha}^{\beta}\left(1-\delta_{i}^{j}\right)}_{\text {intra-layer dynamics }} \\ \nonumber
&+\underbrace{\sum_{\beta=1}^{L} \sum_{j=1}^{N} f_{i \alpha}^{j \beta}(\mathbf{X}(t))\left(1-\delta_{\alpha}^{\beta}\right) \delta_{i}^{j}+\sum_{\beta=1}^{L} \sum_{j=1}^{N} f_{i \alpha}^{j \beta}(\mathbf{X}(t))\left(1-\delta_{\alpha}^{\beta}\right)\left(1-\delta_{i}^{j}\right)}_{\text {inter-layer dynamics }} \\ \nonumber
&=\underbrace{f_{i \alpha}^{i \alpha}(\mathbf{X}(t))}_{\text {self-interaction }}+\underbrace{\sum_{j \neq i} f_{i \alpha}^{j \alpha}(\mathbf{X}(t))}_{\text {endogenous interaction }}+\underbrace{\sum_{\beta \neq \alpha} \sum_{j \neq i} f_{i \alpha}^{j \beta}(\mathbf{X}(t))}_{\text {exogenous interaction }}+\underbrace{\sum_{\beta \neq \alpha} f_{i \alpha}^{i \beta}(\mathbf{X}(t))}_{\text {intertwining }} \\ \nonumber
&=\mathbb{S}_{i \alpha}(\mathbf{X}(t))+\mathbb{N}_{i \alpha}(\mathbf{X}(t))+\mathbb{X}_{i \alpha}(\mathbf{X}(t))+\mathbb{I}_{i \alpha}(\mathbf{X}(t))
\end{align}
where $\mathbf{X}(t) \equiv\left(\mathbf{x}_{11}, \mathbf{x}_{21}, \ldots, \mathbf{x}_{N 1}, \mathbf{x}_{12}, \mathbf{x}_{22}, \ldots, \mathbf{x}_{N 2}, \ldots, \mathbf{x}_{1 L}, \mathbf{x}_{2 L}, \ldots, \mathbf{x}_{N L}\right)$ and we did not use the tensorial formalism to make explicit the contributions of each term in the governing equation. In fact, this equation could also be compactly written as
\begin{eqnarray}
\dot{x}_{j\beta}(t) = F[x_{j\beta}(t)] = \mathbb{S}[x_{j\beta}(t)] + \mathbb{N}[x_{j\beta}(t)] + \mathbb{X}[x_{j\beta}(t)] + \mathbb{I}[x_{j\beta}(t)].
\end{eqnarray}
We call this equation ``dynamical $\mathbb{SNXI}$ decomposition''.
Similarly to the structural decomposition in Eq.~(\ref{eq:sexi-decomp-structure}), we have decoupled the different contributions of intra-layer and inter-layer dynamics, allowing us to classify different dynamical processes in terms of the corresponding dynamical $\mathbb{SNXI}$ components.
The peculiar behavior of the interdependent processes is ascribed to the exogenous and intertwining components. 
Although the most-studied examples come from mixed spreading processes, which are crucial for understanding phenomena such as the spreading dynamics of two concurrent diseases in two-layer multiplex networks~\citep{Sanz2014,salehi2015spreading,Dickison2012epidemics,Cozzo2013social,DeArruda2017}, and the spread of diseases coupled with the spread of information or behavior~\citep{wang2015coupled,funk2015,Granell2013,lima2015disease,Funk2009,granell2014competing}, other types of dynamical interdependence are attracting a growing interest.

\paragraph{Coupling diffusion with synchronization}

An illustrative and pedagogical example has been proposed by~\citep{nicosia2017collective}. In this work, the authors examine the interdependent dynamics of the two processes presented in the two previous sections, namely, diffusion and synchronization. They propose a model that mimics the interplay between the neural activity and energy transport in brain regions, from which a rich collection of behaviors emerges.
These processes evolve in the two layers of a multilayer network and are related to each other by the correspondence between layers, which in this case is realized through the functional relation between the parameters governing the two processes and the state variables. In particular, the dynamics of the entire system is governed by the following equations:
\begin{equation}
   \left\{
    \begin{aligned}
    \dot{x}_{i} &=F_{\omega_{i}}\left(\mathbf{x}, A^{[1]}\right) \\
    \dot{y}_{i} &=G_{\chi_{i}}\left(\mathbf{y}, A^{[2]}\right)
    \end{aligned} \quad i=1,2, \ldots, N
\right.
\end{equation}
where $\mathbf{x}=\left\{x_{1}, x_{2}, \ldots, x_{N}\right\} \in \mathbb{R}^{N}$ and $\mathbf{y}=\left\{y_{1}, y_{2}, \ldots, y_{N}\right\} \in$
$\mathbb{R}^{N}$ denote the states of the two dynamical processes, while the topologies of the two layers are encoded in the adjacency matrices $A^{[1]}=\left\{a_{i j}^{[1]}\right\}$ and $A^{[2]}=\left\{a_{i j}^{[2]}\right\},$ respectively, such that $a_{i j}^{[1]}=1\left(a_{i j}^{[2]}=1\right)$ if a link exists between nodes $i$ and $j$ in the first (second) layer, and $a_{i j}^{[1]}=0$ $\left(a_{i j}^{[2]}=0\right)$ otherwise. The evolution of the system depends on a set of parameters $\omega$ and $\chi$, which in turn depend on the state of the nodes in the adjacent layer:
\begin{equation}\label{coupling}
    \begin{array}{l}
    \dot{\omega}_{i}=f\left(\omega_{i}, y_{i}\right) \\
    \dot{\chi}_{i}=g\left(\chi_{i}, x_{i}\right)
    \end{array} \quad i=1,2, \ldots N
\end{equation}
The authors assign to functions $F_{\omega_{i}}$ and $G_{\chi_{i}}$ a Kuramoto dynamic and a continuous-time random walk, respectively. Subsequently, they assign the functions $f$ and $g$ in Eqs.~\eqref{coupling}, respectively, relating the frequency $\omega_{i}$ of the oscillator $i$ at layer 1 to the state $y_{i}$ at layer 2 , and the bias property $\chi_{i}$ of the random walkers at layer 2 to the oscillator phase $x_{i}$ at layer $1$. The natural frequency $\omega_i$ of the oscillator $i$ evolves relaxing to values proportional to the fraction of random walkers at the replica node $i$ in the other layer. Analogously, the random walks are biased toward (away from) strongly synchronized nodes. 

\begin{figure}[t]
\centering\includegraphics[width=0.9\textwidth]{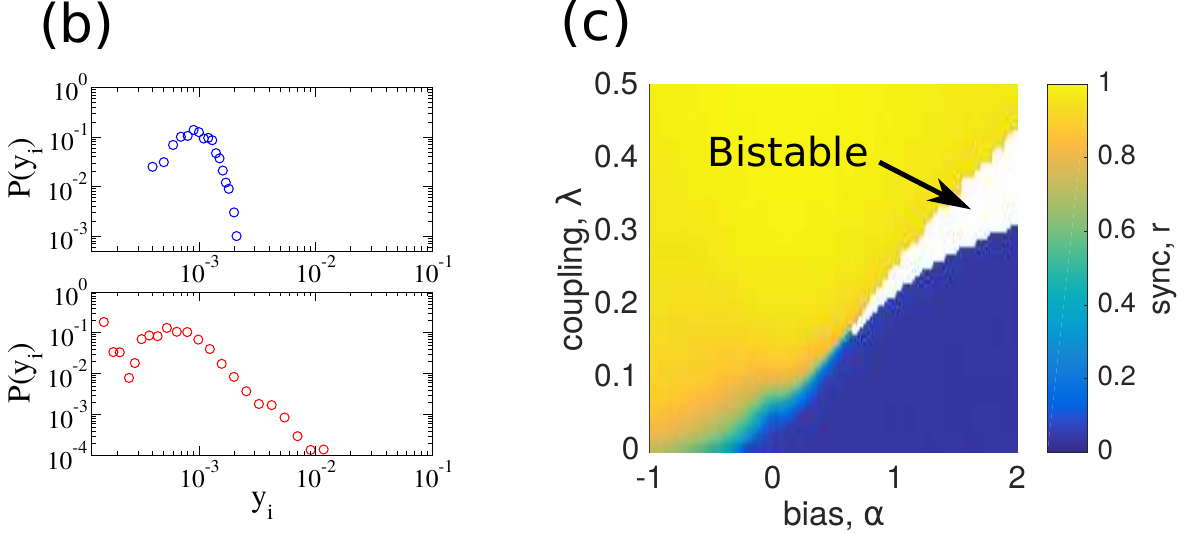}
\caption{\label{fig:empirical_interdependence} Coupling diffusion with synchronization dynamics. a) Distribution $P(y_i)$ of steady-state random walker fractions $y_i$ at layer $2$ for $\alpha=1.0$, when the oscillators at layer $1$ are incoherent ($\lambda=0.1$, top, blue) and synchronized ($\lambda=0.4$ bottom, red). b) Synchronization phase diagram showing the level of synchronization  as a function of coupling $\lambda$ and bias exponent $\alpha$. The bistable region is colored in white. Figure adapted from \citep{nicosia2017collective}.}
\end{figure}

Note that the coupling between the two layers is tunable through two parameters $\lambda$ and $\alpha$ that represent, respectively, the intensity of the influence of the random walk on the oscillators and vice versa. This setup completely defines an interdependent process on a multilayer network: depending on the coupling strengths $\lambda$ and $\alpha$ the collective behavior exhibits special dynamics (see Fig.~\ref{fig:empirical_interdependence}).
The random walkers are homogeneously distributed in the incoherent state, while in the synchronized state the distribution is heterogeneous. Conversely, at certain values of the tuning parameters the system encounters an \textit{explosive synchronization}, or a bistability region characterized by a hysteretic behavior. Similar phase transitions can also be observed in single-layer networks under certain conditions, as reported in \cite{acebron2005kuramoto, martens2009exact, buendia2021broad}. Importantly, the multilayer network model offers a parsimonious explanation of the emergence of these collective phenomena, considering explicitly the intertwined nature of the dynamics.

\paragraph{Coupling epidemics spreading with awareness diffusion}

Many different phenomena in nature can be described, in their essence, by the results of constructive or destructive relations between two or numerous parts. For instance, the mutualistic or competitive relation between dynamical systems gives rise to a wealth of fascinating behaviors. 

The dynamics occurring in one layer can have positive or negative feedbacks on another: for example, human behavior (e.g., information awareness) can inhibit the spread of disease; social mixing between classes and mobility may produce abrupt changes in the critical properties of the epidemic onset; cooperation emerges where the classical expectation was defection. 

A generalization of these results can be found in~\citep{danziger2019dynamic}, in which the authors describe some universal features of interdependent systems with coupled dynamics. There is an interesting relation between the collective behavior emerging from percolation processes and from the ones arising in interdependent systems. In such processes, abrupt transitions and critical dynamics may arise in certain circumstances and, in dynamically coupled systems, other interesting phenomena may be also observed, such as hysteresis and multistability, with functionality of nodes in one layer influencing the functionality of their replicas in the other layers. In \citep{danziger2019dynamic} the functionality is quantified by an order parameter, which plays a role in the coupling strength between the nodes in different layers, with the order of a node dynamics affecting the order of its neighbors. This fact is the dynamical counterpart of interdependent percolation, where the functionality of a node is related to its belonging to the mutual giant connected component (see Sec.~\ref{sect:perc} for details). It turns out that the dynamic interdependence increases the vulnerability of the system~\citep{danziger2016vulnerability}, as in the case of percolation processes.

\begin{figure}[t!]
\centering\includegraphics[width=\textwidth]{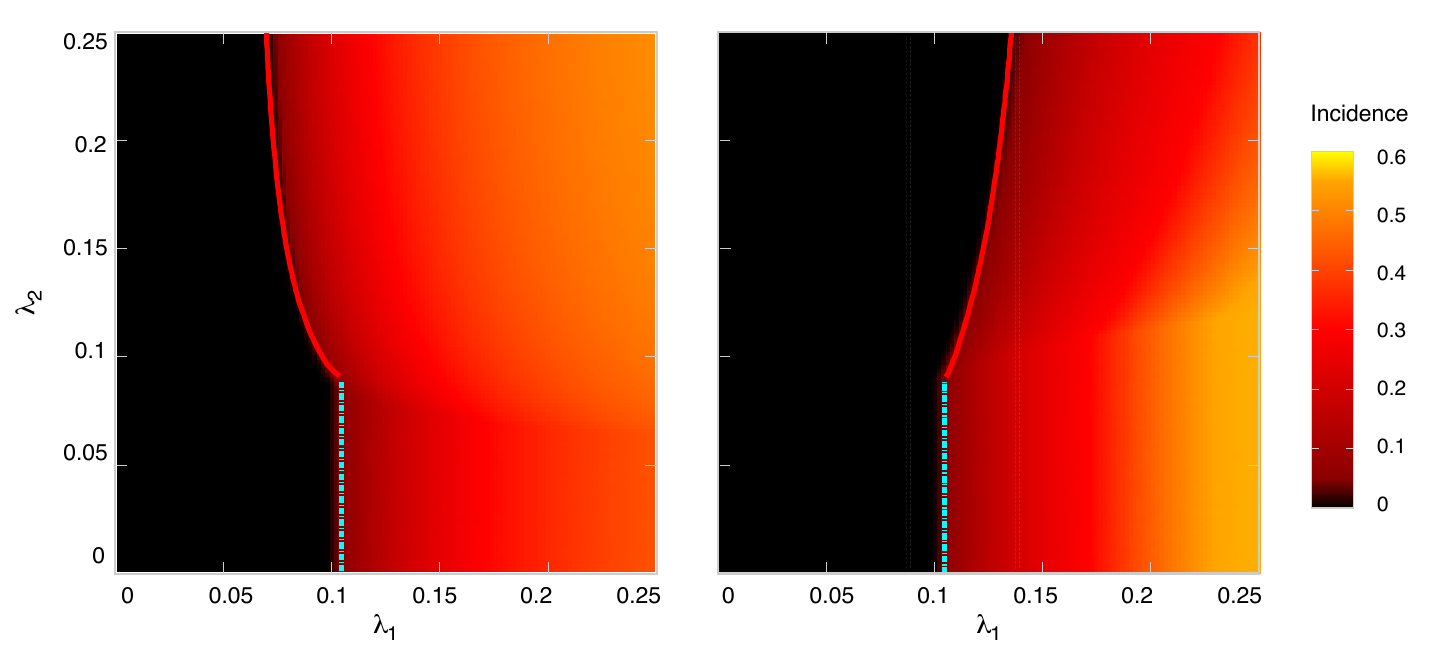}
\caption{\label{fig:critical_curve} Two (left) reciprocally enhanced and (right) reciprocally inhibited disease-spreading processes of susceptible–infected–susceptible type. The colors in the figure represent the prevalence levels of the diseases at a steady state of Monte Carlo simulations. Note the emergence of a curve of critical points (at a ``metacritical point'') in which the spreading in one layer depends on the spreading in the other. Figure from~\citep{de2016physics}.}
\end{figure}

Epidemic models are another important example of systems in which interdependencies play a crucial role (see~\citep{wang2015coupled,de2016physics} for a review) and, also in this case, hysteretic behavior with abrupt transitions may occur. This means that, for instance, explosive pandemics with no early-warning can suddenly appear and, in a similar implosive way, can disappear. Unfortunately, the value of the infection rate which can eradicate the disease must be much lower than the value that triggered it~\citep{danziger2019dynamic}. Other interesting behaviors arise in the case of competitive diseases, where mutual exclusion or endemic coexistence may spontaneously occur. 

Figure~\ref{fig:critical_curve} shows two phase diagrams of disease incidence of reciprocally enhanced (right) and inhibited (left) disease-spreading processes. In the figure is highlighted the existence of a curve of critical points that separate endemic and non-endemic phases of the disease. Moreover, the curve that separates the endemic and non-endemic phases, is in turn divided into two parts: one in which the critical properties of one spreading process are independent of the other (straight dashed line), and one in which the critical properties of one spreading process do depend on those of the other layer (solid curve). The point at which this crossover occurs is called a \textit{metacritical point}.

\begin{figure}[t!]
\centering\includegraphics[width=\textwidth]{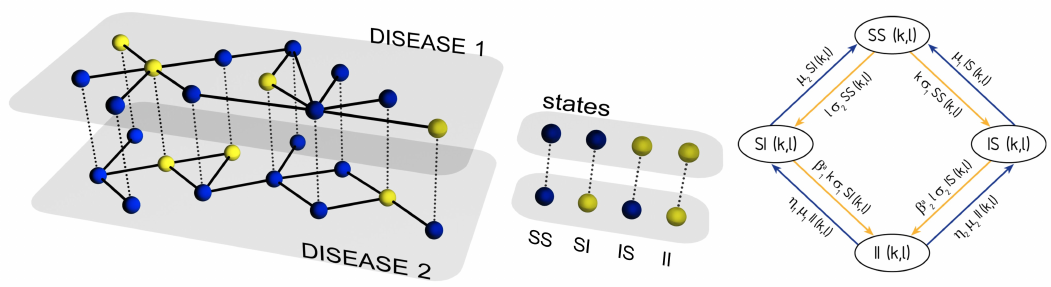}
\caption{\label{fig:concurrent_diseases} SIS-SIS interacting diseases model. Left: multiplex representation of the diseases spreading. Each individual is present in both the layers and can be infected by one (or both) of the diseases, as indicated in the central panel. Right: possible transitions between the different states of the SIS model. The variables represent the densities of individuals of each type having degree equal to $k$ in the first layer and degree $l$ in the second. Figure from~\citep{Sanz2014}.}
\end{figure}

In fact, traditional epidemic models can describe in details the spreading of a single disease in different realistic situations, whereas the multilayer representation can extend the possibilities to spreading processes with interacting diseases. In~\citep{Sanz2014} the authors propose a framework to describe the spreading dynamics of two concurrent diseases~(see Fig.\ref{fig:concurrent_diseases}). Using SIS-SIS and SIR-SIR models with appropriate interlayer couplings describing the influence of one disease on the other, they derive the epidemic threshold for the two interacting diseases, showing that the onset of a disease's outbreak is conditioned to the prevalence levels of the other disease. 

\begin{figure}[t!]
\begin{minipage}[c]{.4\textwidth}
\includegraphics[width=\textwidth]{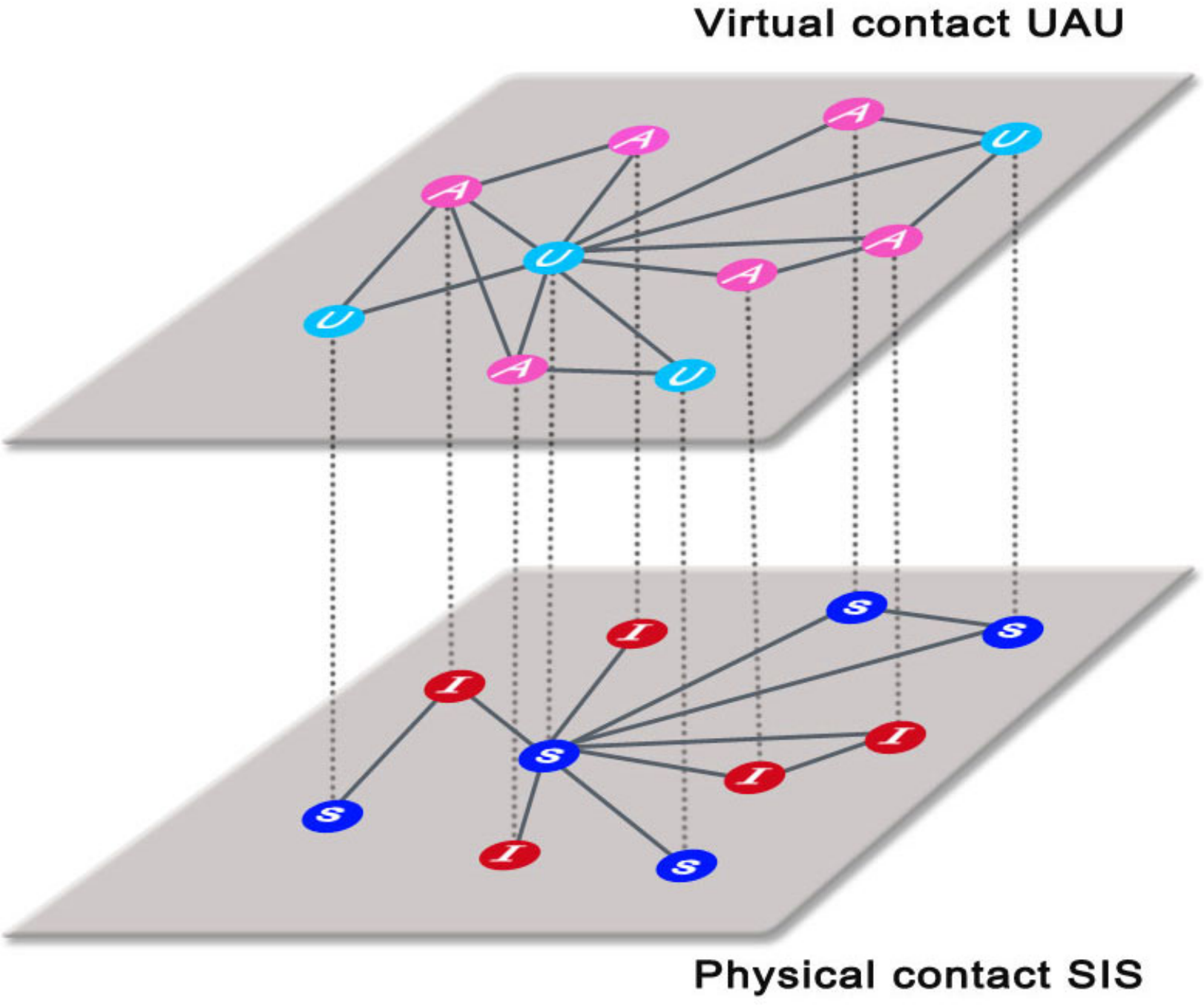} \hfill
\end{minipage} 
~
\begin{minipage}[c]{0.55\textwidth}
\caption{\label{fig:uau}Multiplex representation of awareness-disease spreading. The spreading of awareness occurs in the upper layer, whereas the spreading of the diseases takes place in the lower layer. Figure from~\citep{Granell2013}. }
\end{minipage}
\end{figure}

Epidemic thresholds can be influenced not only by the presence of a second disease, but also by the individual's change of awareness about an ongoing epidemic. In~\citep{Granell2013}, a multilayer network couples the dynamics of disease spreading in a social network and the diffusion of awareness among actors. The two layers are coupled in a multiplex as illustrated in Fig.~\ref{fig:uau}. The process of spreading of awareness (UAU) is akin to an SIS process, where in place of susceptible (S) there are unaware (U) and in place of infected (I) there are aware (A) actors. The probability of being infected is influenced by the state of awareness of the individuals. The authors found the existence of a \textit{metacritical} point in which the state of awareness of the individuals can control the onset of an epidemic.

\paragraph{Coupling game dynamics with opinion dynamics}

\begin{figure}[!ht]
\centering\includegraphics[width=0.8\textwidth]{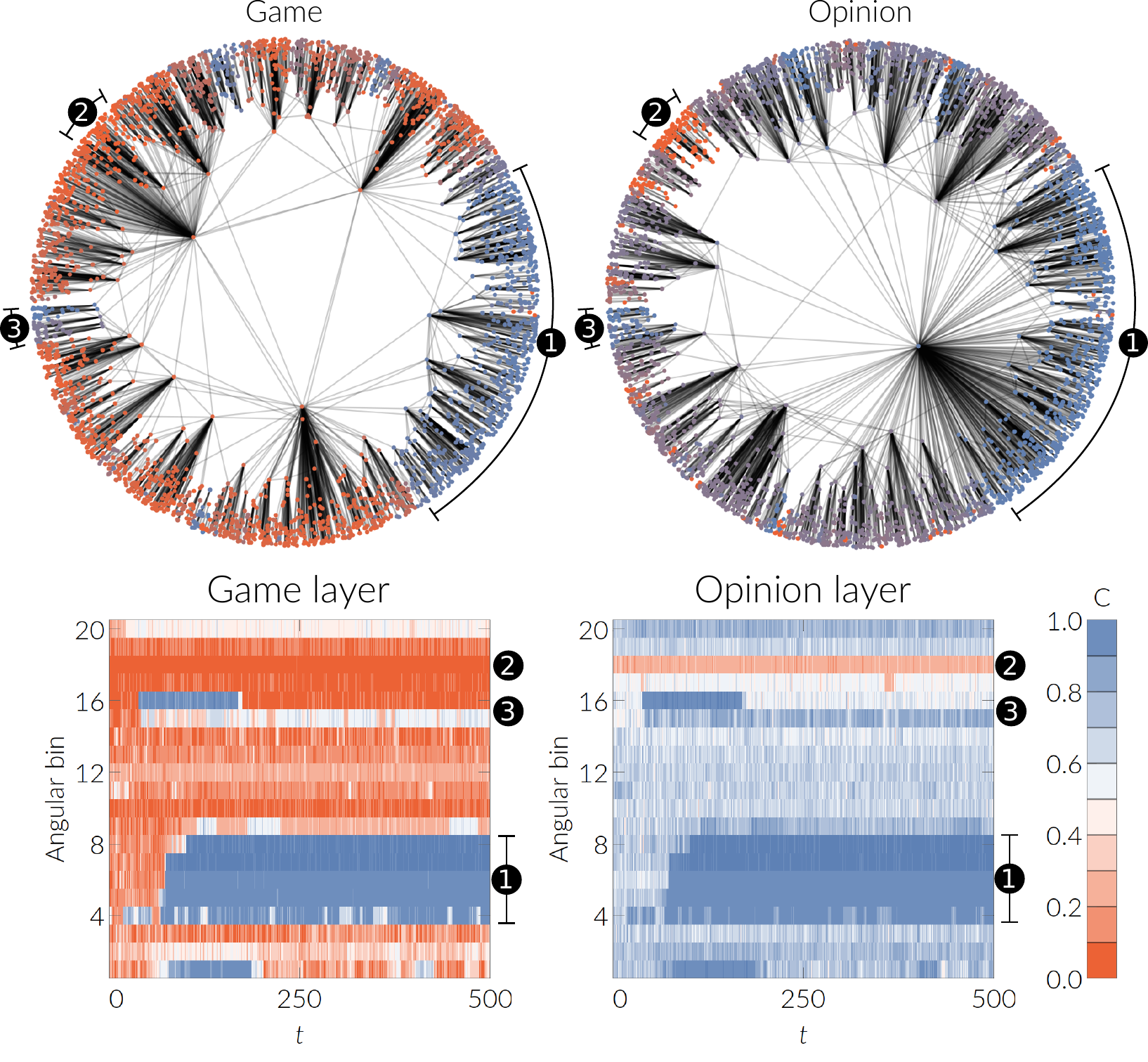}
\caption{\label{fig:game_opinion}Polarization of opinions and strategies in a multiplex with 5,000 nodes, in the presence of angular correlations between the layers. Top: comparison between the two layers; the color is the time average of the state of each node. Bottom: evolution of the density of cooperators in the angular bins used to compute the interlayer correlations. Numeric labels indicate the clusters of nodes that adopt the same strategy. Figure from~\citep{amato2017interplay}.}
\end{figure}

The multilayer representation of the dynamics of complex systems can also be used to shed light on real world social dilemmas. The influence of factors acting on different layers might explain the emergence of particular patterns of cooperation between social agents. In \citep{amato2017interplay} the authors present a model coupling evolutionary game dynamics and opinion dynamics, regarded as two processes evolving on distinct layers of a multiplex network.

To model the game dynamics on the first layer the authors adopt a \textit{replicator}, in which the individuals (nodes) copy the strategy of one of their neighbors with a probability that is so much higher the higher the neighbors' payoff compared to themselves. The possible states are "cooperate" and "defeat". The opinion dynamics on the second layer is modeled using the \textit{voter model}, where individuals (nodes) adopt the opinion of a randomly selected neighbor with a certain probability. The opinions of the nodes can be "cooperate" or "defeat", as in the game layer. The authors assume that imitating a cooperative opinion is more likely than imitating a defection. This can be interpreted as the influence of media campaigns or broadcasting agents. 
Depending on the specific parametrization of both the game and the opinion dynamics, this model gives rise to fascinating dynamical behaviors in which equilibria of different types exist for the game dynamics.
The impact of social influence on the decision of individuals is conveyed by the interlayer coupling, which is encoded in the parameter $\gamma$, representing the tendency of individuals to act in agreement with their proclamations: the nodes in one layer copy their own state from the other layer with probability $\gamma$.

The main result is that this model is sufficient to explain the emergence of cooperation in scenarios where the pure game dynamics predicts defection. This is due both to the intertwined dynamics and to the multilayer structure itself. In fact, the authors proved that the geometric correlation between layers has a significant impact on the stability of the system. Importantly, under certain conditions of correlation between layers, the system can reach a polarized metastable state~(see Fig.~\ref{fig:game_opinion}). This result can explain the observed polarization in real world social systems.

Ultimately, the emergence of cooperation in unexpected conditions is due to the interplay between the coupled dynamics of strategies and opinions, the complex topologies of the networks upon which the dynamics exert, and the structural relations between the layers. Missing one of these elements could hinder the right interpretation of the complex behavior of such systems.

\paragraph{Coupling epidemics spreading with social integration}

A multilayer perspective on epidemic spreading can be a valuable support to design more educated strategies to reduce disease risk. A recent example comes from \citep{bosetti2020heterogeneity}, which integrates social dynamics, human mobility and epidemic spreading to assess the risk of measles outbreak in Turkey. During the past decade, Turkey has received more than 3.5 million refugees coming from Syria. The levels of immunization of the two populations are considerably different. The outbreak risk is analysed through a multilayer transmission model, which takes into account the different levels of immunization in the two populations, along with the mobility pattern and the level of social integration. 
The main result of the study is that, in the case of heterogeneous immunization, high levels of social interaction can drastically reduce the spatial spread and incidence of a disease. This apparently counter-intuitive result is due to the fact that the high immunization coverage of one population (Turkish citizens) can shield the other (Syrian refugees) from getting exposed to the infection, as an effect of herd immunity. 
The network structure is defined by dividing Turkey into patches corresponding to administrative areas that represent the nodes. The layers of the network are encoded by Turkish $(T)$ and Syrian refugee $(R)$ populations. On this network, social dynamics, mobility and epidemic spreading happen simultaneously. 
\begin{figure}[t!]
\centering\includegraphics[width=\textwidth]{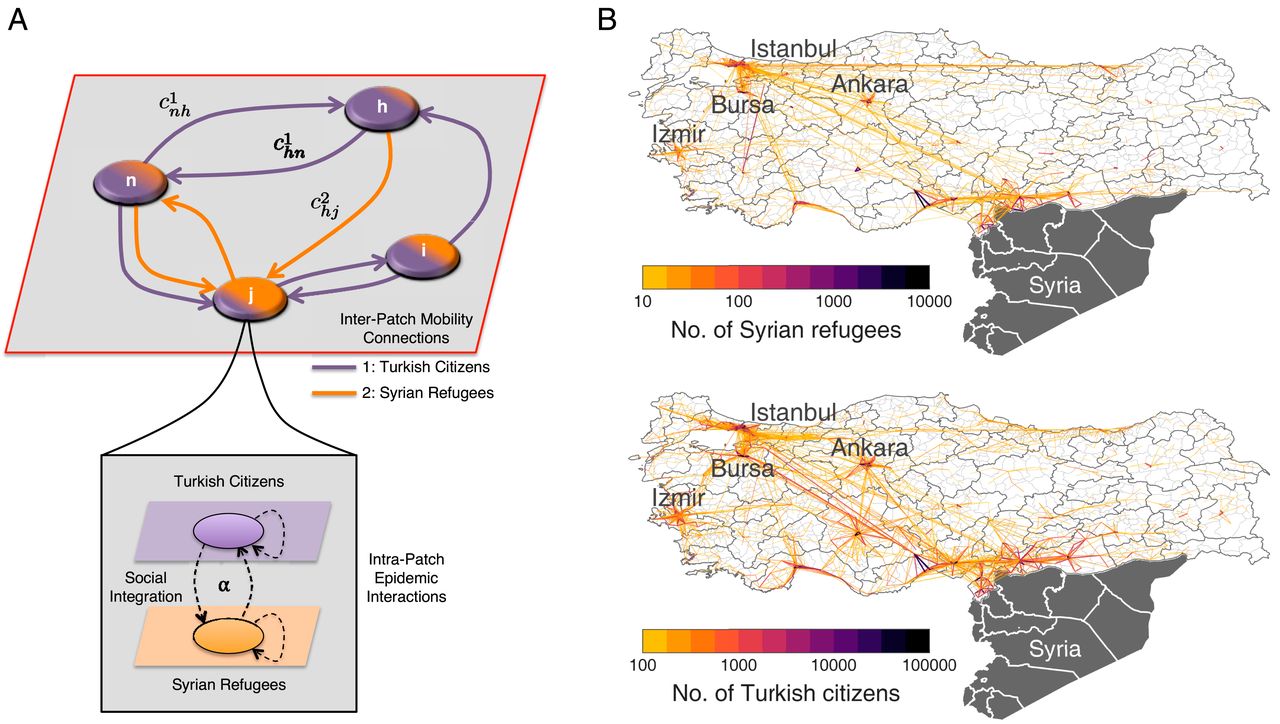}
\caption{\label{fig:multiTurkey}(\textbf{A}) Model scheme. Prefectures of Turkey are the nodes of a network of geographic patches. Turkish and Syrian populations are encoded by two colors and move between patches following the inferred mobility pathways. The two populations encode two layers, where social dynamics and epidemic spreading happen simultaneously. (\textbf{B}) Mobility of Syrian refugees (upper) and Turkish citizens (lower) inferred from mobile phone data. Figure from \citep{bosetti2020heterogeneity}.}
\end{figure}

The authors denotes with $c_{k i}^{(p)}(p \in T, R)$ the elements of a matrix $\mathbf{C}^{(p)}$ encoding the number of people belonging to population $p \in \{T,R\}$ traveling from patch $k$ to patch $i$, and with $\alpha$ the fraction of Syrian contacts with Turkish citizens. The force of infection for each population in the $i$-th patch depends on the contribution of all patches in the country:
\begin{eqnarray*}
 \begin{array}{l}
\lambda_{i}^{(T)}\left(\alpha, \mathbf{C}^{(T)}, \mathbf{C}^{(R)}\right)=\beta_{T} \sum\limits_{k=1}^{L}\left[\underbrace{c_{k i}^{(T)} \frac{I_{k}^{(T)}}{N_{k}^{(T)}}}_{\text {Endogenous }}+\underbrace{\alpha c_{k i}^{(R)} \frac{I_{k}^{(R)}}{N_{k}^{(R)}}}_{\text {Exogenous }}\right] \\
\lambda_{i}^{(R)}\left(\alpha, \mathbf{C}^{(T)}, \mathbf{C}^{(R)}\right)=\beta_{R} \sum\limits_{k=1}^{L}\left[\underbrace{\alpha c_{k i}^{(T)} \frac{I_{k}^{(T)}}{N_{k}^{(T)}}}_{\text {Exogenous }}+\underbrace{c_{k i}^{(R)} \frac{I_{k}^{(R)}}{N_{k}^{(R)}}}_{\text {Endogenous }}\right]
\end{array}
\end{eqnarray*}
where $\beta_{p}=\beta / P_{i}^{(p)}(\alpha, c)$ is the transmission rate for population $p$ and $P_{i}^{(p)}(\alpha, c)$ is an appropriate normalization factor. From this equation we can easily recognise the structure of Eq.~\eqref{eq:sexi_dyn}, where the contributions to the force of infection come from an endogenous term, accounting for the infectivity due to individuals from the same population, and an exogenous term, accounting for the infectivity due to the other population. The parameter $\alpha$ is the level of social mixing and can be changed according to different social integration scenarios. This is the parameter which plays the role of the coupling between the layers. An illustrative representation of the model is shown in 
Fig.~\ref{fig:multiTurkey}.

The epidemic transmission dynamics is eventually regulated by the following SIR dynamical model
\begin{eqnarray*}
\left\{\begin{aligned} \dot{S}_{i}^{1} &=-\lambda_{i}^{1}(\alpha, c) S_{i}^{1} \\ \dot{S}_{i}^{2} &=-\lambda_{i}^{2}(\alpha, c) S_{i}^{2} \\ \dot{I}_{i}^{1} &=\lambda_{i}^{1}(\alpha, c) S_{i}^{1}-\gamma I_{i}^{1} \\ \dot{I}_{i}^{2} &=\lambda_{i}^{2}(\alpha, c) S_{i}^{2}-\gamma I_{i}^{2} \\ \dot{R}_{i}^{1} &=\gamma I_{i}^{1} \\ \dot{R}_{i}^{2} &=\gamma I_{i}^{2} \end{aligned}\right.,
\end{eqnarray*}
whose analysis suggests that the incidence of the measles can be reduced up to 90\% in the case of very high levels of integration~\citep{bosetti2020heterogeneity}.

Despite the different contexts, the multilayer dynamics can have positive or negative feedbacks, leading to interdependence between the corresponding critical points of the dynamics. As a consequence, two different regimes exist: i) one in which the critical properties of one process depend on those of the other, and (ii) in which the critical properties are independent of the other. The two regimes are separated by a metacritical point, where a crossover occurs (see a recent review from~\citep{de2016physics} and Fig.~\ref{fig:critical_curve}). This regime-shift is analogous to the one occurring in percolation processes, which is presented in the next section.

\subsection{Percolation \label{sect:perc}}

In this section we switch our attention to percolation~\citep{stauffer2018introduction}, where the goal is to analyse how different properties of the network change as we remove some of its nodes or links. The properties we are interested in are typically topological, such as the size of the largest connected component or the distribution of small connected components, and they are used as a first proxy to assess the functionality of a system exposed to failures or attacks. Failures are modelled as random removals, whereas attacks assume some a priori knowledge of the network, where the elements are ranked given some criterion ---frequently topological (degree, betweenness, etc.)~\citep{albert2000error, cohen2001breakdown}, albeit not strictly necessary~\citep{artime2020effectiveness, artime2021percolation}--- and removed accordingly. We aim at presenting the basic mathematical framework to address the problem of multilayer percolation and showcase some applications. Good reviews to expand on what we present here can be found in~\citep{bianconi2018multilayer,lee2015towards}.

The first step towards a description of multilayer percolation processes is to consider the generalization of the generating function methodology, frequent in single-layer networks~\citep{newman2001random}. In~\citep{leicht2009percolation}, random percolation is studied in general multilayer networks described by the set of degree distributions $\lbrace p_{k_1 k_2 \ldots k_L}^{\alpha} \rbrace$, where $\alpha$ is the layer label and $k_{\beta}$ denotes the number of links toward nodes in layer $\beta$. Expressions for the size of the giant component of each layer $S^{\alpha}$ can be written as a function of the generating functions. This framework allows us to consider percolation on correlated multilayer networks, leading to non-trivial results. For example, in~\citep{lee2012correlated} correlated duplexes of Erd{\H{o}}s--R{\'e}nyi networks are studied. When degrees across layers are maximally anti-correlated -- i.e., hubs in one layer are low degree nodes in the other layer -- the giant component appears at a link density considerably higher than the value for which it appears in uncorrelated duplexes. The giant component exists, though, for any nonzero link density if interlayer degrees are maximally positively correlated. In other words, the more correlated the degrees are, the fewer edges are needed to make a macroscopic structure emerge. When it comes to attacks on the most connected nodes, the latter statement is true up to a certain intra-layer mean degree (assuming it is the same for both layers), after which the behavior is reversed and maximally negative correlated networks become more robust~\citep{min2014network}.

This framework disregards any functional characteristics of the layers and, from a phenomenological point of view, continuous phase transitions are always found. This is no longer true if the nature of the layers is taken into account, for example via the interdependence of the nodes or antagonistic interactions. These more realistic scenarios define new conditions for a node to remain in the network, and need alternative, more function-oriented metrics to describe the state of the system. A widely accepted choice is the mutually connected component~\citep{buldyrev2010catastrophic, son2012percolation} already defined in Sec.~\ref{sec:conncomp}, which is the set of nodes that are connected within each and every layer simultaneously. The most striking result is that when the giant mutual component (GMC) is computed in interdependent networks, the percolation phase transition changes its nature, becoming a discontinuous one~\citep{son2012percolation}. This has serious implications for the robustness of the system, since the disintegration occurs abruptly, i.e., it is hard to anticipate. Mathematically, the idea is as follows. 

\begin{figure}
    \centering
    \includegraphics[width=0.9\textwidth]{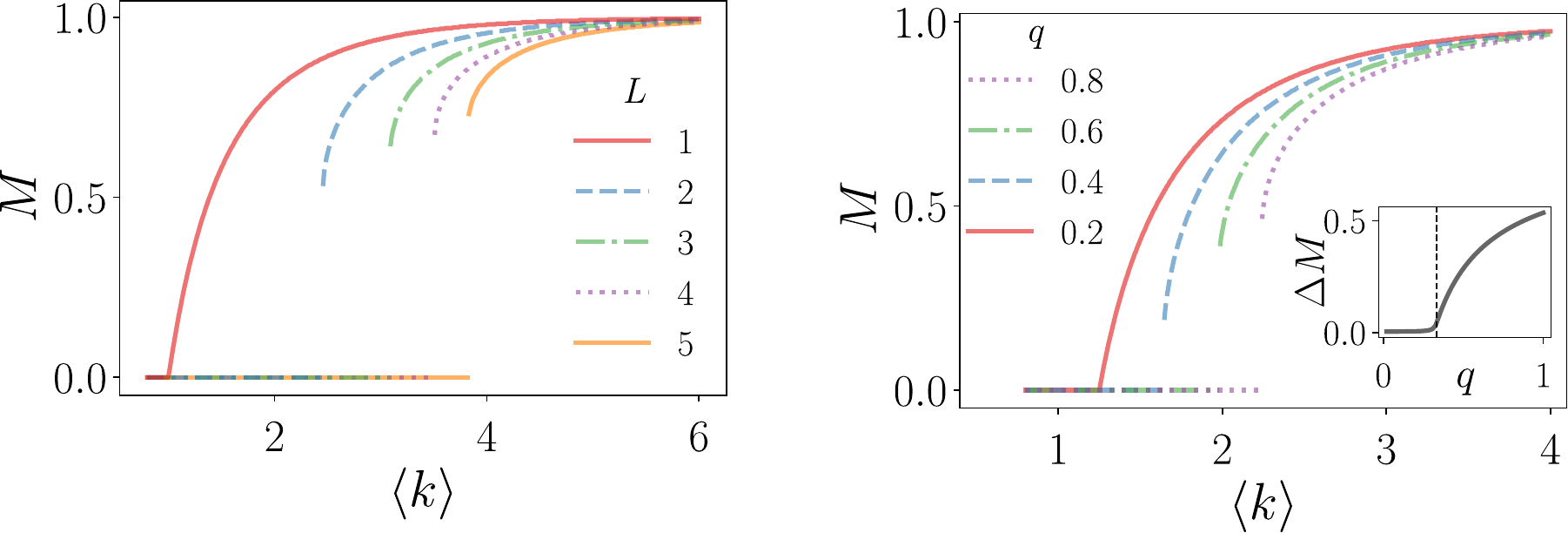}
    \caption{Percolation phase transition in a multiplex formed by intralayer Erd{\H{o}}s--R{\'e}nyi networks, with the same mean degree $\langle k \rangle$. On the left, we display $M$, which is the solution of $M = \phi (1 -  e^{ - \langle k \rangle M})^L$, as a function of the mean degree, for different number of layers, with the occupation probability fixed to $\phi = 1$. On the right, we study the emergence of the giant mutual component for partial multiplexes, where $q$ is the fraction of interdependent nodes. In this case $M$ is the solution of $M = \phi (1 -  e^{ - \langle k \rangle M}) (1 -  qe^{ - \langle k \rangle M})$. In the inset we show the height of the jump of the order parameter at the transition point, along with the theoretical value of the tricritical point (vertical dashed line).}
    \label{fig:chap_dyn_GMC}
\end{figure}

Let us assume an edge-colored multigraph, let $p_{k_1 \ldots k_L}$ be the probability that a node has degree $k_{\alpha}$ to other nodes within the layer $\alpha$, and let $q_{k_1 \ldots k_L}$ be the corresponding excess degree distribution. We indicate by $w_{\alpha}$ the probability that a node does not belong to the GMC via a link in layer $\alpha$. Hence, $w_{\alpha}^{k_{\alpha}}$ gives the probability that the node does not belong to the GMC via any of its neighbors in layer $\alpha$. The condition to belong to the GMC is that the node has to be connected to it in all the layers, i.e., the size of the GMC is proportional to $\prod_{\alpha=1}^L ( 1 - w_{\alpha}^{k_{\alpha}} )$. We just need to rescale by the occupation probability $\phi$ and average over the degree distribution, yielding
\begin{align}
    M & = \phi \sum_{k_1=0}^{\infty} \ldots \sum_{k_L=0}^{\infty} p_{k_1 \ldots k_L} \prod_{\alpha=1}^L ( 1 - w_{\alpha}^{k_{\alpha}} ).
    \label{eq:GMC}
\end{align}
To compute $w_{\alpha}$, we first note that $1 - w_{\alpha}$ is the probability that a node at the end of a link in layer $\alpha$ belongs to the GMC. For this to happen any of its remaining $k - 1$ neighbors in layer $\alpha$ are in the GMC as well. Moreover, due to the condition of mutual connectivity, in every  other layer, the node needs to belong to the GMC via any of its neighbors. Rescaling by $\phi$ because the node needs to be present in the network, and averaging, we obtain
\begin{equation}
    1 - w_{\alpha} = \phi \sum_{k_1=0}^{\infty} \ldots \sum_{k_L=0}^{\infty} q_{k_1 \ldots k_L} \left(1 - w_{\alpha}^{k_{\alpha} - 1} \right) \prod_{\substack{\beta = 1 \\ \beta \neq \alpha}}^{L} ( 1 - w_{\beta}^{k_{\beta}} ).
    \label{eq:GMC2}
\end{equation}
By inserting the solutions $\{ w_{\alpha} \}$ of this system of equations into Eq.~(\ref{eq:GMC}) we readily obtain the size of the giant mutual component. See Fig.~\ref{fig:chap_dyn_GMC} to appreciate the emergence of the abrupt transition of the GMC for multiplex systems with Erd{\H{o}}s--R{\'e}nyi networks in the layers. The discontinuity is also present when considering multiplexes of scale-free networks but, at odds with single-layer scale-free percolation, the occupation probability $\phi_c$ is finite, making them more vulnerable to random failures than the single-layer network~\citep{baxter2012avalanche}. Targeted interventions in the network can be easily modeled as well by including a degree-dependent occupation probability $\phi_{k_1 \ldots k_L}$ inside the sums~\citep{min2014network, zhao2016robustness}.

The dependence of the mutual component $M$ shown in Fig.~\ref{fig:chap_dyn_GMC} for Erd{\H{o}}s--R{\'e}nyi multiplexes is shared for other intra-layer topologies as well. That is, when the number of layers increases, the value of the mean degree at which the discontinuity appears becomes larger, thus broadening the parameter region where the network is not functional. Moreover, the height of the discontinuity jump becomes larger too, thus making the transition harder to anticipate when going from the supercritical to the subcritical region. In light of these results, the more layers, the more fragile the system is, which might seem paradoxical from an evolutionary point of view: why would a system organize itself in a layered structure if that reduces its robustness? In~\citep{radicchi2017redundant}, Radicchi and Bianconi provided a potential answer to this conundrum by proposing a model of multilayer percolation that relaxes the condition for functionality of the nodes. They argue that a node can be functional, i.e., it is not removed, as far as it is functioning in at least a pair of layers. This new condition for functionality allows them to conclude that the addition of extra layers boosts the robustness of the system.

The abrupt nature of the transition induced by Eq.~\eqref{eq:GMC2} holds as far as $L > 1$, see Fig.~\ref{fig:chap_dyn_GMC}. When $L = 1$ the mutual component coincides with the standard giant component, so the transition is continuous. We cannot interpolate continuously from $L = 1$ to $L = 2$ to understand how the nature of the phase transition changes. However, there are other variables that we can tune to go from an effective single-layer system to a  multiplex.

The first of these variables is the multiplexity parameter. It might occur that in real multiplex networks only a fraction $q$ of all nodes shares the functional dependency across layers, hence a natural question is what is the nature of the transition as a function of $q$. Does it suffice to have a non-zero fraction of dependent nodes to observe the discontinuity, or, on the contrary, is there a finite threshold, only above which we observe an abrupt transition? To answer this question, let us focus on duplexes.  If a node in one of the layers has a dependency link, which occurs with probability $q$, then the condition to belong to the GMC is the same as discussed earlier. Instead, if the node does not have a dependency link, which occurs with probability $1-q$, then it belongs to the GMC as far as any of its neighbors of its very same layer belong to the component. Therefore, we can write
\begin{align}
    M_{\alpha} & = q \phi \sum_{k_1=0}^{\infty} \ldots \sum_{k_L=0}^{\infty} p_{k_1 \ldots k_L} \prod_{\beta=1}^L ( 1 - w_{\beta}^{k_{\beta}} ) \nonumber \\ 
    & + (1-q) \phi \sum_{k_1=0}^{\infty} \ldots \sum_{k_L=0}^{\infty} p_{k_1 \ldots k_L} w_{\alpha}^{k_{\alpha}}.
    \label{eq:GMC1_Part}
\end{align}
The set of probabilities $\{ w_{\alpha} \}$ obeys the equations
\begin{align}
    1 - w_{\alpha} & = q \phi \sum_{k_1=0}^{\infty} \ldots \sum_{k_L=0}^{\infty} q_{k_1 \ldots k_L} \left(1 - w_{\alpha}^{k_{\alpha} - 1} \right) \prod_{\substack{\beta = 1 \\ \beta \neq \alpha}}^{L} ( 1 - w_{\beta}^{k_{\beta}} ) \nonumber \\
    & + (1-q) \phi \sum_{k_1=0}^{\infty} \ldots \sum_{k_L=0}^{\infty} q_{k_1 \ldots k_L} (1 - w_{\alpha}^{k_{\alpha} - 1}).
    \label{eq:GMC2_Part}
\end{align}
Focusing again on a duplex of Erd{\H{o}}s--R{\'e}nyi networks, we see in Fig.~\ref{fig:chap_dyn_GMC} that there is a finite tricritical point $q_c$ at which the transition changes its order, confirming that, in general, a multiplex needs a certain level of interdependency between layers to experience the discontinuous transition. Since many infrastructural networks are embedded in a two-dimensional space, in a similar fashion one might ask what type of transition we encounter when coupling low-dimensional networks, that alone show continuous transitions, in a multiplex. It turns out that in this case the transition does not change its order~\citep{son2011percolation, berezin2013comment}. 

The second variable that allows us to interpolate between multiplexes and monoplexes is the link overlap across layers. For complete overlapping all layers are equal and the problem is reduced to percolation of a single-layer, showing a continuous transition. How much overlap do we need to observe the abrupt transition? Different approaches and techniques have been used to answer this question, and accordingly slightly different phenomenology has been discovered. Although the details of the phase diagram depend on the number of layers and the degree distributions, Hu and coathors have found that the transition stays abrupt as far as we do not have complete overlapping~\citep{hu2013percolation}. In~\citep{cellai2013percolation} it has been reported that a critical value of the edge overlap exists that changes the nature of the phase transition from a hybrid first-order to a continuous one. See~\citep{cellai2016message} for the generalization of the theory to an arbitrary number of layers. If the edge overlap is combined with other topological correlations, the phase diagram displays multiple and recursive hybrid phase transitions~\citep{baxter2016correlated}. In~\citep{min2015link}, the problem of link overlap has been formulated in a way that the discontinuous transitions display hysteresis. 
The role of the edge overlap, together with other topological correlations, has been also explored in the problem of multilayer optimal percolation~\citep{osat2017optimal}, that is, in the identification of the smallest set of nodes that, when removed, cause the largest damage in the network~\citep{santoro2020optimal}. This problem is known to be NP-hard in single-layer networks~\citep{kempe2003maximizing}, and although there is no equivalent proof in multilayer architectures, the intuition indicates that it is so as well. On this basis, heuristic approaches to identify the critical subset of nodes predominate. In~\citep{santoro2020optimal}, the efficiency of $20$ of these strategies is evaluated. The authors find that when no structural correlations exist, a family of Pareto-efficient strategies based on both structural descriptors and multiobjective optimization is the best at dismantling the network. However, when evaluated in real multilayer networks that present non-trivial correlations, the variability in performance changes from one dismantling strategy to another, suggesting that a fair assessment of multilayer robustness requires a comparison between strategies.

Beyond node interdependency as a condition for functioning, other interesting mechanisms have been considered in the literature. One of them is antagonistic relations, where a node in one layer is functional --- that is, it has not been removed --- only if its replicas are not~\citep{zhao2013percolation}. Examples of this kind of competitive or non-cooperative relations might be relevant, for example, in biological networks. Interestingly, it has been found that when percolation is considered in this scenario, the abrupt transition shown in what follows persists, but displays as well, at variance with the transition of the mutual component introduced earlier, the typical hysteresis and bi-stability behaviors of equilibrium thermal first-order phase transitions~\citep{goldenfeld2018lectures}. Other mechanisms have followed to include these behaviors too, e.g., in~\citep{min2014multiple,min2015link}.

Note that all of these results derived with the machinery of generating functions come with some underlying assumptions that are very rarely met when studying real networks: (i) the network is tree-like, (ii) it has infinite size and (iii) the quantities of interest are computed as an average over the ensemble of networks with given degree distribution, although, in reality, we only have access to one supra-adjacency matrix, among all the possible ones that a graph model could generate. As a consequence, the analytical predictions might deviate from the actual process of percolation, obtained for example via simulations. Analytical approaches have been proposed to overcome points (ii) and (iii) in~\citep{radicchi2015percolation, bianconi2016percolation}, where a percolation theory of multiplex and interdependent networks is introduced that takes as input the adjacency matrix instead of the degree distribution. Furthermore, in a real network we may need to assess the robustness under perturbation scenarios related to node metadata. For example, in~\citep{baggio2016multiplex} it is addressed, among other things, how a multiplex network capturing the flow of subsistence-related goods and services among households in several Alaskan Natives communities responds to perturbations involving targeted removals of specific resources by category, e.g., terrestrial, marine or riverine, as a representation of natural disasters. An analytical framework to take into account  non-topological features in the robustness of a network has been recently developed for single layers~\citep{artime2021percolation}, but, at present, there is not a generalization for multilayer networks.

\subsection{Cascade failures}\label{sec:cascades}

The percolation model has the limitation that failures and attacks are treated statically. A more complete description of multilayer robustness and resilience would include their time evolution, in order to better understand under which conditions small perturbations can trigger global network-wide effects, the so-called \emph{cascades}. In this section we review some of the most emblematic models for multilayer cascade failures.

\begin{figure}[!th]
    \centering
    \includegraphics[width=0.75\textwidth]{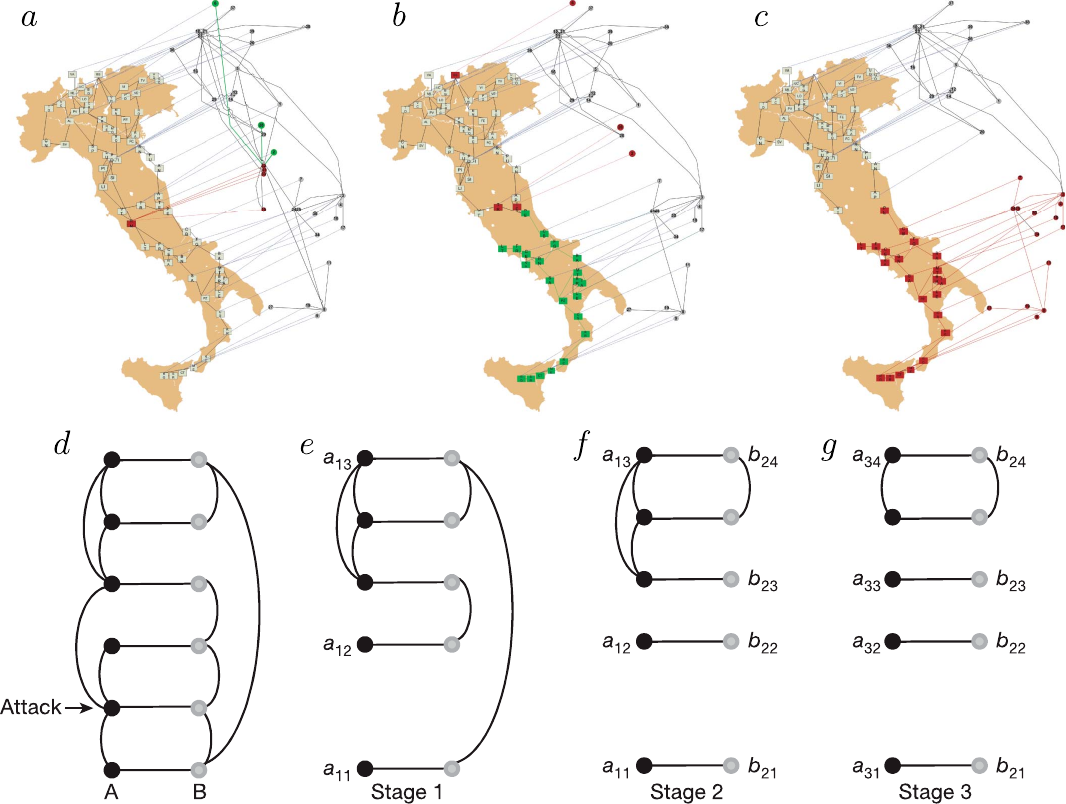}
    \caption{Top: evolution of a cascade of failures in a real interdependent system composed by a power network (on the map) and an Internet network (shifted above) and involved in the Italian blackout of September 2003. In \textbf{a}, failure of a power station makes the Internet nodes to stop functioning, indicated in red. In green is shown the nodes that disconnect from the largest connected component in the next cascade step. These nodes will fail immediately after ($b$), introducing a feedback of failures back in the power network. Those nodes that have been isolated (green nodes in $b$) are the ones that will fail in the next event $(c)$, inducing further interdependent failures in the Internet network. Bottom: we sketch this process for further clarification. In \textbf{d}, an initial attack to one node in network \textbf{$\mathsf{A}$} occurs. In \textbf{e}, the attacked node is removed, along with its links, and the corresponding interdependent node in network \textbf{$\mathsf{B}$}, with its links, is also removed. In \textbf{f}, the actual cascade starts. All the \textbf{$\mathsf{B}$}-links between \textbf{$\mathsf{B}$}-nodes connected interdependently to different \textbf{$\mathsf{A}$}-clusters are removed. In \textbf{d}, the same rule applies but for links in network \textbf{$\mathsf{A}$}: all \textbf{$\mathsf{A}$}-links between \textbf{$\mathsf{A}$}-nodes are deleted if the interdependent nodes do not belong to the same \textbf{$\mathsf{B}$}-cluster. Steps \textbf{f} and \textbf{g} are repeated, propagating the failures back and forth, until the cascade cannot further evolve. The remaining connected components at the end of the propagation of failures coincide with the mutually connected components introduced in Sec~\ref{sect:perc}. Figures from~\citep{buldyrev2010catastrophic}.}
    \label{fig:chap_dyn_casc1}
\end{figure}

The seminal work of Buldyrev and colleagues was one of the first to deal with the propagation of failures across interdependent multiplexes~\citep{buldyrev2010catastrophic}. The motivation for such a study was the 2003 Italy blackout, for which it was hypothesized that the power grid and the Internet network, the latter acting as a supervisory control and data acquisition system, were interdependent, and that failures in the power stations hampered the Internet communication and further propagated the malfunction across the system, see top panels of Fig.~\ref{fig:chap_dyn_casc1}).

However, the interest in the amplification of small perturbations throughout the network is much more general, finding eventual applications in areas such as biochemistry, where metabolic pathways interact in complex ways with key tissues, or finance, where different banks might share the same asset in their balance sheet~\citep{huang2013cascading}, among many other examples. In~\citep{buldyrev2010catastrophic} was introduced the concept of a mutually connected component from a dynamical perspective, at odds with its static definition presented in the previous section, see bottom panels of Fig.~\ref{fig:chap_dyn_casc1}. It was shown that these cascades evolving in multiplex networks yield a discontinuous phase transition in the size of the giant mutual component~\citep{buldyrev2010catastrophic}, meaning that the failure of one single node from an apparently healthy, functional infrastructure can generate the emergence of a global cascade that collapses the entire system. It was later shown that these cascading failures can be mapped to static percolation in multiplexes~\citep{son2012percolation}. In fact, many of the results presented in Sec.~\ref{sect:perc} are recovered at the final state of the cascade propagation of this type of model. 

Despite the fact that the dynamical propagation of cascades needs a more convoluted mathematical treatment than percolation, these techniques have been very flexible at adapting to variations of the original work of Buldyrev, so that the propagation of cascades is modeled in more realistic scenarios. We briefly review some of them in the following, and refer the interested reader to the more complete reviews~\citep{kenett2014network,shekhtman2016recent,valdez2020cascading}. 

After~\citep{buldyrev2010catastrophic}, the problem of dependency-based cascades has been generalized to $L$ layers in~\citep{gao2012networks}, allowing analytical, closed solutions in certain interdependent setups. The problem of failure propagation in multiplexes with partial interdependency, where not all nodes have dependency connections, has been addressed in~\citep{Parshani2010reducing}. If the fraction of interdependent nodes is decreased enough, the transition is no longer abrupt but becomes second-order. In order to narrow the range of parameters for which the abrupt transition occurs, different strategies to choose the autonomous nodes, those without interdependencies, have been proposed~\citep{schneider2013towards}. It turns out that selecting those with the highest degree or highest betweenness, significantly reduces the likelihood of an abrupt collapse~\citep{schneider2013towards}. Intra-layer correlations can be included in the analysis as well. In~\citep{huang2013robustness}, interdependent multiplexes with a tunable average number of single links and an average number of triangles per node are analyzed, finding that, for fixed average degree, the higher clustered networks are less robust than those with lower clustering. Intra-layer correlations might be coupled as well to inter-layer ones: in~\citep{reis2014avoiding} a flexible model including both types of correlations is studied, and the maximization of robustness is addressed as a function of the tunable strength of the correlations and the failure propagation dynamics, showing that the theoretical results match well with the experimental results of coupled functional brain modules.

Beyond cascades driven by topological failures, there are other mechanisms via which small perturbations can drive a multilayer network to collapse. In many problems related to social sciences, such as the adoption of fads, the diffusion of norms and innovations, and the changes in the collective attention in a population, models of behavioral contagion are used to model the decisions of the agents. Agents need to decide between two alternative options/actions and the influence of their neighbors is crucial in the final choice. A simple way to encompass this influence is by setting an activation threshold. Initially all nodes are inactive but a controlled small fraction. An inactive agent changes her state and becomes active only when the fraction of her active neighbors is larger than a threshold. This process might require several activation steps before reaching a frozen state. In~\citep{watts2002simple} was given the range of parameters for which such global cascades, measured as the number of active users when the process stops, emerge in monoplexes, turning out that they are only possible when the network is neither too sparse nor too dense. The generalization of this threshold model in multiplexes was addressed in~\citep{brummitt2012multiplexity}, where a propagation rule is proposed such that an agent activates as far as the fraction of active neighbors in at least one layer exceeds the threshold. Following this rule, multiplexity widens the ranges of parameters in which global cascades can occur, therefore increasing the network's vulnerability. Interestingly, isolated layers in which, owing to their topological properties, global cascades would not be observed, when multiplex-coupled, cooperate in a way that facilitates agent activation and therefore lead to cascades. 

\begin{figure}[!th]
    \centering
    \includegraphics[width=0.65\textwidth]{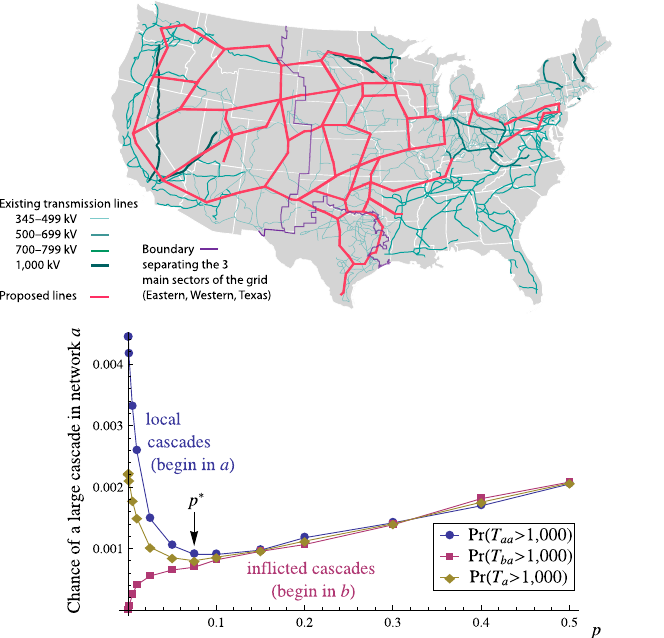}
    \caption{The power grid is a paradigmatic example of a network susceptible to overload cascades of failures due to the energy transported along the links. We see on the top an aggregated representation of the multilayer power grid of the USA, where each layer (link color) corresponds to different voltage ranges. In red is depicted the planned interconnections to transport wind power. On the bottom, the Bak-Tang-Wiesenfeld sandpile model is simulated in a duplex of random regular graphs. For a cascade starting in one of the networks (network $a$), is shown the probability of observing a final cascade size in $a$, $T_{aa}$, and in network $b$, $T_{ab}$, as a function of the probability of interconnections $p$ between nodes in $a$ and in $b$. Both $T_{aa}$ and $T_a = 1/2(T_{aa}+T_{ab})$ display a minimum, indicating that an isolated network can reduce the damage caused by a cascade by interconnecting to another network, but only up to a certain level of interconnectedness $p^*$. Figures readapted from~\citep{brummitt2012suppressing}.}
    \label{fig:chap_dyn_casc2}
\end{figure}

\begin{figure}[!th]
\centering\includegraphics[width=0.99\textwidth]{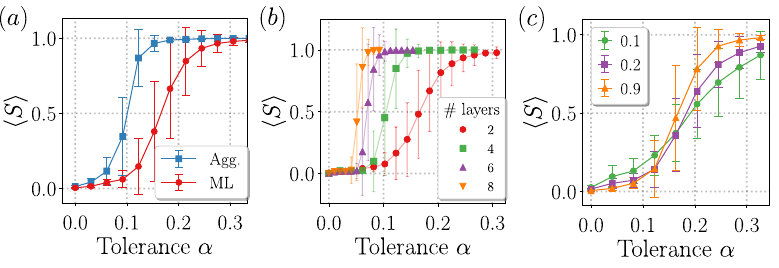}
\caption{\label{fig:chap_dyn_casc3}Robustness of multilayer networks exposed to overload failures due to nonlocal load redistribution. We show the behavior of $\langle S \rangle$, the size of the largest connected component of the network at the end of the cascade, as a function of the tolerance parameter $\alpha$ (see text). The faster $\langle S \rangle$ grows, the more robust is the system because there the range of tolerance values for which it disintegrates is smaller. Three mechanisms to increase the robustness are presented. In $(a)$, the comparison between the multilayer network and its aggregated counterpart. In $(b)$, dependence on the number of layers. In $(c)$, $\langle S \rangle$ for different values of the multiplexity parameter (fraction of nodes participating simultaneously in a duplex), with the value indicated in the legend. Figure readapted from~\citep{artime2020abrupt}.}
\end{figure}

Another mechanism that can induce cascades is load redistribution due to overloads. Arguably, the most famous stylized model accounting for this type of dynamics is the Bak-Tang-Wiesenfeld sandpile model~\citep{bak1988self}, a paradigmatic example of self-organized criticality used in a variety of contexts~\citep{jensen1998self}. Each node has an internal, discrete variable, the load. At each unit of time the load of a node selected uniformly at random increases in one unit. When the load reaches a threshold, assumed to be degree, the node redistributes its load to its neighbors. The neighbors might, in turn, exceed their capacity and reallocate their load to their own neighbors, hence propagating the cascade. Once all nodes have their own load below the threshold, the random addition of load again is restarted. To avoid inundation of the system, a frequently used strategy is to dissipate load at a certain rate when it is reallocated. In~\citep{brummitt2012suppressing}, the BTW model is used in the context of multilayer power grids, shedding light on the benefits and dangers of the level of interconnectivity between the layers. Surprisingly, it is found that there is an optimal value of the interconnectivity for which the chance to observe large cascades is minimum, see Fig.~\ref{fig:chap_dyn_casc2}. This is because the addition of interlayer links between highly isolated power grids helps mitigate the cascades by creating a reservoir that absorbs excess load, a sort of alternative dissipative mechanism. This behavior, though, is valid up to a certain point, from which the further addition of interlayer links is no longer beneficial since it creates more loaded systems due to larger thresholds and, therefore, chances of larger cascades, as well as it creates a positive feedback due to the reentrant new paths of load into a layer. Regarding the critical properties of the BTW model, it has been found that multiplexity does not alter the mean-field scaling behavior observed in monoplexes~\citep{lee2012sandpiles}.

To close this section, we discuss the relevant, but still quite unexplored, case of nonlocal cascade propagation. One might argue that nonlocality can be realized in spatially extended systems by allowing interlayer dependencies between different locations of the space~\citep{li2012cascading}. Yet, under this description the failures propagate via first neighbors, in the interdependent sense. In fact, all types of cascades discussed earlier evolve locally via first-neighbors, something that need not to be true in real systems, as happened in the 1996 disturbance of the Western Systems Coordinating Council (WSCC) system~\citep{NERC2002system}, in the 2003 blackout in the northeastern USA \citep{nerc2004technical} or in the air-traffic disruption due to the eruption of the Icelandic volcano Eyjafjallaj{\"o}kull \citep{eurocontrol2010ashcloud}. A plausible description of this phenomenon is based on the load-capacity model of Motter and Lai~\citep{motter2002cascade}, where a load, defined as the number of shortest paths crossing a node, and a constant capacity, a factor $1 + \alpha$ larger than the initial load, are assigned to every node. When the load of a node exceeds its capacity, the node fails. An initial perturbation in the form of node removal is applied to the network, that globally modifies the loads and allows subsequent failures to occur not necessarily close to the prior failure. If during this process nodes that overload, they also fail, propagating the cascade. This model has been investigated recently in~\citep{artime2020abrupt}, with the finding that the size of the largest connected component at the end of the cascade suffers an abrupt jump when the tolerance parameter $\alpha$ is increased. Moreover, since the load redistribution depends significantly on the topology of the network, the average path length is identified as a metric that correlates well with robustness. Based on this, the article proposes different strategies to increase the robustness of the network, such as adding new layers or reducing the level of multiplexity (see Fig.~\ref{fig:chap_dyn_casc3}). Unlike the other cascades phenomena introduced in this section, which can be analytically treated with generating functions or multi-type branching processes, nonlocal cascades do not have a solid mathematical machinery to be described, and this certainly represents a challenge for the future.

%%%%%%%%%%%%%%%%%%%%%%%%%%%%%
%%%%%%%%%%%%%%%%%%%%%%%%%%%%%
%%%%%%%%%%%%%%%%%%%%%%%%%%%%%
\section{Frontiers}
\subsection{Kinematic geometry} 

The geometric approach~\citep{boguna2020network} to network analysis has garnered growing interest in the past two decades. 
Here we focus on the geometry of network-driven processes on multilayer networks, a family of kinematic geometries generalizing the diffusion geometry introduced in \citep{de2017diffusion}.

The structure of a wide variety of real-world complex systems is modular and hierarchical~\citep{guimera2005functional} and the effect of these large scale properties on the dynamics of such systems has been studied, during the past decade. It has been shown that complex systems with such a mesoscale organization~\citep{fortunato2010community} are characterized by topological scales~\citep{arenas2006synchronization}, exhibiting the emergence of functional clusters which might be different from topological ones. 

In~\citep{de2017diffusion} the authors investigate the multiscale functional geometry of monoplexes to characterize functional clusters. This approach defines the diffusion distance between any pair of units in a networked system, based on random walk dynamics, shown to correspond, among others, to the phase deviation of coupled oscillators close to metastable synchronization state and consensus dynamics. The diffusion distance for single-layer networks is a key concept to define a kinematic geometry based on network-driven processes and it can be calculated as the $L^2-$norm of the difference between rows of the propagator $e^{-t\tilde{\mathbf{L}}}$:
\begin{equation}\label{eq:diffu-dist-monoplex}
 D_t(i, j) = \norm{\mathbf{p}(t | i) - \mathbf{p}(t | j)}_2.
\end{equation}

Exploiting the fact that diffusion geometry is based on random walk dynamics, it is possible to extend it to the realm of multilayer networks. 
At variance with walks on edge-colored networks presented in Sec.~\ref{sec:diffproc}, where one can obtain the transition rules for the multigraph as a weighted average of the transition probabilities in each distinct layer, see Eq.~\eqref{eq:tmatrix-ec}, on a multilayer network the walk type determines the probability of a random walker jumping across and within layers (see Eq.~\ref{eq:multilayer_random_walk} and Tab.~\ref{tab:rws}).
Consequently, in the edge-colored case diffusion distances between nodes can be obtained directly from Eq.~\eqref{eq:diffu-dist-monoplex}, while for multilayers we have to introduce a diffusion distance between state nodes.

Regardless of the random walk type, let us indicate the probability of finding a random walker at a given node and layer at time $t$ by $p_{j \beta}(t)$. 
Similarly to Eq.~(\ref{eq:diffu-dist-monoplex}) we define the diffusion distance between state nodes $(i, \alpha)$ and $(j, \beta)$ as
\begin{equation}\label{eq:diffu-dist-ml}
 D_{t}^2((i, \alpha), (j, \beta))= \sum\limits_{k, \gamma}(p_{k\gamma}(t|(i, \alpha)) - p_{k\gamma}(t|(j, \beta)))^2,
\end{equation}
where probabilities are conditional to using the corresponding state nodes as the origins of random walkers at time $t=0$. 
One can summarize this supra-distance matrix $\mathbf{D}_{t}=\left(D_t((i, \alpha), (j, \beta))\right)$ into an $N \times N$ matrix by encoding the diffusion distance among the physical nodes, across all the layers. 
Intuitively, resembling the parallel sum of resistances in electrical circuits leading to the equivalent resistance, the \emph{equivalent diffusion distance} can be written as
\begin{equation}\label{eq:equiv-ddm}
  D_t^{\text{eq}}(i, j) = \left(\sum_{\alpha=1}^L \frac{1}{D_t((i, \alpha), (j, \alpha))} \right)^{-1}.
\end{equation}

Usually, the functional distance (supra-)matrices are rescaled in $[0, 1]$, normalizing each by its maximum value to allow for comparisons. If, additionally, one is interested in the most persistent patterns, the diffusion distance is averaged over time and called the average diffusion distance. The corresponding supra-distance matrix is indicated by $\bar{\mathbf{D}}_t$.

\begin{figure*}[!htb]
\centering
\includegraphics[width=.95\textwidth]{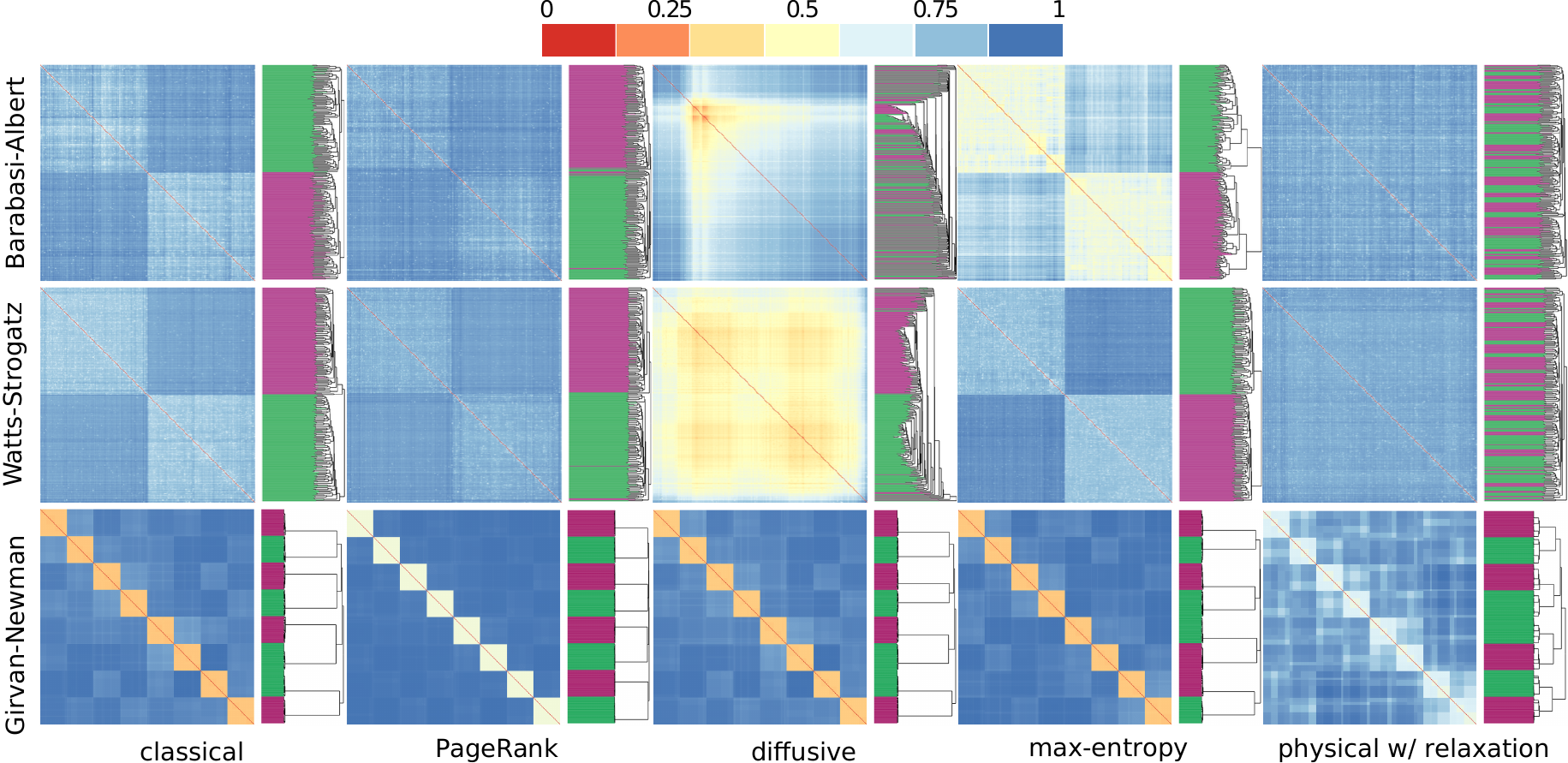}
\caption{Average diffusion distance supra-matrices $\bar{\mathbf{D}}_t$ for different combinations of multilayer topologies and random walk dynamics (see the text for details). Figure reprinted with permission from~\citep{bertagnolli2020diffusion}. Copyright (2021) by the American Physical Society.}
\label{fig:mux_geom_overlap}
\end{figure*}

The diffusion distance between nodes is highly dependent on the type of random walk dynamics, its propagation time, the topology of layers and the layer-layer correlations. To better understand the effects of dynamics and topological variations, let us consider three classes of two-layer networks, namely:
\begin{itemize}
    \item Barabasi-Albert layers with preferential attachment of 4 links;
    \item Watts-Strogatz layers, with rewiring probability 0.2;
    \item Girvan-Newman model layers, where the inter-community connectivity probability is 1, whereas cross-group connections exist with probability 0.05.
\end{itemize}
Additionally, we consider the five random walks of Tab.~\ref{tab:rws}. For the three cases, the link overlap across layers -- i.e., the fraction of links present in both layers among the same pairs of nodes~\citep{de2015structural}--- is fixed to 10\%. As shown in Fig.~\ref{fig:mux_geom_overlap}, the diffusive walk on a scale-free topology leads to a high level of mixed pathways across layers, while in small-world systems, nodes tend to shape more distinct functional clusters.

The functional geometry framework has also been used to analyze empirical systems in~\citep{bertagnolli2020diffusion}, where the authors highlighted dissimilarities in the diffusion spaces of the public transportation of London and of the social network of Noordin terrorists, by looking at the projections of the spaces and at the results of a Mantel's test~\citep{mantel1967detection} applied to supra-distance matrices.

Finally, it is worth mentioning that one can use other walks instead of random ones: this is the case of walks that result from powers of the adjacency matrix and provide the basis for \emph{communicability}, a measure recently used to introduce a geometric framework with applications to single and multiplex networks~\citep{Estrada2019}.

\subsection{Statistical physics of multilayer systems} 

\subsubsection{Classical ensembles}

To study the properties of an observed multilayer network, comparison with null models is often essential. The maximum entropy approach is one of the standard ways to obtain the required null models, in terms of ensembles of networks exhibiting one (or more) specific property of the observed network, while being maximally random with respect to the other properties.
For instance, in the case of single layer networks, one of the most famous null models is the configuration model (CM)~\citep{molloy1995critical} which is an ensemble of networks with the same degree sequence as the observed one. The unbiased probability distribution of the members of this ensemble is indicated as $P(G)$ and must maximize the Shannon information entropy
\begin{equation}\label{eq:shannon-entropy}
    S = -\sum\limits_{G \in \mathcal{G}} P(G) \log{P(G)}.
\end{equation}

It is worth mentioning that such a probability distribution does not correspond to a physical process related to the second law of thermodynamics. However, the discussed entropy maximization approach is mimicking the mathematical machinery of statistical physics and, in the case of a fixed degree sequence, it leads to a microcanonical ensemble where all members of the CM have equal probability.

\begin{figure}[!t]
\centering
\begin{minipage}[c]{0.3\textwidth}
\includegraphics[width=\textwidth]{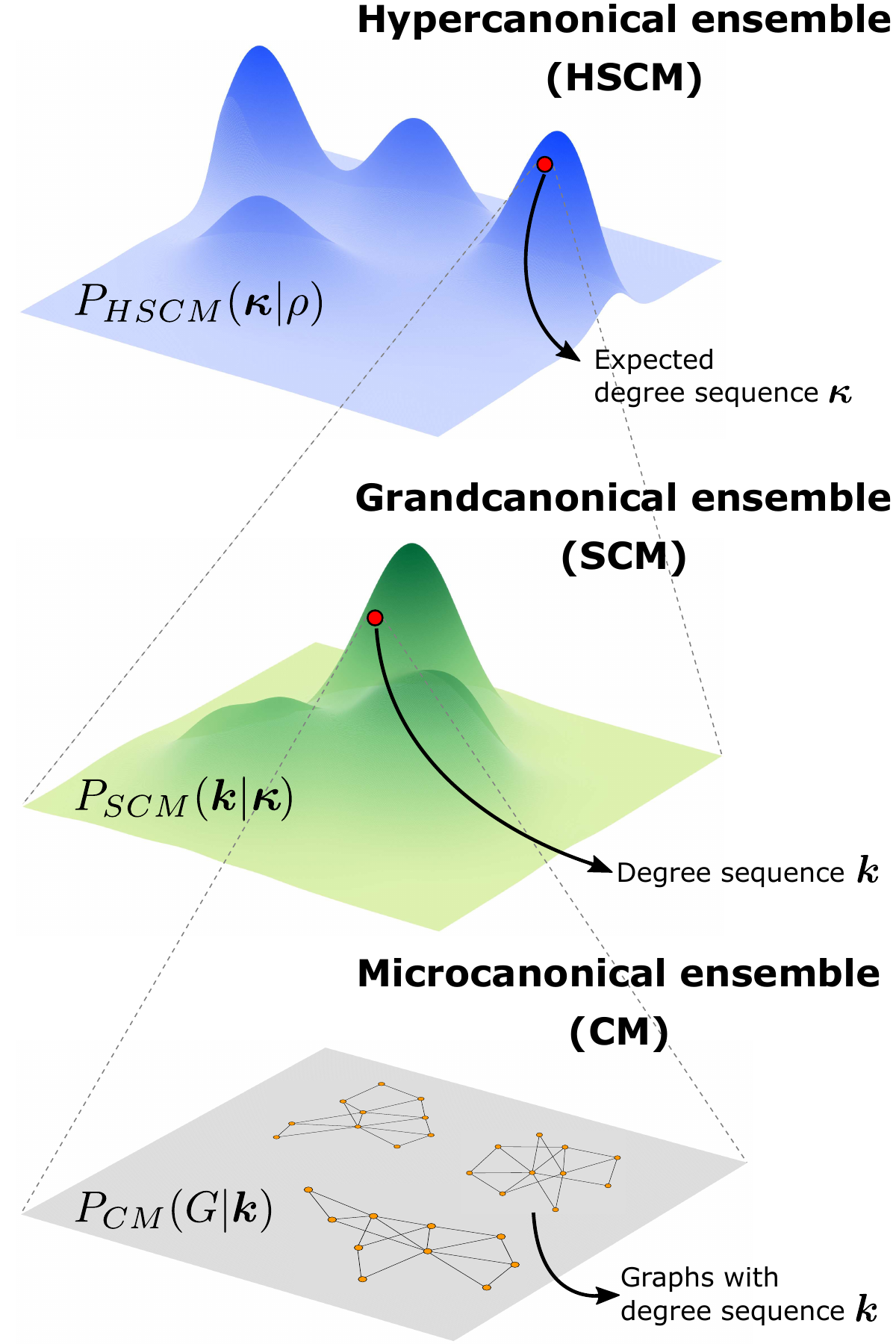}
\end{minipage}
\qquad
\begin{minipage}[c]{0.5\textwidth}
\caption{\label{fig:classical_ensembles} Schematic of sampling the network, represented by $G$, from hypercanonical, grandcanonical, and microcanonical  ensembles discussed in the text. Figure from~\citep{Krioukov2020MaxEntropy}.}
\end{minipage}
\end{figure}

Depending on the type of study, the mentioned constraint of a fixed degree sequence can be relaxed to obtain other ensembles~\citep{cimini2019}. For instance, by fixing the average degree instead of the full degree sequence, we achieve the grand-canonical ensemble of random graphs. Furthermore, if our knowledge is limited to the degree distribution from which the degree sequence is sampled, one obtains the hypersoft configuration model ~\citep{Krioukov2020MaxEntropy} which is a hypercanonical ensemble of random networks all drawn from the fixed distribution (see Fig.~\ref{fig:classical_ensembles}). 

Similarly, maximum entropy approaches have been used to study ensembles of non-interconnected multiplex networks. Assume an edge-colored multigraph $M$ with $L$ layers, each forming a network $G^{(\ell)}$ ($\ell=1,2,...N$). Remarkably, when there is no correlation between the layers, the probability of observing $M$ can be obtained as the product of the probabilities of the layers:
\begin{eqnarray}
P(M) = \prod\limits_{\ell = 1}^{L} P(G^{(\ell)})
\end{eqnarray}
and the corresponding Shannon entropy becomes the summation of Shannon entropies of layers~\citep{bianconi2013statistical}:
\begin{eqnarray}
S = \sum\limits_{\ell = 1}^{L} S^{(\ell)}.
\end{eqnarray}
By imposing the soft constraints -- e.g., fixing the average degree instead of degree sequence -- and using the Lagrangian multipliers method, one can obtain the canonical multiplex ensemble with probability $P_{C}(M)$ maximizing the Shannon entropy:

\begin{eqnarray}
P_{C}(M) = \frac{e^{-\sum\limits_{\mu} \lambda_{\mu}F_{\mu}(M)}}{Z_{C}} = \prod\limits_{\ell = 1}^{L} P_{C}(G^{(\ell)}),
\end{eqnarray}
where $\lambda_{\mu}$ corresponds to the Lagrangian multipliers and $F_{\mu}(M)$ determine the constraints on the network (e.g., average degree). Note that, in presence of layer-layer correlations, we have $P(M) \neq \prod\limits_{\ell = 1}^{L} P(G^{(\ell)})$ leading to other probability distributions extensively studied in~\citep{bianconi2013statistical}.

\subsubsection{Quantum-like ensembles}\label{sec:qensembl}

Complex systems include a wide variety of physical attributes and dynamics. The interactions between their units vary, from the electrochemical signals traveling among neurons in the human brain to the transport of goods between different areas of a urban ecosystem and the spreading of an infectious pathogen between individuals of a society. Regardless of the nature of these interactions, they can be described as information exchange. In fact, complex systems resemble one another in the way they handle information. Therefore, to understand how these systems operate, from a physical point of view, investigating their information dynamics is crucial.

It is important to note that the information flow between the units is regulated by the underlying structure, depending on the local neighborhood in certain classes of networks and on long-range communication between distant components in other topologies. At the same time, it is essential to consider the coupling between the structure and the dynamical processes governing the flow of information, such as diffusion, random walks, synchronization and consensus and, since it is often difficult to understand a system in terms of microscopic features, a framework providing a statistical description of the system might be relevant for applications.

Recently, a statistical field theory of complex information dynamics has been introduced to unify a range of dynamical processes governing the evolution of information on top of static or time-varying structures~\citep{de2020SFT}. This framework describes the interactions among the units in terms of \emph{information streams}, a set of operators that determines the direction of information flow and provides a statistical ensemble to construct the statistical physics of complex information dynamics. In fact, the formalism allows for defining density matrices--- i.e., statistical average of streams--- from which a variety of descriptors can be derived. Of course, density matrix formalism comes from quantum mechanics, where vectors were found insufficient to represent the mixed states and encode the pairwise coherence between quantum states. Similarly, properties of networked systems cannot be fully described by vectors or distribution functions, without information loss. For instance, a system's structure is often encoded in two dimensional data structures, being the adjacency matrix, except for trivial symmetries like chains or paths. Therefore, the counterpart of the quantum density matrix has been introduced for complex networks, where off-diagonal elements provide a proxy for interactions between the nodes, instead of the coherence between the states. So far, this framework has only been used to analyze classical networks. In the following, we provide a direct application of the theoretical framework to multilayer systems.

\begin{figure}[!t]
\centering\includegraphics[width=\textwidth]{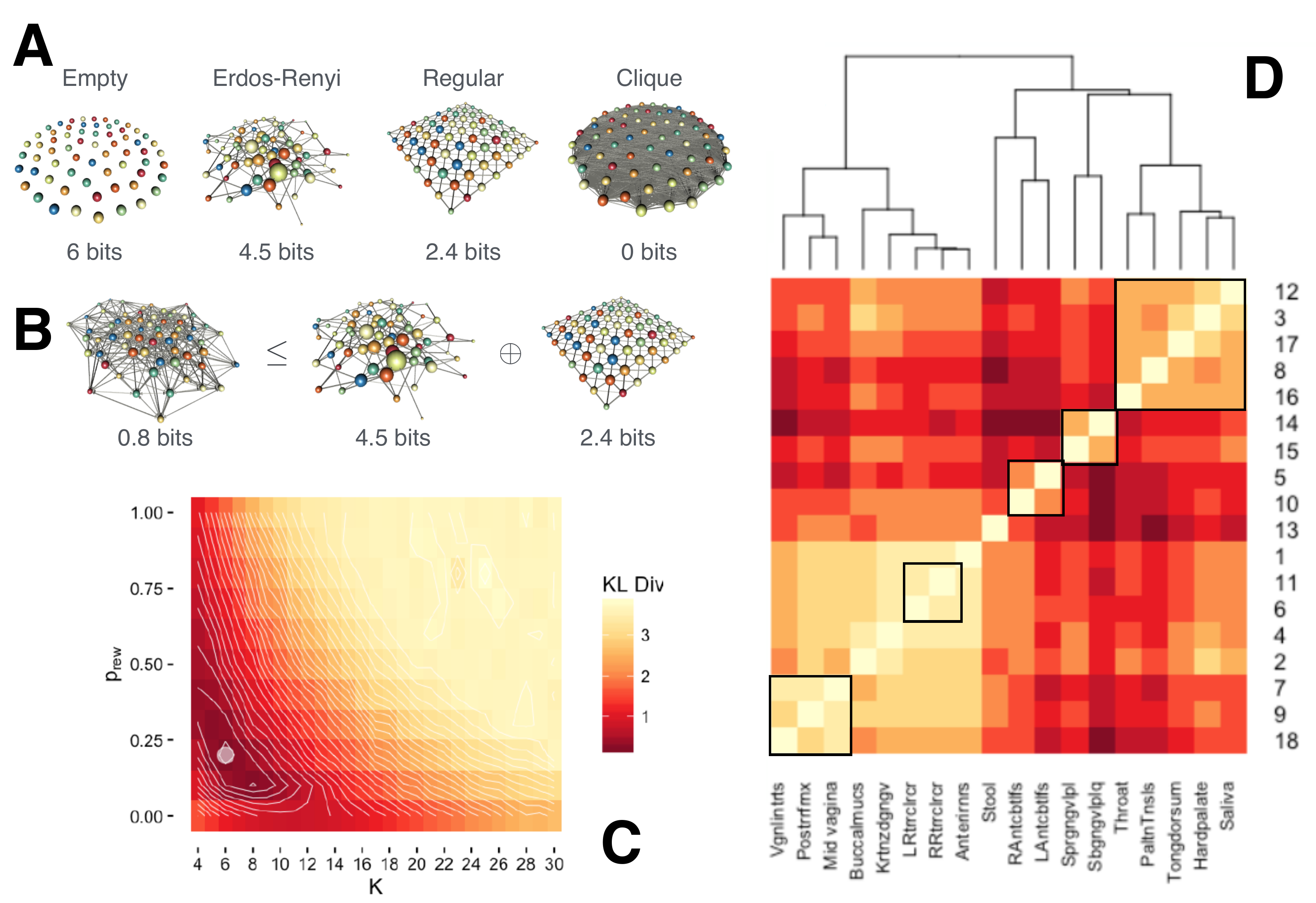}
\caption{\label{fig:chap_spectral_entropy} Schematic illustration of the spectral entropy for different classes of networks has been presented. Remarkably, the entropic distances provided by the framework can be used to compare network similarity and characterize the network topology with high accuracy. Figure from~\citep{de2016spectral}.}
\end{figure}

\paragraph{Redundancy and reducibility of multilayer systems}

As discussed previously, the constituents of complex systems must exchange information efficiently in order to function properly. However, a deep understanding of the structure's role in enhancing or hindering the transport properties -- such as navigability~\citep{boguna2009navigability} -- continues to elude us, especially in the case of multilayer systems~\citep{gao2012networks,de2013mathematical,gomez2013diffusion,radicchi2013abrupt,kivela2014multilayer,boccaletti2014structure,de2016physics} where an efficient flow might be hindered by the lack of synchronization between different layers~\citep{gallotti2014anatomy} or the redundancy of pathways across the layers. There are a number of methods to reduce multiplex networks by structurally merging their most similar layers, based on information-theoretic frameworks~\citep{de2015structural}. Despite their success, these methods are mostly based on heuristics and have been proved inaccurate under specific circumstances~\citep{de2020EnhanceTransport}. 

\begin{figure}[!t]
\centering\includegraphics[width=\textwidth]{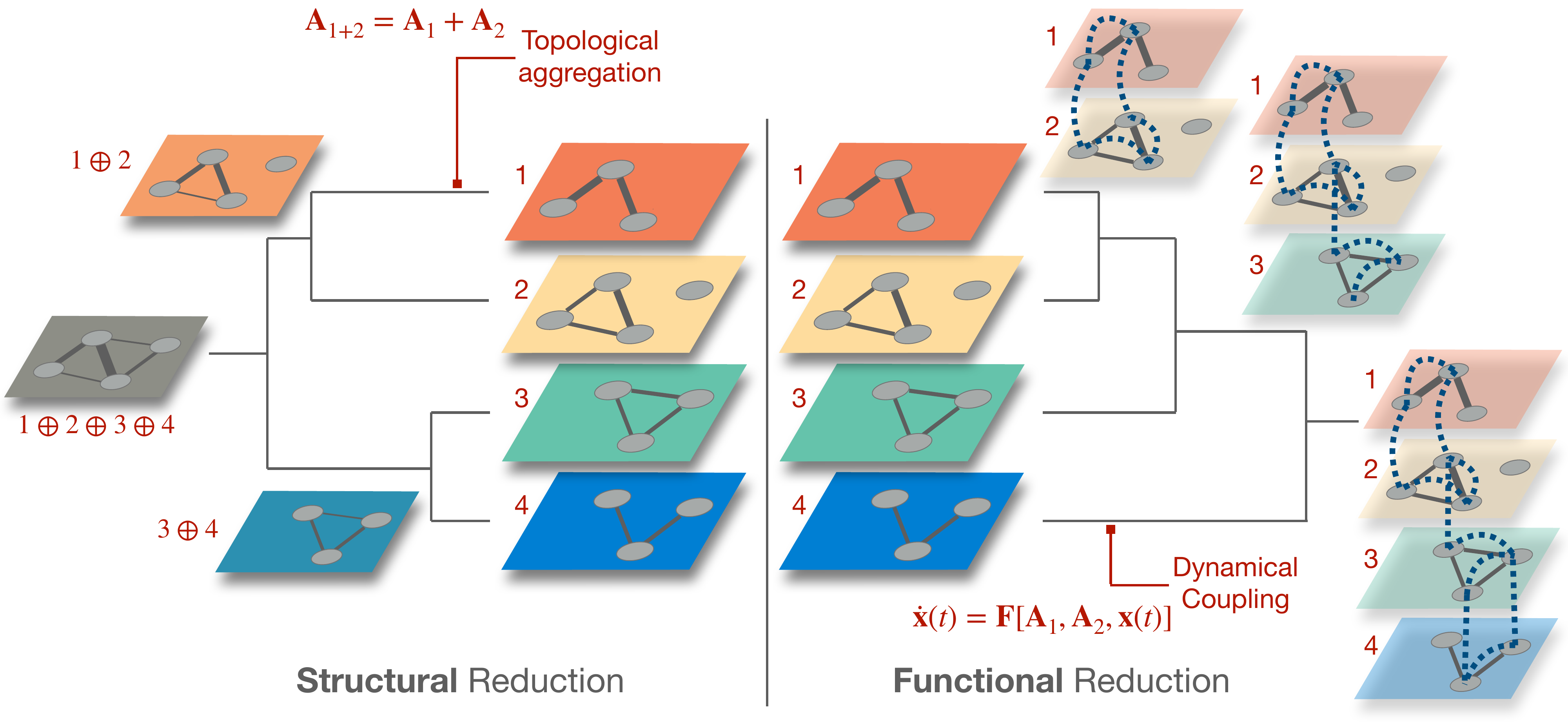}
\caption{\label{fig:chap_gui_func_reducibility} Schematic illustration comparing structural against functional reduction of a multilayer network consisting of $L=4$ layers. The procedure is similar for both approaches, but the relevant difference is that while structural reducibility alters the topology of the system, function reducibility allows us to functionally couple layers without altering their structure. Figure from~\citep{de2020EnhanceTransport}.}
\end{figure}

Enhancing the flow distribution in multilayer systems is challenging, since adding links to the layers (e.g., highways, tubes, flights, synapses, etc.)  comes with a cost. Interestingly, for multilayer networks with interlinks, changing the weights of interlinks can enhance the diffusion on top of the networks \citep{gomez2013diffusion}. When acting on the structure is not an option, it has been shown that one can still enhance the transport properties (see Fig.~\ref{fig:chap_gui_func_reducibility}) using the statistical physics of complex information dynamics by coupling layers dynamically, in a way that a dynamical process can not distinguish the functionally coupled layers -- e.g., airlines proposing shared flights to their customers -- while evolving. The functional reduction includes coupling the layers with high similarity that are responsible for the redundant diffusion pathways in the system, in order to identify the (sub)set with maximally diverse layers~\citep{de2020EnhanceTransport}.

Diffusive processes, such as random walks, have been used extensively to model the information transport within complex structures~\citep{masuda2017random}. Here, we consider random walk dynamics governed by the normalized Laplacian $\hat{\tilde{\mathbf{L}}}=\langle \hat{\tilde{\mathbf{L}}}^{(\ell)} \rangle$ given by $\hat{\tilde{\mathbf{L}}}=\hat{\mathbf{I}}- \langle \hat{\mathbf{T}}^{(\ell)} \rangle$ playing the role of the quasi-Hamiltonian (see Eq.~(\ref{eq:tmatrix-ec})). The density matrix can be obtained for random walks on multiplex networks as $\hat{\boldsymbol{\rho}}(t) = \frac{e^{-t \hat{\tilde{\mathbf{L}}}}}{Z(t)}$.

Interestingly, it has been shown that the partition function, which is encoding the amount of the trapped field, is proportional to the average return probability of random walk dynamics: $Z(t)=N\mathcal{R}(t)$, where $\mathcal{R}(t)=N^{-1}\sum\limits_{i=1}^{N}e^{-t \lambda_{i}}$ is the average return probability and $\lambda_{i}$ is the $i$--th eigenvalue of $\hat{\tilde{\mathbf{L}}}$. Intuitively, the average return probability is high when the structural symmetries and abundance of redundant diffusion pathways slows down the information propagation between the units. Thus, it is expected that breaking the structural regularities by adding long-range interactions~\citep{watts1998collective} or increasing the diversity of diffusion pathways across layers can lead to faster information flow and and lower $Z(t)$.

Multiplexity of interactions among units generates non-trivial dynamical correlations between layers that have no counterpart when layers are considered in isolation. This important difference can be characterized in terms of average entropy distance between the multiplex and its layers. This measure, named intertwining, is defined by
\begin{eqnarray}
\label{eq:avgdiv}
\mathcal{I} &=& \langle \mathcal{D}_{KL}(\hat{\boldsymbol{\rho}}||\hat{\boldsymbol{\rho}}^{(\ell)}) \rangle =\frac{1}{L} \sum\limits_{\ell=1}^{L} \mathcal{D}_{KL}(\hat{\boldsymbol{\rho}}||\hat{\boldsymbol{\rho}}^{(\ell)})
\end{eqnarray}
where $\mathcal{D}_{KL}(\hat{\boldsymbol{\rho}}||\hat{\boldsymbol{\rho}}^{(\ell)})=Tr[\hat{\boldsymbol{\rho}}(\log_2 \hat{\boldsymbol{\rho}} - \log_2 \hat{\boldsymbol{\rho}}^{(\ell)})]$ is the quantum-like Kullback-Leibler (KL) divergence between layer $\ell$ and a multiplex system as a whole.

Directly from intertwining, a fundamental inequality between the partition function of a multiplex system as whole and the partition functions of its layers can be derived. This inequality is important, as it relates the transport phenomena of multiplex system and layers, through average dynamical trapping (i.e., the partition function):
\begin{eqnarray}
\label{eq:fundrel}Z(t) \leq \prod\limits_{\ell=1}^{L} Z^{(\ell)}(t)^{1/L},
\end{eqnarray}
where equality holds if and only if all the layers are the same. Using dynamical trapping as a measure of transport, Eq.~(\ref{eq:fundrel}) shows that a multiplex network has better transport properties than the geometric average of layers, adding an advantage to multilayer structures. 

Furthermore, being reminiscent of statistical physics of particles, the equality defines the non-interacting scenario where layer-layer correlations do not alter the underlying dynamics: the entropy $S^{(\ell)}(t)$ of each layer is calculated separately and the overall entropy is given by their average $\mathcal{S}_{nint}(t)=\langle S^{(\ell)}(t)\rangle$. Conversely, any topological alteration of the non-interacting scenario introduces a dynamical correlation between layers, requiring the exploration of layers to gather more information about the system: in this case the network consists of interacting layers, where the diffusion dynamics on the whole multiplex network is considered to measure the entropy $\mathcal{S}_{int}(t)$. To obtain another form of Eq.~(\ref{eq:avgdiv}), a mean-field approximation of the Von Neumann entropy is given by
\begin{eqnarray}
\label{eq:entrmf}
\mathcal{S}^{MF}(t) = \frac{1}{\log 2}\left(t\frac{Z(t)-1}{Z(t)}+\log Z(t)\right),
\end{eqnarray}
and can be used to prove that layer-layer interactions lower the system's entropy ($\mathcal{S}_{int}(t)\leq \mathcal{S}_{nint}(t)$) and partition function $Z(t)$. Normalizing the intertwining by its upper bound, for values of time $t$ sufficiently large and in absence of isolated state nodes, Eq.~(\ref{eq:avgdiv}) reduces to the relative intertwining:
\begin{eqnarray}
\mathcal{I}^{*}(t) =1 -\frac{\mathcal{S}_{int}(t)}{\mathcal{S}_{nint}(t)},
\end{eqnarray}
which is bounded between 0 (i.e., the layers are redundant) and 1 (i.e., the layers are diverse and the system is irreducible).

We can show how intertwining is proportional to the functional diversity of layers. The Laplacian matrix of the multiplex network is given by $\hat{\tilde{\mathbf{L}}}=\langle \hat{\tilde{\mathbf{L}}}^{(\ell)}\rangle$. Therefore, the Laplacian matrix of each layer $\hat{\tilde{\mathbf{L}}}^{(\ell)}$ can be written as a perturbation of multiplex Laplacian $\hat{\tilde{\mathbf{L}}}^{(\ell)}=\hat{\tilde{\mathbf{L}}}+\Delta \hat{\tilde{\mathbf{L}}}^{(\ell)}$, reflected in its eigenvalues as $\lambda_{i}^{(\ell)}=\lambda_{i} + \Delta\lambda_{i}^{(\ell)}$ $(i = 0,1,...N)$. It is straightforward to show that $\frac{1}{N}\sum\limits_{i=1}^{N}\Delta\lambda_{i}^{(\ell)}=\overline{\Delta\lambda^{(\ell)}}=0$ and that $\overline{ \Delta\lambda^{(\ell)^2}} \geq 0 $, the latter quantifying the influence of the perturbation to each layer. The average of the variance across all layers $\overline{\langle(\Delta\lambda^{(\ell)})^{2} \rangle}$ provides a measure of the overall spectral diversity of layers which, interestingly, is demonstrated to be proportional to the relative intertwining:
\begin{eqnarray}
\label{eq:ivsvariance}
\mathcal{I}^\star(t) &\approx& \frac{t^{2}}{2} \overline{\left\langle(\Delta\lambda^{(\ell)})^{2} \right\rangle}.
\end{eqnarray}
This proves the sensitivity of intertwining to the spectral diversity of layers.
Furthermore, the partition functions of layers can be written in terms of perturbations such as $Z^{(\ell)}(t)=Z(t)+\Delta Z^{(\ell)}(t)$, leading to
\begin{eqnarray}
\label{eq:ivsz}
\mathcal{I}^{\star}(t) &\approx& \frac{\overline{\Delta Z^{(\ell)}}(t)}{ Z(t) - 1 };\ \overline{\Delta Z^{(\ell)}}(t)=\frac{1}{L}\sum\limits_{\ell=1}^{L} \Delta Z^{(\ell)}(t).
\end{eqnarray}
Equations~(\ref{eq:ivsvariance}) and (\ref{eq:ivsz}) provide a fundamental result: they show that by minimizing the partition function of the system one maximizes the relative intertwining while favoring the maximum functional diversity of layers. Additionally, it has been shown that an inverse proportionality holds between the diffusion time ($1/\lambda_{2}$) and intertwining, as further evidence for the role of intertwining in characterizing the transport properties.

These theoretical findings have been applied to a broad range of synthetic and empirical systems, providing a transparent framework for coupling the most similar layers of multiplex networks, in order to improve their transport properties including dynamical trapping, diffusion time, and navigability~\citep{de2020EnhanceTransport}.

%%%%%%%%%%%%%%%%%%%%%%%%%%%%%
%%%%%%%%%%%%%%%%%%%%%%%%%%%%%
%%%%%%%%%%%%%%%%%%%%%%%%%%%%%
\section{Conclusions}
Network Science is one of the greatest achievements of the 21st century, paving the way toward a mathematical approach for the analysis of disparate complex systems, and allowing us to find regularities in apparently disordered connectivity patterns. The past decade has seen the flourishing of analytical techniques and models exploiting, or characterizing, the inherent multidimensionality of empirical systems, from multiplexity -- i.e., the existence of distinct types of relationships among the same set of actors or units -- to interdependency -- i.e., the existence of structural or functional connections among sets of actors or units of a different nature. Such multiple dimensions are nowadays easily encoded into layers of information.

\begin{figure}[!b]
\centering
\begin{minipage}[c]{.45\textwidth}
\includegraphics[width=\textwidth]{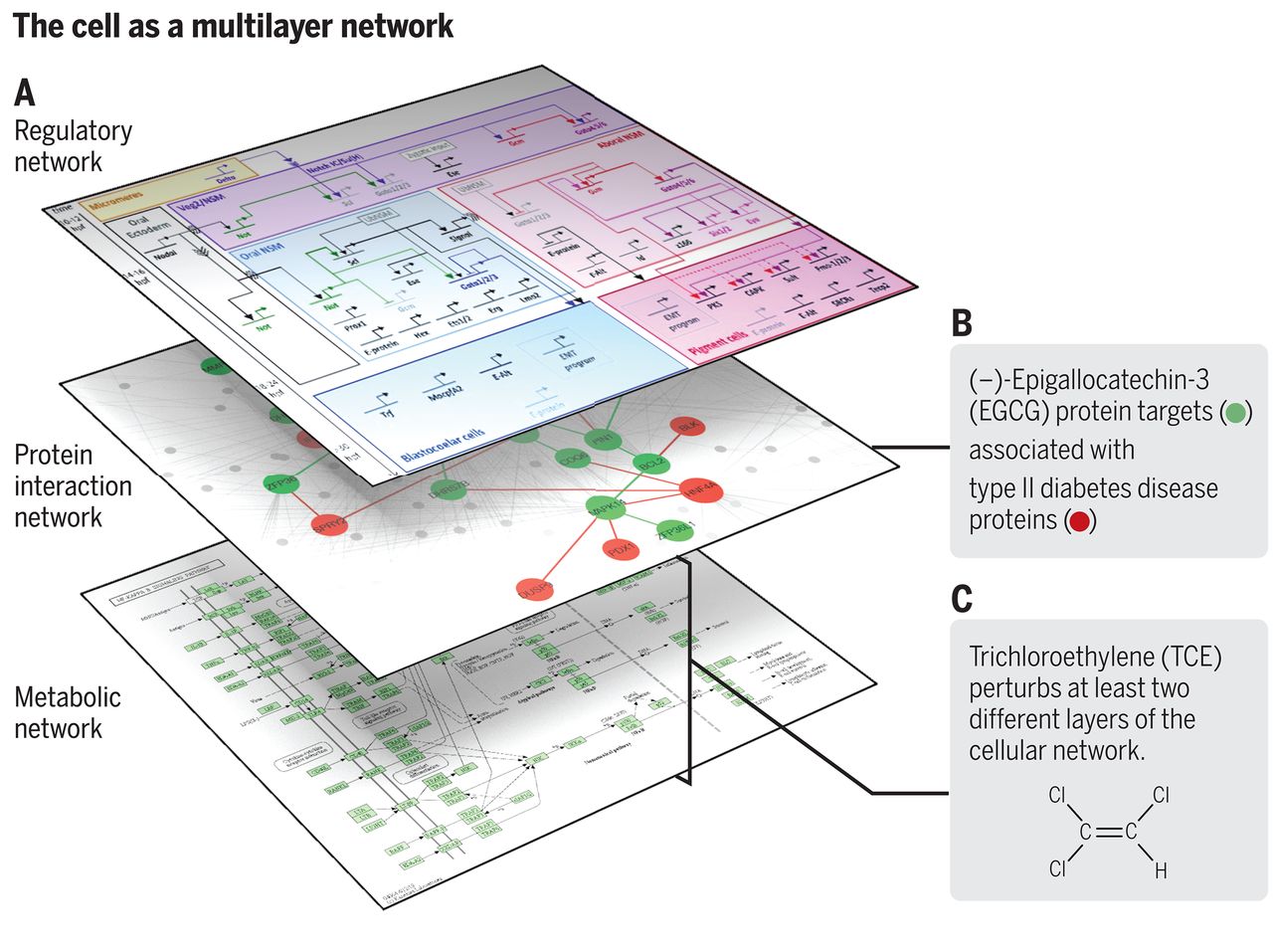}
\end{minipage}
\quad
\begin{minipage}[c]{.5\textwidth}
\caption{\label{fig:chap_concl_cell}Multilayer representation of a cell. Layers: i) regulatory interactions involving RNA and protein expression, ii) protein-protein interactions involved in signaling and responsible cell function, iii) metabolic interactions with reactions and pathways crucial for cell function. Figure from~\citep{vermeulen2020exposome}.}
\end{minipage}
\end{figure}

Most of the advances in this direction are described in some detail or referred to in this work. Starting from the mathematical representation of multilayer networks (Chap.~2) we have introduced structural descriptors for units, layers and the whole system (Chap.~3), providing an overview of the micro- and meso-scale organization of such systems. We have discussed the rich spectrum of phenomena, with no classical counterparts, related to dynamical processes on the top of the networks and their intertwining (Chap.~4), guiding the reader towards two promising research areas for the future, namely network geometry and information dynamics (Chap.~5), although many other exciting sub-field are emerging, e.g., higher-order modeling and analysis~\citep{lambiotte2019networks,battiston2020networks}.

At this point, the reader should be sufficiently familiar with multilayer network science and, to conclude, we would like to make a quick journey through the most recent applications of its paradigm, moving across different spatial scales and ranging from cells to societies.

The first stop of this journey is exactly a cell, which can vary in diameter between $10^{-6}$~m and $10^{-4}$~m (note that a DNA double helix is about $10^{-8}$~m wide). The cell can be seen as a multilayer system consisting of three interdependent layers (see Fig.~\ref{fig:chap_concl_cell}). Here, applications are mostly related to the emerging field of systems biology and network medicine, promising to use biomolecular interactions to develop a deeper knowledge of biology across scales with the ultimate goal to better understand diseases, to prevent them, and to treat them with medical drugs that reduce side effects.

The second stop requires a big jump of about three orders of magnitude, to explore one of the most famous multicellular organisms with a small-scale neural system: the Caenorhabditis elegans, a nematode of about $10^{-3}$~m. The nervous system of this small worm consists of synaptic and neuropeptide interactions which can be seen as interconnected layers. Here, the multilayer perspective provides the opportunity to better understand how the integration between hard-wired synaptic or junctional circuits and extrasynaptic signals can modulate the large-scale behavior of the worm~\citep{bentley2016multilayer}.

\begin{figure}[!ht]
\centering
\begin{minipage}[c]{.5\textwidth}
\includegraphics[width=\textwidth]{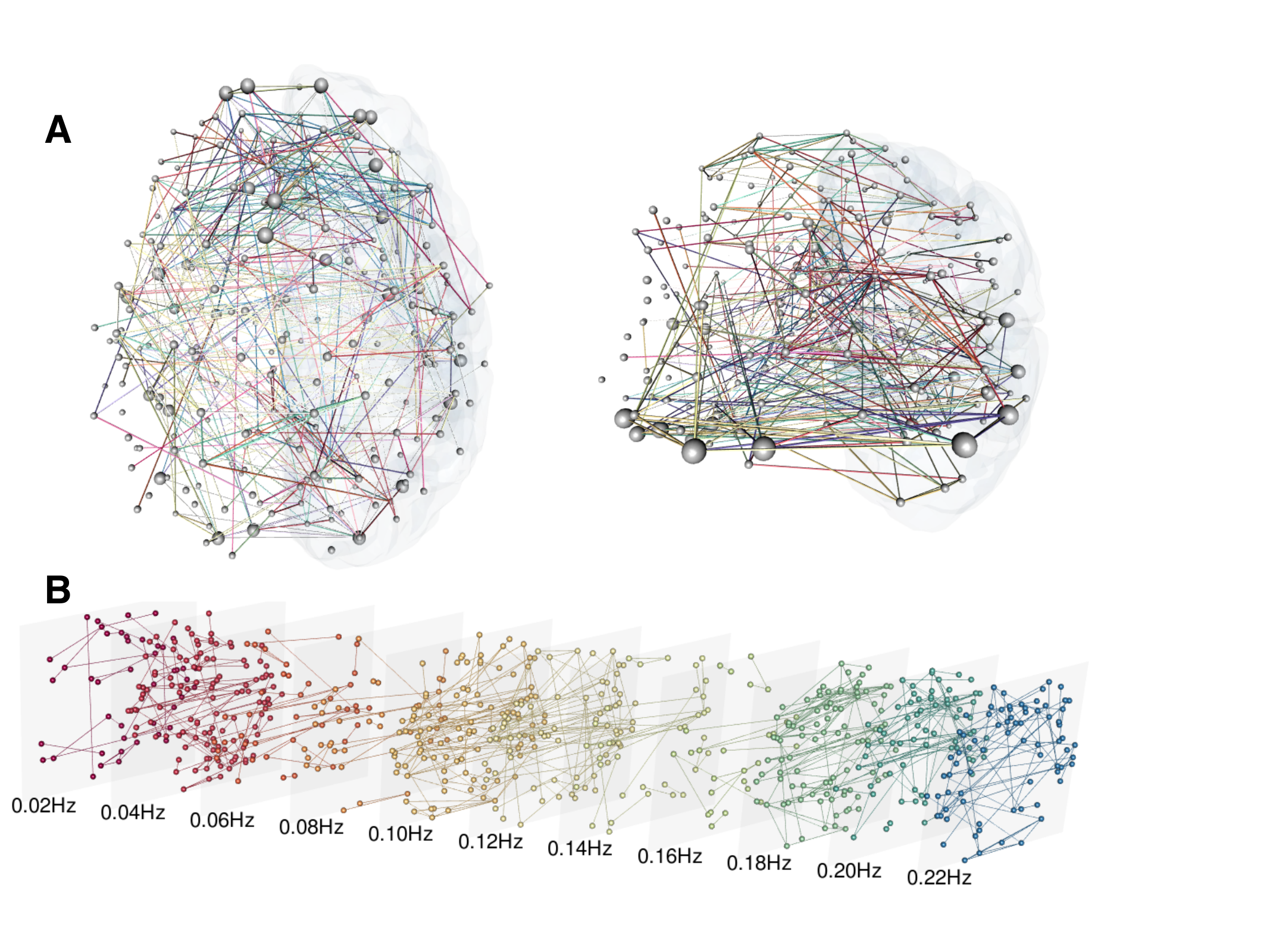}
\end{minipage}
\quad
\begin{minipage}[c]{.45\textwidth}
\caption{\label{fig:chap_concl_brain}Multilayer representation of a human brain. Both the 3D and layered visualization encode the functional brain of a schizophrenic subject (11 non-overlapping frequency-band layers between 0.01 and 0.23 Hz). Figure from~\citep{de2017multilayer}, readapted from~\citep{de2016mapping}.}
\end{minipage}
\end{figure}

At a larger scale, around $10^{-1}$~m, another emblematic neural system, namely the human brain, is currently being characterized in terms of how its structural and functional connectivity evolves over time, e.g., while performing a specific task, or across groups, unraveling the existence of modular and hierarchical structures which would remain hidden under the lens of less sophisticated models and analytical techniques~\citep{bassett2017network}. In parallel, the functional connectivity of the human brain can be stratified by frequency bands (see Fig.~\ref{fig:chap_concl_brain}) where specific correlations or causal relationships between regions of interests appear~\citep{de2017multilayer}. Multilayer analysis can be used to better characterize the relative importance of all brain regions~\citep{williamson2021multilayer} and to enhance the accuracy in discriminating between healthy and unhealthy patients starting from imaging information~\citep{de2016mapping}.

\begin{figure}[!ht]
\centering
\begin{minipage}[c]{.6\textwidth}
\includegraphics[width=\textwidth]{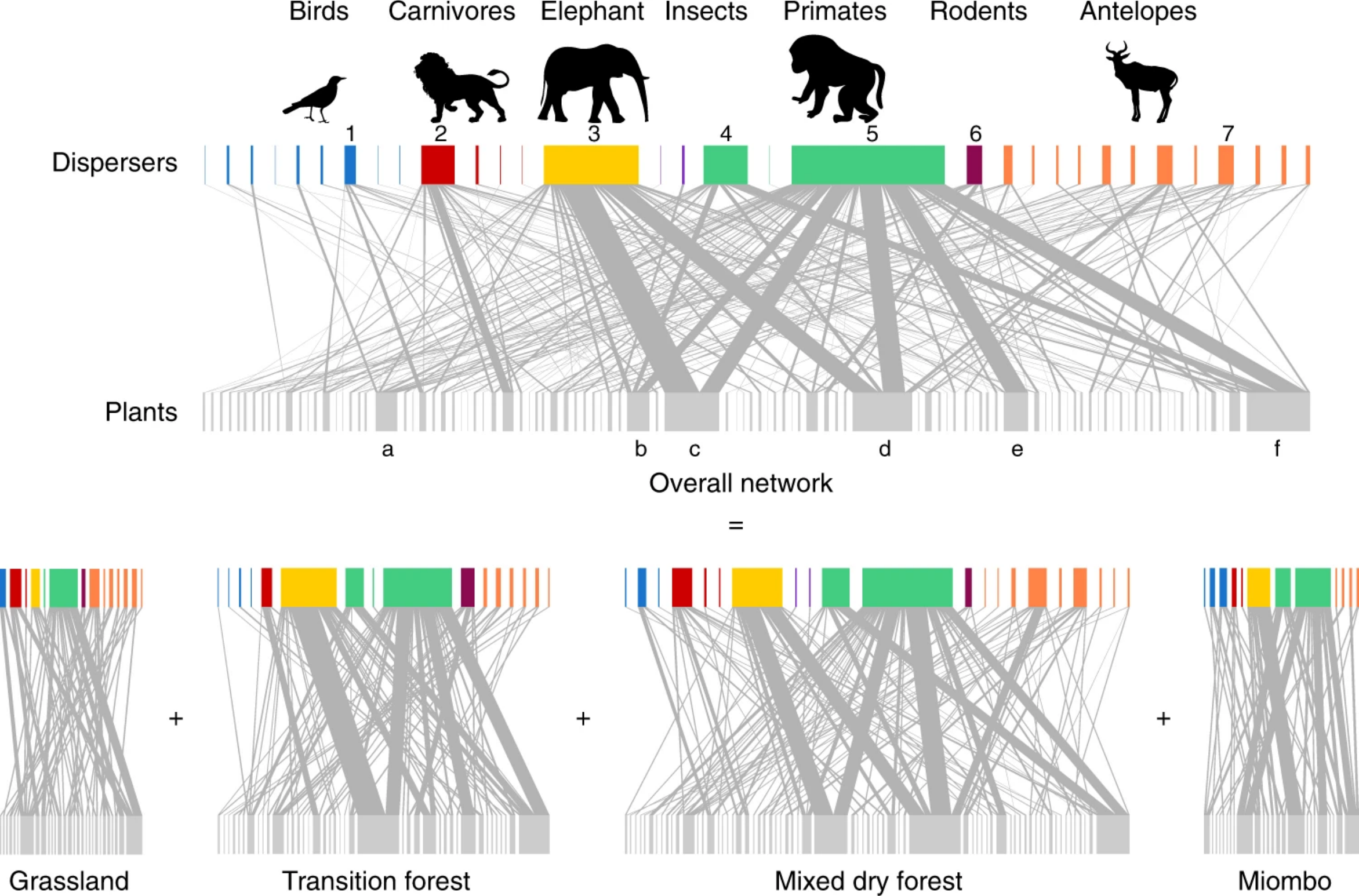}
\end{minipage}
\quad
\begin{minipage}[c]{.35\textwidth}
\caption{\label{fig:chap_concl_ecomux_timoteo} Aggregate (top) and multilayer (bottom) representations of a real seed-dispersal network. Reproduced from~\citep{timoteo2018multilayer} under Creative Commons Attribution 4.0 International License.}
\end{minipage}
\end{figure}

At larger scales, on the order of hundreds of meters ($10^2$m), cooperative systems like that of dolphins, can be characterized with respect to distinct types of interactions allowing us to get unprecedented insights about the underlying social organization and group dynamics. At similar spatial scales, another emblematic example, accounting for the socio-spatial interdependence typical of many other networks, concerns the organization of ecological systems (Fig.~\ref{fig:chap_concl_ecomux_timoteo}): for instance, it has been recently shown that the importance of dispersers for an ecosystem is better captured by multilayer measures of importance rather than by standard metrics~\citep{timoteo2018multilayer}.

At the human scale, relatively small-scale social systems ($10^4~$m) have been studied to better understand the impact of external shocks on social structure and dynamics. For instance, it has been shown that multilayer modeling can disentangle the different social ties that conform to the proximity interaction networks of extant hunter-gatherer societies, identifying social relationships with a key role in the spread and accumulation of culture~\citep{migliano2017characterization}. In another study, the analysis of mixed economies in three villages of Alaska unraveled that factors related to climate change, such as global warming, have a non-negligible effect on household structure, but the most important factors for vulnerability are indeed due to social shift rather than resource depletion~\citep{baggio2016multiplex}.

Let us jump by three more orders of magnitude and discuss systems at the planetary scale: the ones based on information exchange thanks to large-scale communication infrastructures, such as the Internet. At this scale the activity of a complex system might be very frenetic; think about an online social media platform, where millions or billions of users worldwide continuously produce content to be shared, which quickly travels and bounces from one country to another. For instance, it is interesting to study how rumors and memes spread on these systems, as recently proposed in~\citep{d2019spreading} to capture the behavior of users who post information from one social media platform to another and to provide a plausible explanation for the heavy-tailed distributions of meme popularity that is usually observed in empirical data.

\begin{figure}[!ht]
\centering
\begin{minipage}[c]{.55\textwidth}
\includegraphics[width=\textwidth]{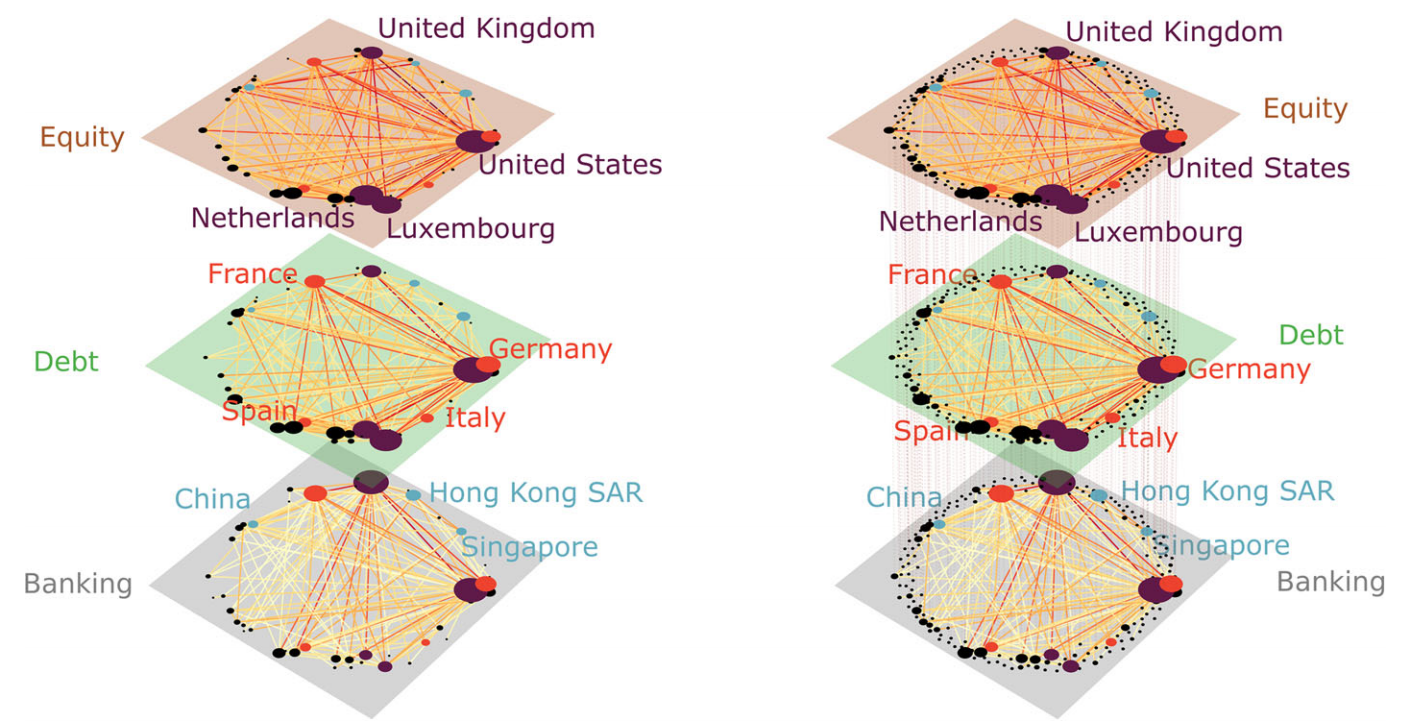}
\end{minipage}
\quad
\begin{minipage}[c]{.4\textwidth}
\caption{\label{fig:chap_concl_financial}Multilayer representation of a global financial network. Layers are asset types, nodes are countries and links are cross-country financial relationships. Figure from~\citep{del2020multiplex}.}
\end{minipage}
\end{figure}

Recent applications also include the analysis of trade networks and their nested and modular structure~\citep{torreggiani2018identifying,a2018unfolding,alves2019nested}. Additionally, as anticipated, multilayer models of financial networks have been proposed very recently, to better explain how the financial distress of one country can ignite a global financial crisis, like the one in 2008. By considering distinct asset types as layers where countries (nodes) exchange financial assets (links), \citep{del2020multiplex} have highlighted the importance of both intra-layer and inter-layer connectivity contributions to the propagation of contagions (Fig.~\ref{fig:chap_concl_financial}).

The long journey summarized in the last few pages of this work allowed us to stress the importance of multilayer modeling across a broad spectrum of disciplines, including cell biology, neuroscience, ecology and social sciences, spanning 10 orders of magnitude from a spatial scale of about $10^{-6}$~m to one of about $10^4$~m. The theoretical and computational framework presented in this work provides researchers and practitioners with a versatile and unified tool kit to shed light and gain new insights on the complexity of natural and artificial systems.

%%%%-----------------------
\newpage
\appendix  %% DO NOT DELETE
%%%%-----------------------

\section{Master Stability Function (MSF) formalism}\label{sec:MSF appendix}

To test the stability of the synchronized state $\textbf{s}$ we study how the perturbation error $ \delta \textbf{x}_i(t) = \textbf{x}_i(t) - \textbf{s}(t) $ evolves. By assuming small perturbations, we can write the variational equation for $\delta \textbf{x}_i$:
\begin{equation}
\delta \textbf{x}_i = J\textbf{F}(\textbf{s}) \delta \textbf{x}_i - 
\sigma J\textbf{H}(\textbf{s}) \sum_{j=1}^{N} L_{ij} \delta\textbf{x}_j
\label{syn_eq_3}
\end{equation}
where $J\textbf{F}$ and $J\textbf{H}$ are the Jacobian of $\textbf{F}$ and $\textbf{H}$, respectively.

To find the solution of Eq. (\ref{syn_eq_3}) we have to project $ \delta \textbf{x}$ into the eigenspace formed by the eigenvectors of the Laplacian matrix $\textbf{L}$, obtaining a decomposition of the time evolution of perturbation error into $N$ decoupled eigenmodes:

\begin{equation}
 \dot{\boldsymbol{\xi}_i} = [J\textbf{F}(\textbf{s})  - \sigma \lambda_i J\textbf{H}(\textbf{s})] \boldsymbol{\xi}_i , \quad i=1,...,N
\label{syn_eq_4}
\end{equation}
where $\boldsymbol{\xi}_i$ is the eigenmode associated with the eigenvalue $ \lambda_i$ of $\textbf{L}$. 
By indicating as $\Lambda_{max}$ the maximum Lyapunov exponent associated with the system of equations (\ref{syn_eq_4}), then we can write the time evolution of $\boldsymbol{\xi}$ as $|\boldsymbol{\xi}|\sim e^{\Lambda_{max}t} $ and, finally, we found that a necessary condition for the stability of a synchronized state is that $\Lambda_{max}<0$. The expression of $\Lambda_{max}$ as a function of a generic parameter $\alpha_i=\sigma \lambda_i$ is named the Master Stability Function (MSF), and it usually assumes negative values in an interval $\alpha \in (\alpha_1, \alpha_2)$. It means that, for a fixed coupling strength $\sigma$, a network can reach and maintain complete synchronization only if its structure, defined by its Laplacian matrix, is such that:
 
 \begin{equation}
\alpha_1 < \sigma \lambda_2 \leq \sigma \lambda_ \leq ... \leq \sigma \lambda_N <\alpha_2.
\label{syn_eq_5}
\end{equation}
or, equivalently,

 \begin{equation}
R \equiv \frac{\lambda_N}{\lambda_2}<\frac{\alpha_2}{\alpha_1},
\label{syn_eq_6}
\end{equation}
used, for instance, in~\citep{sole2013spectral} to unravel the existence of an optimal value for the synchronizability of a multilayer system.

\section{Kuramoto model on networks}\label{sec:Kuramoto}

The first approach to model collective synchronization considers a population of coupled limit-cycle oscillators whose natural frequencies are drawn from some prescribed distribution and exert a phase-dependent influence on each others. In formulae, we can write the Kuramoto model (KM)~\citep{Strogatz2000,Arenas2008} as a system of $N$ oscillators, whose instantaneous phases $\theta_i$ are described by the equation:
\begin{equation}
 \dot{\theta_i}=\omega_i + \frac{\sigma}{N}\sum_{j=1}^{N}\sin(\theta_j-\theta_i),
\label{syn_eq_11}
\end{equation}
where $\sigma$ is the coupling constant and $\omega_i$ is the natural frequency of oscillator $i$, chosen from an unimodal distribution $g(\omega)$.

We can use a order parameter that describes the transition: it is defined as a macroscopic measure that quantifies the collective rhythm produced by the whole population:
 \begin{equation}
r(t)e^{i\Phi(t)}= \frac{1}{N}\sum_{j=1}^{N}\sin(\theta_j),
\label{12}
\end{equation}
where $\Phi(t)$ is the average phase, and $0\leq r(t) \leq1$ measures the phase coherence, where the two extreme values correspond to phase locked ($r=1)$ or incoherent oscillators.
Equation~(\ref{syn_eq_11}) is then rewritten in terms of the order parameter, and the equation for the instantaneous phase reduces to
 \begin{equation}
 \dot{\theta_i}=\omega_i + \sigma r \sin(\theta_i), 
\end{equation}
which has two types of long-term behaviours. Oscillators for which $|\omega_i|\leq \sigma r$ are phase-locked and form a mutually synchronized cluster. Oscillators with frequencies in the tails of $g(\omega)$ distribution, where $|\omega_i|> \sigma r$ holds, are drifting with respect to the synchronized cluster.
 
The Kuramoto model is then generalized on networks by including in Eq.~(\ref{syn_eq_11}) information about network connectivity:
\begin{equation}
 \dot{\theta_i}=\omega_i + \sum_{j=1}^{N} \sigma_{ij}A_{ij}\sin(\theta_j-\theta_i).
\label{syn_eq_103}
\end{equation}

\section{Transitions in multilayer systems}

\begin{table}[!ht]
\centering
\scalebox{0.9}{
\begin{tabular}{p{80mm}p{10mm}p{20mm}}
\textbf{Type of phase transition} & \textbf{Mode} & \textbf{Reference} \\
\hline
Enhanced diffusion & C & \citep{gomez2013diffusion} \\
Emergence of multiplexity & D & \citep{radicchi2013abrupt} \\
\hline
\end{tabular}
}
\caption{\label{tab:table_app1} Phase transitions (C: continuous, D: discontinuous, H: hybrid) in multilayer networks due to the algebraic, topological and simple dynamics described in Sec.~\ref{sec:diffproc}.}
\end{table}

\begin{table}[!ht]
\centering
\scalebox{0.9}{
\begin{tabular}{p{80mm}p{10mm}p{20mm}}
\textbf{Type of dynamics} & \textbf{Mode} & \textbf{Reference} \\
\hline
Synchronization  & C & \citep{Gambuzza2015} \\
Explosive synchronization with local order parameter & D & \citep{Zhang2015} \\
Synchronization of oscillators with random walks  & C \& D & \citep{nicosia2017collective}\\
Interacting diseases  & C &  \citep{Sanz2014} \\
Epidemic onset control by raising of awareness  & C &  \citep{Granell2013} \\
Explosive pandemics with no early-warning & D & \citep{danziger2019dynamic} \\
Cooperative behaviour in coupled competitive games -- social influence & C & \citep{amato2017interplay}  \\
Cooperative behaviour in Prisoner's Dilemma game & C & \citep{gomez2012evolution}  \\
Cooperative behaviour in Public Good game & C & \citep{battiston2017determinants}  \\
\hline
\end{tabular}
}
\caption{\label{tab:table_app2} Phase transitions in multilayer networks due to the simple and couple dynamics described in Sec.~\ref{sec:synchro}, Sec.~\ref{sec:coopdyn} and Sec.~\ref{sec:interdepproc}.}
\end{table}

\begin{table}[!ht]
\centering
\scalebox{0.9}{
\begin{tabular}{p{70mm}p{15mm}p{25mm}}
\textbf{Type of network} & \textbf{Mode} & \textbf{Reference} \\
\hline
Multilayer network & C & \citep{leicht2009percolation} \\
Multiplex  & D & \citep{son2012percolation} \\
Partial Multiplex & C \& D  & \citep{son2012percolation} \\
Multiplex with overlap & C, D \& H & \citep{hu2013percolation, baxter2016correlated, cellai2016message} \\
Multiplex with spatially embedded networks & C \& D & \citep{son2011percolation,bashan2013extreme,grassberger2015percolation} \\
Bond Perc. on Multiplex & C & \citep{hackett2016bond} \\
Multiplex \& Local cascades & D & \citep{buldyrev2010catastrophic} \\
Multiplex \& Non-local cascades & D & \citep{artime2020abrupt} \\
\hline
\end{tabular}
}
\caption{\label{tab:table_app3} Type of transitions for different scenarios in percolation and cascades propagation.} 
\end{table}

%footnotesize
%\begin{small}
%	\bibliography{refs}
%\end{small}

\end{document}